\documentclass[a4paper,11pt]{article}
\pdfoutput=1 

\usepackage{jhep_like}

\title{\boldmath Entanglement entropies of equilibrated pure states in quantum many-body systems and gravity}

\author{Hong Liu}
 \author{and Shreya Vardhan}
\affiliation{Center for Theoretical Physics, Massachusetts Institute of Technology,\\Cambridge, MA 02139, U.S.A}

\emailAdd{hong\_liu@mit.edu}
\emailAdd{vardhan@mit.edu}

\preprint{MIT-CTP/5227}

\abstract{We develop a universal approximation for the Renyi entropies of a pure state at late times in a non-integrable many-body system, which macroscopically resembles an equilibrium density matrix. The resulting expressions are fully determined by properties of the associated equilibrium density matrix, and are hence independent of the details of the initial state, while also being manifestly consistent with unitary time-evolution. For equilibrated pure states in gravity systems, such as those involving black holes, this approximation gives a prescription for calculating entanglement entropies using Euclidean path integrals which is consistent with unitarity and hence can be used to address the information loss paradox of Hawking.
 Applied to recent models of evaporating black holes and eternal black holes coupled to baths, it provides a derivation of replica wormholes, and elucidates their mathematical and physical origins. In particular, it shows that replica wormholes can arise in a system with a fixed Hamiltonian, without the need for ensemble averages. }

\usepackage{tensor}
\usepackage{braket}
\usepackage{comment}

\newcommand{\bid}{{\mathbf 1}}

\usepackage[normalem]{ulem}
\usepackage{amsmath}
\usepackage{enumerate}
\usepackage{epsfig}
\usepackage{mathbbol}

\def\Tr{\mathop{\rm Tr}}

\newcommand\half{{\ensuremath{\frac{1}{2}}}}

\newcommand\vev[1]{{\ensuremath{\left\langle{#1}\right\rangle}}}

\newcommand{\be}{\begin{equation}}
\newcommand{\ee}{\end{equation}}
\newcommand{\bea}{\begin{eqnarray}}
\newcommand{\eea}{\end{eqnarray}}
\newcommand{\bega}{\begin{gather}}
\newcommand{\eega}{\end{gather}}

\newcommand{\bi}{\begin{itemize}}
\newcommand{\ei}{\end{itemize}}
\newcommand{\ben}{\begin{enumerate}}
\newcommand{\een}{\end{enumerate}}
\newcommand{\bca}{\begin{cases}}
\newcommand{\eca}{\end{cases}}
\newcommand{\bln}{\begin{align}}
\newcommand{\eln}{\end{align}}
\newcommand{\bst}{\begin{split}}
\newcommand{\est}{\end{split}}
\def\ie{\begin{equation}\begin{aligned}}
\def\fe{\end{aligned}\end{equation}}
\newcommand{\bma}{\le(\begin{matrix}}
\newcommand{\ema}{\end{matrix}\ri)}
\newcommand{\bwt}{\begin{widetext}}
\newcommand{\ewt}{\end{widetext}}

\newcommand\al{{\alpha}}
\def\b{{\beta}}
\newcommand\ep{\epsilon}
\newcommand\sig{\sigma}

\newcommand\lam{\lambda}

\newcommand\de{{\ensuremath{{\delta}}}}
\newcommand\De{{\ensuremath{{\Delta}}}}

\newcommand\da{{\dagger}}

\newcommand\ov{\over}
\newcommand\ha{{\half}}

\def\le{\left}
\def\ri{\right}

\newcommand\sC{{\ensuremath{{\mathcal C}}}}

\newcommand\sE{{\ensuremath{{\mathcal E}}}}

\newcommand\sI{{\ensuremath{{\mathcal I}}}}

\newcommand\sH{{\ensuremath{{\mathcal H}}}}

\newcommand\sO{{\ensuremath{{\mathcal O}}}}
\newcommand\sP{{\ensuremath{{\mathcal P}}}}

\newcommand\sS{{\mathcal S}}
\newcommand\sT{{\mathcal T}}

\newcommand\sZ{{\mathcal Z}}

\begin{document}

\maketitle

\section{Introduction}

Consider a quantum many-body system initially in a far-from-equilibrium pure state $\ket{\Psi_0}$. 
If the system is non-integrable, it should eventually approach a thermal equilibrium, in the following sense. For times  $t \gg t_s$, where $t_s$ is a thermalization time scale,  $\ket{\Psi (t)} = U(t) \ket{\Psi_0}$ can be associated with macroscopic thermodynamic quantities such as temperature, entropy, and free energy which obey the usual thermodynamic relations, and measurements 
of generic few-body observables in $\ket{\Psi(t)}$ exhibit the same behavior as in the equilibrium density matrix $\rho^{\text{(eq)}}$ with those macroscopic parameters.


Even as the state equilibrates in the above sense, under unitary time-evolution it must go to a pure state, and therefore cannot become equal to the mixed state $\rho^{(\rm eq)}$. A natural question then is how we can tell an equilibrated pure state apart from an equilibrium density matrix. For instance, this question arises in the context of Hawking's information loss paradox~\cite{Hawking, Hawking2}, in trying to understand whether the evolution of a black hole formed from the gravitational 
collapse of a pure state is unitary. For this purpose, we can use the Renyi entropies $S_n^{(A)}$ for $n=1,2,\cdots$\footnote{Here $n=1$ corresponds to the von Neumann entropy and is the $n\rightarrow 1$ limit of \eqref{enry}.}
\be\label{enry}
S_n^{(A)} (t) = -{1 \ov n-1} \log \Tr \rho_A^n (t) , \quad
\rho_A (t) = {\rm Tr}_{\bar A} \rho (t) , \quad \rho  (t) = \ket{\Psi (t)} \bra{\Psi (t)} \ .
\ee
In a pure state, one must have for any subsystem $A$ and its complement $\bar{A}$
\be \label{unit}
S_n^{(A)} (t) = S_n^{(\bar A)} (t) , \quad n=1,2, \cdots  \ .
\ee
In an equilibrium density matrix, equation~\eqref{unit} is not satisfied. 

Other than brute-force numerical simulations of individual cases,  we currently do not have an efficient method for   calculating $S_n^{(A)} (t)$ for an arbitrary initial state $\ket{\Psi_0}$ in a general system.   For a finite-dimensional Hilbert space with no energy constraint, where we expect equilibration to the thermal state at infinite temperature, a valuable insight comes from calculating the averages $\overline{S_n^{(A)}}$ over all pure states with the Haar measure~\cite{Page_entropy,lubkin,Lloyd,nadal,sam}: 
\be \label{sehn}
\overline{S_n^{(A)}} =  \overline{S_{n}^{(\bar A)}} = {\rm min} \, (\log d_A, \log d_{\bar A})  , \quad 
n=1,2, \cdots  \ .
\ee
Here $d_A$ and $d_{\bar A}$ are respectively the dimensions of the subsystems $A$ and $\bar A$, and the above expression is exact only in the limit where one of $d_A, d_{\bar A}$ is much larger than the other.  When $d_A \ll d_{\bar{A}}$, these are equal to the entanglement entropies of a thermal state at infinite temperature. It can be checked that when the dimension of the Hilbert space is large, the standard deviation about the average~\eqref{sehn} is small. Thus the right-hand side of \eqref{sehn} should provide a good approximation for the entanglement entropies of a  typical pure state. 

This observation has yielded many important results, such as the prediction of the Page curve for black evaporation~\cite{page}. Similarly, a Haar average over pure states has been used to predict when information in a black hole can be transferred to its Hawking radiation~\cite{Hayden:2007cs}. 
The random average idea, however, cannot be directly applied in cases where we expect equilibration to a finite temperature, or in field theories and other systems with infinite Hilbert space dimension, where a canonical and calculable average such as the Haar average does not seem to exist. 
 Even for a finite-dimensional system that equilibrates to infinite temperature, in order to apply~\eqref{sehn} when we have a fixed initial state and time-evolution operator, we need to make the highly non-trivial assumption that the system can evolve to a typical pure state. It would be useful to better understand the physical basis for this assumption, and have a systematic procedure with which we can improve upon~\eqref{sehn} for such cases. 

In this paper, we develop a general approximation method for calculating  $S_n^{(A)}$ for equilibrated pure states in systems with a fixed initial state and time-evolution operator, in the limit where the effective dimension of the Hilbert space (roughly, the dimension of the accessible part of the Hilbert space from the initial state) is large. The approximation scheme, which we will refer to as the equilibrium approximation, can be applied to finite temperatures, systems with infinite Hilbert space dimension, and field theories. 

The method we propose builds on an  observation in~\cite{adam} that the Haar average in a finite-dimensional Hilbert space can be seen as a projection into a subspace $P$ of the replica Hilbert space used for computing Renyi entropies. This means that for a system with a fixed Hamiltonian, the result~\eqref{sehn} can be considered  an approximation in which at leading order one ignores the contribution from the orthogonal subspace to $P$. The generalization to finite temperatures or systems with infinite-dimensional Hilbert spaces then boils down to identifying the appropriate subspace of the replica Hilbert space to project into to obtain the leading contribution. Equivalently, the method can be thought of as identifying the contributions from the most important subset of configurations in the Lorentzian path integrals for 
the Renyi entropies.  Since we drop certain well-defined contributions to the time-evolved quantities in making this approximation, it can in principle be systematically improved by adding these contributions back.  We also develop a self-consistent criterion to demonstrate the validity of the approximation. 

Some important general features of the results from our approximation method are:
\ben

\item The expressions for $S_n^{(A)} (t)$ in an equilibrated pure state are time-independent, and can be expressed solely in terms of partition functions and entropies of an equilibrium density operator $\rho^{(\rm eq)}$. They are thus independent of details of the initial state and capture the effects of equilibration. 

\item The expressions are manifestly compatible with the constraint from unitarity in~\eqref{unit}. 

\item While $S_n^{(A)} (t) $ are defined in terms of Lorentzian path integrals, the approximate expression for $S_n^{(A)} (t)$ for an equilibrated pure state can be expressed in terms of 
a sum of Euclidean path integrals when $\rho^{(\rm eq)}$ has a Euclidean path integral representation.  Each term in the sum has $n$ replicas of the Euclidean path integrals of $\rho^{(\rm eq)}$ connected in a certain way, determined by an element of the permutation group $\sS_n$. The approximation thus provides a general physical mechanism for how Euclidean path integrals associated with the equilibrium density operator can arise as the dominant subset of contributions from  intrinsically Lorentzian path integrals.

\een

Since the only input that goes into our approximation method is information about the equilibrium density matrix $\rho^{(\rm eq)}$, it can be used to obtain universal results for the entanglement entropies of a variety of quantum many-body systems when the initial state equilibrates to a given type of ensemble. We explicitly obtain the universal expressions for the microcanonical and canonical ensembles.

 One important motivation for studying the entanglement entropies of equilibrated pure states is to understand the unitarity of black hole evolution. 
Consider a situation where the initial state $\ket{\Psi_0}$  describes a star, which under time-evolution collapses to 
form a black hole. For all practical purposes, a black hole looks like a thermal state: it emits thermal radiation at a certain temperature, and has an entropy which satisfies the standard thermodynamic relations. Furthermore, correlation functions 
of a finite number of few-body observables in the black hole geometry have the same behavior as in a thermal state. 
The black hole is thus in an equilibrated pure state if the time-evolution in gravity obeys the usual rule of unitarity in quantum mechanics. The formalism we developed can thus be applied to a black hole system to 
obtain its entanglement entropies in a way that respects unitarity. In particular, item 3 above implies that one can get expressions  which are compatible with unitarity using Euclidean gravity path integrals. 

Recently, there has been important progress in understanding the unitarity of black hole evolution through  derivations of the Page curve~\cite{pen,Al1,Al2} (see also~\cite{ Al3,
Rozali:2019day, 
Akers:2019nfi, 
Chen:2019uhq,
Al4, replica_1, 
Almheiri:2019qdq,  Zhao:2019nxk, Bousso:2019ykv, Almheiri:2019psy, Chen:2019iro, Chen:2020wiq, Marolf:2020xie, Verlinde:2020upt, Giddings:2020yes, eth_wormholes, Balasubramanian:2020hfs, Gautason:2020tmk, Anegawa:2020ezn, Hollowood:2020cou, Krishnan:2020oun, Banks:2020zrt, Geng:2020qvw, page_void, dong, karl, gomez, myers, ensemble, Piroli:2020dlx, Engelhardt:2020qpv}).\footnote{See also~\cite{bh_review}  for a review and~\cite{Maldacena:2020ady,Liu:2020rrn} for non-technical reviews.} 
In particular, in order to obtain Renyi and von Neumann entropies compatible with unitarity, 
one needs to include  certain ``island'' contributions~\cite{Al2} in the quantum extremal surface prescription \cite{netta}. In models for an evaporating black hole in \cite{replica_1} and for an eternal black hole coupled to a bath in~\cite{Almheiri:2019qdq}, these island contributions were derived by including Euclidean gravity path integrals with replica wormholes in the calculation. By applying the equilibrium approximation to these models, we provide a derivation of the replica wormholes introduced in these references, explaining how such Euclidean configurations emerge from Lorentzian time-evolution at late times, and why they lead to answers which are consistent with unitarity constraints.  Our discussion also clarifies an issue raised in \cite{replica_1}, due to which the authors there suggested that an averaging procedure may be necessary to explain the results from including replica wormholes. We show that no average over theories is needed, and that the issue can be resolved within the framework of the equilibrium approximation. Applied to more general holographic systems, our results predict new bulk geometries that must be summed over in the calculation of Renyi entropies for states with black holes. 

We also explore the underlying physical mechanism in the Heisenberg evolution of operators that underlies the emergence of the equilibrium behavior of entanglement entropies. In~\cite{page_void}, we showed that for the second Renyi entropy, the random void distribution conjectured for quantum chaotic systems in~\cite{void} leads to the right-hand side of~\eqref{sehn}. Using the equilibrium approximation, we show that the behavior of the Renyi entropies can be seen as a special example of a general behavior of operator growth in chaotic systems, and derive a higher-moment generalization of the random void distribution. 

The plan of the paper is as follows. In section \ref{sec:2}, we develop the equilibrium approximation, discuss its justification and universal consequences, and apply it to a variety of equilibrated pure states. In section \ref{sec:holo}, we apply the approximation to gravity systems with holographic duals, explain how replica wormholes emerge from it, and make comments on the need for averaging based on this derivation. In section \ref{sec:rvd}, we explain how equilibration to infinite temperature can be understood in terms of the random void distribution. In section \ref{sec:conc}, we discuss the applicability of the equilibrium approximation to observables other than the Renyi entropies,  and mention some open questions.

\section{Universal behavior of entanglement entropies in equilibrated pure states}
\label{sec:2}

In this section, we present an approximation for the entanglement entropies of  equilibrated pure states, which can be expressed in a simple, universal form, and applies to a variety of systems. We first explain the physical reasoning behind this approximation and its mathematical structure, and then examine its consequences for a variety of equilibrated pure states.

\subsection{Equilibrated pure states}
\label{sec:examples}

Consider a quantum system in some far-from-equilibrium pure state $\ket{\Psi_0}$ at $t=0$. At 
time $t$, it evolves to
\be \label{ste}
\ket{\Psi} = U \ket{\Psi_0},
\ee
where $U$ is the time-evolution operator for the system. 
 For now, we will assume the system is compact, so that there exists a finite equilibration time scale $t_s$ such that for $t \gg t_s$, macroscopic properties of $\ket{\Psi}$ can be well-approximated by some equilibrium density operator $\rho^{\rm (eq)}$.\footnote{The precise definition of time scale $t_s$ will not be important for our purpose. Since we are interested in quantum-informational quantities in chaotic systems, a likely candidate for $t_s$ is the scrambling time.  In Sec.~\ref{sec:uncompact}, we will discuss uncompact 
systems. 
} We will refer to  $\ket{\Psi}$ as an equilibrated pure state.\footnote{We caution that the fact that  $\ket{\Psi}$ resembles an equilibrium density operator $\rho^{\rm (eq)}$
macroscopically does not mean that $\ket{\Psi} \bra{\Psi}$ is close to $\rho^{\rm (eq)}$ by measures like the trace distance. 
}~\footnote{A special situation is when $\ket{\Psi_0}$ is an energy eigenstate of a chaotic system, which does not evolve with time, but exhibits thermal behavior as postulated by the eigenstate thermalization hypothesis (ETH)~\cite{deutsch, sredniki}.} 


We can write the equilibrium density matrix $\rho_{\rm eq}$ in the form
\be \label{heno}
\rho^{\rm (eq)} = {1 \ov Z (\al)} \sI_\al, \qquad Z(\al)= \Tr \sI_\al 
\ee
where $\sI_\al$ is an un-normalized density operator, and $\al$ is a set of equilibrium parameters such as  temperature, chemical potential, and so on, which can be determined from the expectation values of conserved quantities in $\ket{\Psi_0}$. 
We require that $\sI_\al$ commute with the evolution operator, i.e. 
\be \label{pje1}
U \sI_\al U^\da = \sI_\al , \qquad  U^\da \sI_\al U = \sI_\al,
\ee
which can be viewed as a requirement for $\rho^{\rm (eq)}$ to be an equilibrium state.

Here are some specific examples of $\sI_\al$:  
\ben 

\item The system has a finite-dimensional Hilbert space, and there is no constraint on accessible states from $\ket{\Psi_0}$. 
 In this case, the associated equilibrium state $\rho_{\rm eq}$ does not need to be labelled with any parameters $\al$, and $\sI$ is the identity operator,
\be \label{infT}
\sI = \bid , \quad Z = d
\ee
where $d$ is the dimension of the Hilbert space. Below, we will refer to this case as the infinite temperature case.

\item  The system has a time-independent Hamiltonian $H$ with energy eigenstates $\ket{n}$, and  $\ket{\Psi_0}$ (and hence $\ket{\Psi}$)  involves only energy eigenstates localized in a narrow energy band $I= (E-\De E, E + \De E)$. In this case,  
\be \label{mien}
\sI_E = \sum_{E_n \in I} \ket{n} \bra{n}, \quad Z (E) = \Tr \sI_E = N_I,  
\ee
where $N_I$ is the number of energy eigenstates in the energy band $I$. 

\item The system has a time-independent Hamiltonian $H$, and $\ket{\Psi_0}$ (and hence $\ket{\Psi}$) involves energy eigenstates with a broader range of energies. In this case,     
\be\label{jen1}
\sI_\b =  e^{- \b H}  , \quad  Z (\b ) = \Tr e^{- \b H}, 
\ee
where the inverse temperature $\b$ is determined by 
requiring $\sI_\b$ has the same energy as $\ket{\Psi_0}$, 
\be 
{1 \ov Z (\b)} \Tr \le(H  e^{- \b H} \ri) = \vev{\Psi|H |\Psi} =\vev{\Psi_0|H |\Psi_0}  \ .
\ee

\een

For~\eqref{infT} and~\eqref{mien}, $\sI_\al$ is a projector, $\sI_\al^2 = \sI_\al$, but this is not true for~\eqref{jen1}. 
In~\eqref{mien}--\eqref{jen1}, one can view $\sI_\al$ as an ``effective identity operator'' defining the accessible part 
of the Hilbert space, and the partition function $Z (\al)$ as the corresponding ``effective dimension.''
It is clear that each choice of $\sI_{\al}$ in~\eqref{infT}--\eqref{jen1} satisfies the requirement~\eqref{pje1} of invariance under $U$. 


\subsection{Renyi entropies as transition amplitudes in a replicated Hilbert space}
\begin{figure} [!h] 
\centering
\includegraphics[width=10cm]{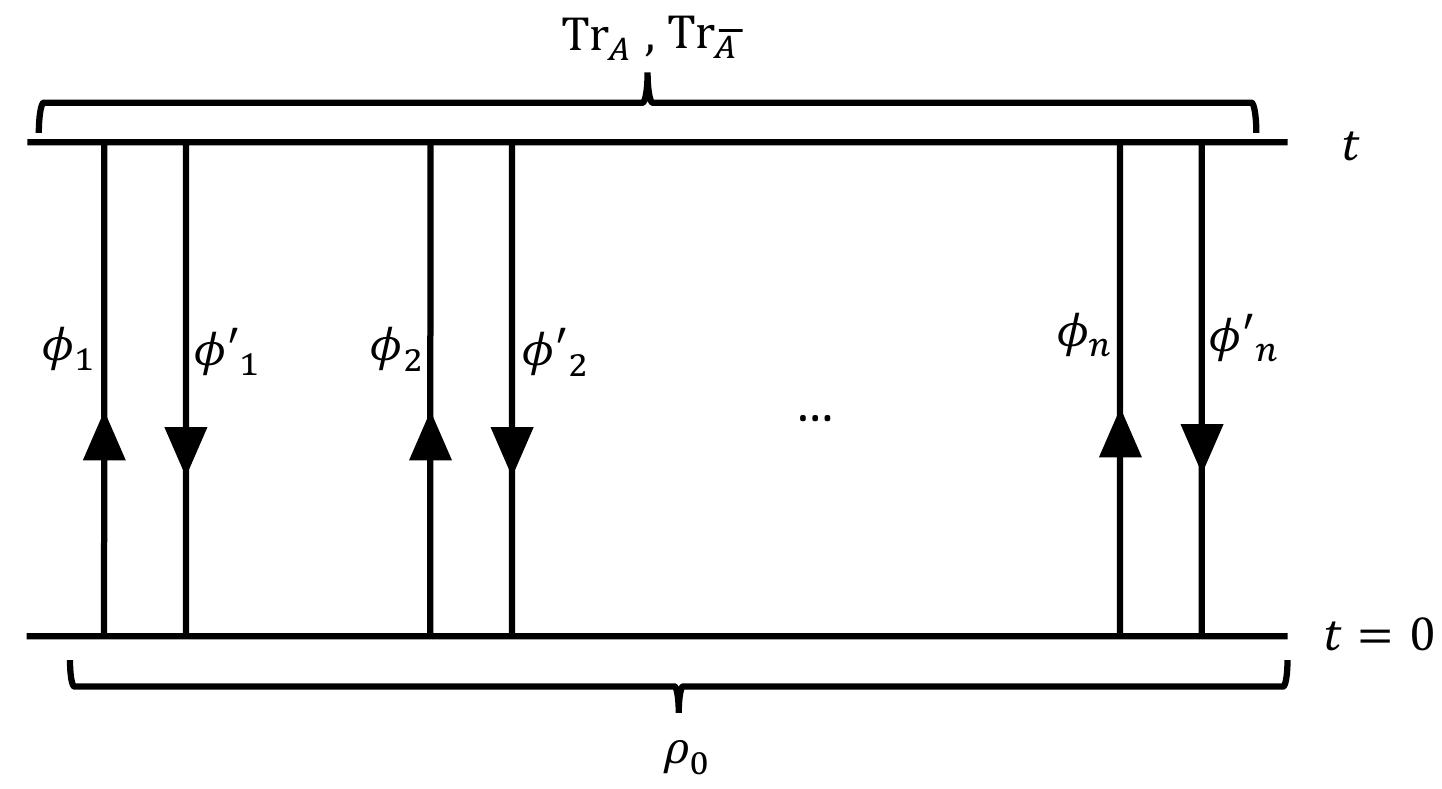}
\caption{The path integrals~\eqref{jh} for~\eqref{rnyi} involve $2n$ integration contours, with those for $U$'s going forward in time and those for $U^\da$ backward in time. $\rho_0$ provides the initial conditions while the contractions implied by the traces in~\eqref{rnyi} define the final conditions for the path integrals. 
}
\label{fig:path}
\end{figure}

We are interested in quantum-informational properties of $\ket{\Psi}$ at time scales $t \gg t_s$. The $n$-th Renyi entropy with respect to a subsystem $A$ is given by 
\be \label{rnyi}
\sZ_n^{(A)} = e^{- (n-1) S_n^{(A)}} =  {\rm Tr}_A  \rho_A^n =   {\rm Tr}_A \le({\rm Tr}_{\bar A} U \rho_0 U^\da \ri)^n  , \quad \rho_0= \ket{\Psi_0} \bra{\Psi_0} \ . 
\ee
Recall that in a quantum system with evolution operator $U$,  the transition amplitude from an initial state $\ket{\Psi_0}$ to a final state $\ket{\Psi_f}$ has the path integral representation 
\be 
\braket{\Psi_f| U | \Psi_0} =\int D \psi D \chi \, \Psi_f^* [\psi] \, \Psi_0 [\chi] \int_{\phi(0)= \chi}^{\phi(t) = \psi} D\phi(t') ~e^{i S[\phi(t')]} 
\ee
where $\phi (t)$ collectively denotes the dynamical variables of the system and $S[\phi (t)]$ is the corresponding action.\footnote{Note that $\phi$ can also collectively represent the dynamical fields in a field theory, for which case we have suppressed spatial dependence.} 
Equation \eqref{rnyi} contains $2n$ $U$'s and thus can
be written in terms of path integrals over $2n$ time integration contours,
\ie \label{jh}
& \sZ_n^{(A)} = \int \prod_{i=1}^n \le( D \psi_{i} \, D  \psi'_{i}  \,
 \de (\psi'_{iA} - \psi_{(i+1)A}) \de (\psi_{i \bar A} -  \psi'_{i \bar A}) \ri) \\
 &\times \int \prod_{i=1}^n D \chi_{i} \, D \chi'_{i} \, \rho_0 [\chi_i, \chi_i']  \, \prod_{i=1}^n \bigg(\int_{\chi_i , \chi'_{i}}^{\psi_i, \psi'_{i}} D \phi_i (t) D \phi'_{i}  (t)\bigg) \,
 \exp \le(i \sum_{i=1}^n (S [\phi_i] - S [\phi'_{i}]) \ri)    \ . 
\fe
Here $\phi_i, \phi'_{i}$ are respectively associated with the $i$-th contours going forward 
and backward in time, as in Fig.~\ref{fig:path}.  $\{\chi_i ,  \chi'_{i}\}$ and $\{\psi_i, \psi'_{i}\}$ denote respectively the initial and final values of the dynamical fields. $\psi_{iA}$ denotes the value of $\psi_i$ restricted to subsystem $A$, and $\psi_{(n+1)A} = \psi_{1A}$. The initial state $\rho_0$ determines the initial conditions for the integrations through its ``wave function'' $\rho_0 [\chi_i, \chi_i'] $ while 
the contractions dictated by $\Tr_A$ and $\Tr_{\bar A}$ determine the final conditions for the integrals. 

From~\eqref{jh}, and also intuitively from Fig.~\ref{fig:path}, we can view~\eqref{rnyi} as a transition amplitude 
in a new ``replica'' quantum system consisting of $2n$ copies of the original system, with an evolution operator given by $(U\otimes U^{\dagger})^n$.\footnote{Such expressions have been used, for instance, in recent studies of random unitary circuits \cite{nahum1, frank, nahum2}.} This way of viewing $\sZ_n^{(A)}$ will 
be particularly convenient for our discussion below. To write down the explicit form for $\sZ_n^{(A)}$ in the replica system,  let us first introduce some notations.


Suppose $\sH$ is the Hilbert space of the original system. The Hilbert space of the replica system can be taken to be 
$(\sH \otimes \sH)^n$.
If we use an orthonormal basis $\{\ket{i}\}$ for the first copy of $\sH$ in each $\sH \otimes \sH$, then it is convenient to use a basis $\{\ket{\bar i}\}$ for the second copy, defined as 
\be 
\ket{\bar i} = T \ket{i}, \quad \vev{\bar i |\bar j} = \vev{j|i} = \de_{ij} , \quad
U_{ij} = \vev{i |U |j} , \quad \braket{\bar i |U^{\dagger} |\bar j}  = U_{ij}^*   
\ee
where $T$ is an anti-unitary operator such that $T U T^{-1} = U^{\dagger}$. For example, $T$ can be taken to be the time-reversal operator in systems with time-reversal symmetry, or $\sC\sP\sT$ in more general systems. 
A basis for $(\sH \otimes \sH)^{n}$
can then be written as $\{\ket{i_1 \bar i_1' i_2 \bar i_2' \cdots i_n \bar i_n'}\}$. 

For any operator $\sO$ acting on $\sH$, we can define a set of $n!$ states $\ket{\sO, \sig} \in (\sH \otimes  \sH)^n$, where $\sig$ is an element of the permutation group $\sS_n$ of $n$ objects, 
\bega \label{aa1}
\vev{i_1 \bar i_1' i_2 \bar i_2' \cdots i_n \bar i_n'| \sO, \sig} = \sO_{i_1 i'_{\sig (1)}} \sO_{i_2 i'_{\sig (2)}} \cdots 
\sO_{i_n i'_{\sig (n)}} , \quad \sO_{ij} = \vev{i |\sO|j} 
\end{gather} 
and $\sig (i)$ denotes the image of $i$ under $\sig$. The states associated with the identity operator $\bid$ are simply denoted as $\ket{\sig}$ 
\bega \label{aa2}
\vev{i_1 \bar i_1' i_2 \bar i_2' \cdots i_n \bar i_n'| \sig} = \de_{i_1 i'_{\sig (1)}} \de_{i_2 i'_{\sig (2)}} \cdots 
\de_{i_n i'_{\sig (n)}}  \ .
\end{gather} 
We note that when $\sH$ is infinite-dimensional, $\ket{\sig}$ are not normalizable and should be viewed as 
formal definitions rather than genuine states in the replica Hilbert space. 

In the discussion below, we will often use the following properties of the inner products among these states:
\ie \label{imy}
\vev{\sO_1, \tau|\sO_2 , \sig} & = \vev{\sO_1, \lam \tau|\sO_2 , \lam \sig} 
 = \vev{\sO_1, \tau \lam |\sO_2 , \sig \lam} \cr
& = \vev{\sO_1, \sig|\sO_2 , \tau} = \vev{\sO_1, \tau^{-1}|\sO_2 , \sig^{-1}} 
 \cr
& =  \Tr \le( \sO_1^\da \sO_2  \ri)^{n_1}  \cdots  \Tr \le(\sO_1^\da \sO_2 \ri)^{n_k} ,
 \quad \forall \sig, \tau, \lam \in \sS_n 
\fe
where $k$ is the number of cycles in the permutation
$\sig \tau^{-1}$ and $n_s, ~s=1, 2, \cdots, k,$ are the lengths of the cycles. 

For $\sH = \sH_A \otimes \sH_{\bar A}$, the associated replica Hilbert space inherits a tensor product structure, and we can define the corresponding states for each tensor factor. 

Using the above notation, we can now write~\eqref{rnyi} as 
\be \label{repl}
\sZ_n^{(A)} = \braket{\eta_A \otimes e_{\bar A}| (U \otimes U^{\dagger})^n| \rho_0,e}
\ee
where $e$ is the identity element of $\sS_n$ and $\eta$ is the element $(n, n-1, \cdots 1)$. 
$\ket{\eta_A}$ is in the space $(\sH_A \otimes  \sH_A)^n$ and similarly $\ket{e_{\bar A}} \in (\sH_{\bar A} \otimes  \sH_{\bar A})^n$. The equivalence between~\eqref{repl} and~\eqref{rnyi} can be checked by inserting complete sets of states 
of $\sH$ and $(\sH \otimes \sH)^{n}$ respectively in~\eqref{rnyi}  and~\eqref{repl}, and using~\eqref{aa1}--\eqref{aa2}.


\subsection{Proposal for a general equilibrium approximation} \label{sec:sche}

Consider the special set of configurations in the path integral~\eqref{jh} that satisfy
\be \label{een}
\phi_i (t) = \bar \phi_{\sig (i)} (t) , \quad i=1, \cdots, n, \quad \sig \in \sS_n \ .
\ee
For such configurations, the phase factor in the exponent of~\eqref{jh} vanishes identically.\footnote{The importance of such configurations has been pointed out earlier in \cite{adam}.}  
Heuristically, one may imagine that for sufficiently late times, the contributions from configurations of $\phi_i, \bar \phi_i$ 
which do not satisfy~\eqref{een} will generically lead to large oscillations of the integrand in the path integral, so that the integral will evaluate to a small value. It is thus natural to expect that configurations satisfying~\eqref{een} dominate. An important feature of the configurations~\eqref{een} is that they lead to contributions which are independent of $t$, which also makes it natural 
for them to describe the behavior of the system after reaching macroscopic equilibrium at $t \gg t_s$.


However, naively setting~\eqref{een} in the path integrals (which also sets $\chi_i = \chi'_{\sig (i)}$ and 
$\psi_i =  \psi'_{\sig (i)}$) leads to divergences, for example even in a simple system of two harmonic oscillators. 
Physically, the divergences come from the fact that~\eqref{een} includes unphysical configurations of arbitrarily large energies. 
Mathematically, such a procedure corresponds to replacing $(U\otimes U^{\dagger})^n$ in \eqref{repl} with a projector onto the set of states $\ket{\sig}, \sig \in \sS_n$ associated with the identity operator $\bid$. In a system of an infinite dimensional Hilbert space, $\ket{\sig}$ is not normalizable.

We will now present a systematic procedure which incorporates the idea of matching the configurations on forward evolutions with those on backward evolutions, but avoids such divergences by restricting to configurations which are accessible to the evolution.
Our discussion does not depend on whether the system has a finite or infinite dimensional Hilbert space. Our proposal builds on an important observation in~\cite{adam}, that the Haar average in a finite-dimensional Hilbert space can be seen as a projection 
into the set of states $\ket{\sig}$ associated with the identity operator. This observation can be seen as the infinite-temperature case of our discussion.

Let us first introduce some further notation. Consider the set of states in $(\sH \otimes  \sH)^n$ associated with the effective identity operator $\sI_\al$ of~\eqref{heno}, 
\be \label{al4}
\vev{i_1 \bar i_1' i_2 \bar i_2' \cdots i_n \bar i_n'| \sI_\al, \sig} =   \prod_{a=1}^n     \vev{i_a  \le| \sI_\al \ri|i_{\sig(a)}' }, \quad \sig \in \sS_n
\ee
which satisfy (using~\eqref{imy}) 
\be \label{uen}
  \vev{\sI_\al, \tau |\sI_\al , \sig} = Z_{2n_1} \cdots Z_{2n_k} , \quad
\vev{\sI_\al, \sig |\sI_\al , \sig} = Z_{2}^n,
\qquad Z_n \equiv {\rm Tr} \sI_\al^n  ,
\ee
where $k$ is the number of cycles in the permutation
$\sig \tau^{-1}$ and $n_s, s=1, 2, \cdots, k$ are the lengths of the cycles. 
For notational simplicity, we have suppressed the $\al$-dependence in $Z_n$, and $Z(\al)$ in~\eqref{heno} is now referred to as $Z_1$. 
It is convenient to define the metric for the normalized states corresponding to $\ket{\sI_{\al}, \sigma}$,  
\be \label{metrc} 
g_{\tau \sig} \equiv  {\vev{\sI_\al, \tau |\sI_\al , \sig}  \ov \le(\vev{\sI_\al, \sig |\sI_\al , \sig} \vev{\sI_\al, \tau |\sI_\al , \tau}\ri)^\ha } = {Z_{2n_1} \cdots Z_{2n_k} \ov Z_2^n}.  
\ee
Note that $g_{\tau \sig}$ is a symmetric matrix,  and from~\eqref{imy} it is invariant under simultaneous multiplication of an element $\lam \in \sS_n$ on $\sig$ and $\tau$ from the left or  
the right, as well as under simultaneously taking inverses of $\sig$ and $\tau$. Moreover, we can show that $g^{\tau \sig}$, the inverse of $g_{\tau \sig}$, is also invariant under each of these operations.

The projector onto the set of 
states spanned by $\{\ket{\sI_\al, \sig}\}$ then has the form 
\be 
P_{\al} =  {1 \ov Z_2^n} \sum_{\sig, \tau} g^{\sig \tau} \ket{\sI_\al , \sig} \bra{\sI_\al, \tau} \ .
\ee
One key physical input in our approximation is that due to the invariance~\eqref{pje1} of $\sI_{\al}$ under the action of $U$, the states $\ket{\sI_\al, \sig}$ are invariant under the action of the time-evolution operator $(U\otimes U^{\dagger})^n$ in the replica Hilbert space,  
\be \label{invc}
(U \otimes U^{\dagger})^n \ket{\sI_\al, \sig} = \ket{\sI_\al,\sig}  , \quad  (U^\da \otimes U)^n \ket{\sI_\al, \sig} = \ket{\sI_\al,\sig}  
\ee
which in turn implies that  
\be 
(U \otimes U^{\da})^n  P_{\al} = P_{\al} (U \otimes U^{\da})^n = P_{\al} \ .
\ee

Then, decomposing the identity on $(\sH \otimes \sH)^n$ as 
\be \label{spl}
\bid = P_{\al} + Q , \qquad P_{\al} Q = Q P_{\al} = 0, \qquad Q^2 = Q
\ee
where $Q$ is the orthogonal projector of $P_{\al}$, we can rewrite the transition amplitude expression for the $n$-th Renyi entropy in \eqref{repl} as 
\ie 
\sZ_{n}^{(A)} & = \vev{\eta_A \otimes e_{\bar A}| (P_{\al} + Q) (U \otimes U^{\da})^{n} (P_{\al}+Q) |\rho_0, e}  \cr
& = \vev{\eta_A \otimes e_{\bar A} |P_{\al} |\rho_0, e} + \vev{\eta_A \otimes e_{\bar A}|Q (U \otimes U^{\da})^{n} Q |\rho_0, e} \cr
& = \sZ_{n,P}^{(A)} +\sZ_{n,Q}^{(A)} \ .
\label{may}
\fe 

Our proposal is that for $t \gg t_s$, for a chaotic system with a large effective dimension $Z_1$, 
 $\sZ_{n,Q}^{(A)}$ is small compared to $\sZ_{n,P}^{(A)}$  and can be ignored, so that 
\be \label{fen}
\sZ_{n}^{(A)} \approx \sZ_{n,P}^{(A)}  =  {1 \ov Z_2^n}  \sum_{\sig, \tau} g^{\tau\sig}  \vev{\eta_A \otimes e_{\bar A} | \sI_\al , \tau} \vev{\sI_\al, \sig|\rho_0,e}, \quad n=1,2,3, \cdots  \ .
\ee
We will refer to this approximation as the equilibrium approximation in the following discussion. In next subsection, we discuss a justification for the approximation.

We have thus replaced the time-evolution operator $(U\otimes U^{\da})^n$ of the replica Hilbert space with the projector onto the set of states spanned by $\{ \ket{\sI_{\al}, \sigma}\}$ in the expression for $\sZ_n^{(A)}$, which can be seen as a realization of the heuristic idea discussed at the beginning of this subsection. Equation~\eqref{fen} is time-independent, which is consistent with the proposal that it captures quantum-informational properties of 
pure state $\ket{\Psi}$ after it has reached a macroscopic equilibrium.\footnote{A pure state never really stops evolving and one expects occasional deviations from the ``equilibrium values''~\eqref{fen}. However, for a macroscopic systems such instances should be very rare. See e.g.~\cite{popescu2}.} We note that each of the steps in equations \eqref{invc}-\eqref{may} applies to any choice of $\sI_{\al}$ that is invariant under the action of $U$. The approximation that $Z_{n, Q}^{(A)}$ is negligible should, however, only be valid for $\sI_{\al}$ chosen such that it corresponds to the late-time equilibration of $\ket{\Psi_0}$. 
 
In particular, we can obtain a self-consistency condition from considering the implication of the approximation \eqref{fen} for $n=1$. In this case, equation~\eqref{rnyi} simply reduces to the trace of $\rho_0$ and we should find 
$\sZ_{n}^{(A)} =1$.  For $n=1$, the only permutation is the identity $e$, and \eqref{fen} then has the form
\be 
\sZ_1^{(A)}  \approx {1 \ov Z_2} \vev{e_A \otimes e_{\bar A}|\sI_\al, e} \vev{\sI_\al, e| \rho_0, e} = 
{Z_1 \ov Z_2} \Tr (\rho_0 \sI_\al)
\ee
where we have used~\eqref{uen} and~\eqref{imy}. 
For this result to reproduce the normalization of $\rho_0$, we thus need
\be \label{r11}
\Tr (\rho_0 \sI_\al)  = {Z_2 \ov Z_1} \ .
\ee 
We will impose equation~\eqref{r11} as a  self-consistency condition for $\ket{\Psi_0}$ to evolve to a state $\ket{\Psi}$ which resembles ${1 \ov Z_1} \sI_\al$ macroscopically.\footnote{From~\eqref{pje1}, we have $\Tr (\rho_0 \sI_\al)  = \Tr (\rho  \sI_\al)$, where $\rho = \ket{\Psi} \bra{\Psi}$ is the equilibrated pure state. Equation~\eqref{r11} is satisfied if {\it inside} the trace $\rho$ can be approximated by ${1 \ov Z_1} \sI_\al$.} 
Let us now use the self-consistency condition \eqref{r11} in our approximation for $Z_n^{(A)}$. First, using~\eqref{imy}, we find that 
\be \label{hem}
\vev{\sI_\al, \sig|\rho_0,e} =  \le(\Tr (\rho_0 \sI_\al)\ri)^n 
= {Z_2^n \ov Z_1^n},
\ee
where we have used that since $\rho_0$ is a pure state, 
\be 
 \Tr \le(\sI_\al  \rho_0  \ri)^{k} = \le(\vev{\Psi_0 \bigl| \sI_\al  \bigr|\Psi_0} \ri)^{k} = \le(\Tr (\rho_0 \sI_\al)\ri)^k \ .
\ee
Equation~\eqref{fen} can then be written in a form independent of the initial state, 
\bega \label{fen1}
\sZ_{n}^{(A)} \approx {a \ov Z_1^n} \sum_{\tau \in \sS_n}   \vev{\eta_A \otimes e_{\bar A} | \sI_\al , \tau} ,
 \quad
a = \sum_{\sig} g^{\tau\sig}  = \sum_{\sigma} g^{e\, (\sigma\tau^{-1})} = \sum_{\sigma} g^{e \sigma} \ .
\end{gather}

From our usual intuition about statistical systems and the comments below \eqref{nel1} about the standard deviation of the Haar average in the infinite temperature case, we expect more generally that to suppress potential contributions from $\sZ_{n,Q}^{(A)}$ in~\eqref{may}, we should always consider the regime that the ``effective dimension'' $Z_1 = \Tr \sI_\al$ is large. 
Since $Z_n$ contains only a single trace, we should then have  
\be\label{hebl}
{Z_n \ov Z_1^n} \lesssim Z_1^{1-n} \ll 1 , \quad n=2,3, \cdots    \ .
\ee
For the choices of $\sI_{\al}$ in~\eqref{infT} and~\eqref{mien}, the first relation in~\eqref{hebl} is in fact an equality.  
For~\eqref{jen1}, we expect the following general behavior 
\be \label{kjn}
Z_1 = \Tr  e^{-\b H} = e^{N g(\b)}  \quad  \to \quad {Z_n \ov Z_1^n}  = e^{N (g(n \b) - n g (\b))} \ll Z_1^{1-n} \ll 1
\ee
where $N$ is proportional to number of degrees of freedom\footnote{The volume of the system is also included in $N$.} and is large, and $g (\b) > 0$ is an $O(1)$ monotonically decreasing function of $\b$.
From~\eqref{metrc} we then find that 
\be \label{idme}
g_{\tau \sig} = \de_{\tau \sig} + O(Z^{-1}_1)  \quad \to \quad a = 1 + O(Z^{-1}_1)  \ .
\ee
We then obtain our final general expression for the approximation to $Z_n^{(A)}$ for an equilibrated pure state:
\be \label{fen11}
\sZ_{n}^{(A)} 
\approx  \sum_{\tau \in \sS_n} \sZ_{n}^{(A)}  (\tau), \qquad 
\sZ_{n}^{(A)}  (\tau) = {1 \ov Z_1^n}    \vev{\eta_A \otimes e_{\bar A} | \sI_\al , \tau} \ .
\ee

We will examine the mathematical structure of \eqref{fen11} further in subsection \ref{sec:gen}. In Sec.~\ref{sec:unit} we show that while~\eqref{fen11} is expressed solely in terms of properties of equilibrium density operator $\sI_\al$,  the unitarity constraint~\eqref{unit} is maintained. The resulting physical properties are discussed in Sec.~\ref{sec:univ}--\ref{sec:uncompact}. 


The equilibrium approximation for the case where the initial state $\rho_0$ is a mixed state is discussed in Appendix \ref{app:general_state}.

\subsection{A justification of the equilibrium approximation}\label{sec:cri}

In the infinite-temperature case~\eqref{infT} with $\sI = \bid$, the equilibrium approximation~\eqref{fen} yields results identical 
with those obtained from the Haar average over unitary matrices $U$ acting on the full system, as it can be checked that~\cite{adam} 
\be \label{nel1}
\overline{ (U \otimes U^{\da})^{n}} = P_{\sI =\bid}  \
\ee
where overline denotes the Haar average. With this interpretation, one can estimate the magnitude of $\sZ_{n,Q}^{(A)}$
by considering the variance of $\sZ_n^{(A)}$ under the Haar average
\be \label{vah}
\overline{\le(\sZ_{n,Q}^{(A)}\ri)^2} = \overline{(\sZ_{n}^{(A)})^2}-\le(\, \overline{\sZ_n^{(A)}} \, \ri)^2 = \overline{(\sZ_{n}^{(A)})^2}-(\sZ_{n, P}^{(A)})^2 \ .
 \ee
If $\overline{\le(\sZ_{n,Q}^{(A)}\ri)^2}  \ll (\sZ_{n, P}^{(A)})^2$,  then for a randomly chosen time-evolution operator $U$, $\sZ_{n,Q}^{(A)}$ has high probability of being small compared with $\sZ_{n, P}^{(A)}$, and the approximation is justified
for most systems, including those where $U$ comes from a fixed Hamiltonian.  

For a general $\sI_\al$, the projection to $P_\al$ in \eqref{may} may not emerge from an over average time-evolution operators. As we will discuss later, our results for $\sZ_{n}^{(A)}$ from the equilibrium approximation for the microcanonical and canonical ensembles agree with previous calculations based on averages over special sets of states \cite{canonical, grover}, but in these cases there does not seem to be a clear way of interpreting the averages over states as averages over time-evolution operators from physically relevant Hamiltonians.  Moreover, in the case of 
an infinite-dimensional Hilbert space, there does not exist a canonical averaging procedure over all physical time-evolution operators analogous to the Haar average.

Here we propose a self-consistent criterion for deciding whether~\eqref{fen} is a good approximation, which does not depend on whether an average exists. 
We consider the following quantity 
\be \label{varR}
\De^2 \equiv \le[\le(\sZ_{n,Q}^{(A)}\ri)^2 \ri]_{\text{eq app}} = \le[\le(\sZ_{n}^{(A)}- \sZ_{n,P}^{(A)} \ri)^2 \ri]_{\text{eq app}} =
\le[\le(\sZ_{n}^{(A)} \ri)^2 \ri]_{\text{eq app}} - \le(\sZ_{n, P}^{(A)} \ri)^2 
\ee
where subscript ``eq app'' denotes that we apply the equilibrium approximation~\eqref{fen}  to the quantity inside the bracket. We will explain more explicitly in appendix \ref{app:A} what is meant by applying the equilibrium approximation to $\le(\sZ_{n}^{(A)} \ri)^2$.
If $\De \ll \sZ_{n, P}^{(A)}$, then the approximation is self-consistent. The criterion can also be interpreted as the question of whether the equilibrium approximation is compatible with factorized form of $\le(\sZ_{n}^{(A)} \ri)^2$. When $\De   \ll \sZ_{n, P}^{(A)}$, it means 
that to a good approximation we have 
\be 
\le[\le(\sZ_{n}^{(A)} \ri)^2 \ri]_{\text{eq app}} \approx \le((\sZ_{n}^{(A)} )_{\text{eq app}}\ri)^2  \ .
\ee

We can also extend the self-consistency criterion to higher powers: for the approximation for be valid, we need 
\be 
\Delta_m \equiv \le[\le(\sZ_{n}^{(A)} \ri)^m \ri]_{\text{eq app}}- \le(\sZ_{n,P}^{(A)} \ri)^m \ll   \le(\sZ_{n,P}^{(A)} \ri)^m, 
 \label{delta_m}
\ee
i.e. the equilibrium approximation is compatible with factorization for any power, 
\be 
\le[\le(\sZ_{n}^{(A)} \ri)^m \ri]_{\text{eq app}} \approx \le((\sZ_{n}^{(A)} )_{\text{eq app}}\ri)^m \ .
\label{higher_m}
\ee

In Appendix~\ref{app:A}, we calculate~\eqref{varR} explicitly, and show that $\De$  is suppressed by at least a factor $Z_1^{-\ha}$ compared with the leading contribution from $\sZ_{n,P}^{(A)}$ in the limit of large $Z_1$. However, note that $\sZ_{n,Q}^{(A)}$ can be comparable to or larger than the next-to-leading term in $\sZ_{n,P}^{(A)}$. We further show that $\Delta_m$ is suppressed by at least a factor $Z_1^{-1}$ compared with the leading contribution from $(\sZ_{n,P}^{(A)})^m$ in the limit of large $Z_1$. Through analytic continuation, this leads to the equilibrium approximation for the Renyi entropies,  
\begin{align}
{[S_n^{(A)}]}_{\text{eq app}} &= -\frac{1}{n-1}\lim_{m\rightarrow 0} \frac{\partial [(\sZ_n^{(A)})^m]_{\rm eq~app}}{\partial  m} \cr 
& \approx -\frac{1}{n-1}\lim_{m\rightarrow 0} \frac{\partial (\sZ_{n, P}^{(A)})^m}{\partial  m} = -\frac{1}{n-1}\log(\sZ_{n,P}^{(A)})\ . 
\end{align}

Similarly, the approximation for the von Neumann entropy can be obtained by analytic continuation as 
\be 
 [S_1^{(A)}]_{\text{eq app}} = - \lim_{n\rightarrow 1}\frac{\partial \sZ_{n,P}^{(A)}}{\partial n}   \ . \label{vn_approx}
\ee

\subsection{Diagrammatic structure and path integral representation} \label{sec:gen} 

\begin{figure} [!h] 
\centering
\includegraphics[width=15cm]{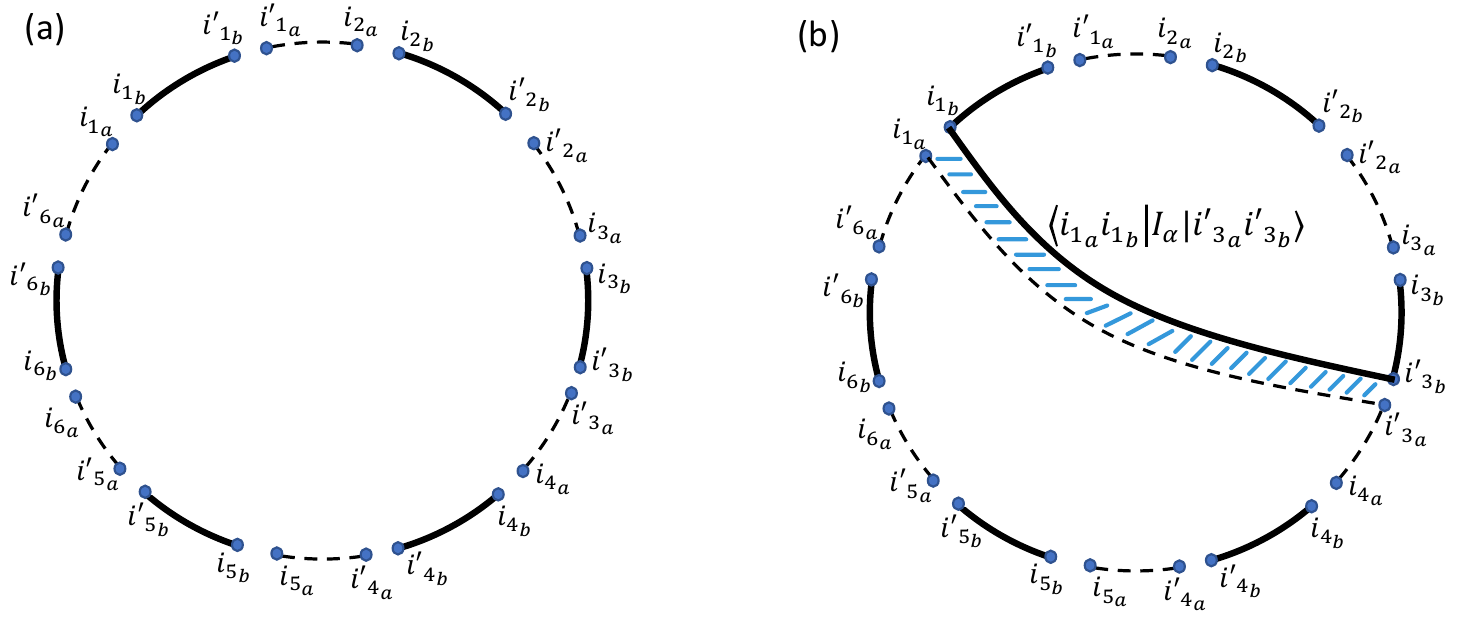}
\caption{(a) shows the ``future conditions" for each of the $\sZ_n^{(A)}(\tau)$, coming from the factor in the first line of  \eqref{zn_t}, for $n=6$. In (b), we show an example of how to connect indices in the interior of the diagram for a permutation $\tau$ such that $\tau(1)=3$, and the factor of $\braket{i_{1_a} i_{1_b}| \sI_{\al}| i'_{3_a} i'_{3_b}}$ that comes from this interior connection.}
\label{fig:ind_circ}
\end{figure}

\begin{figure} [!h] 
\centering
\includegraphics[width=15cm]{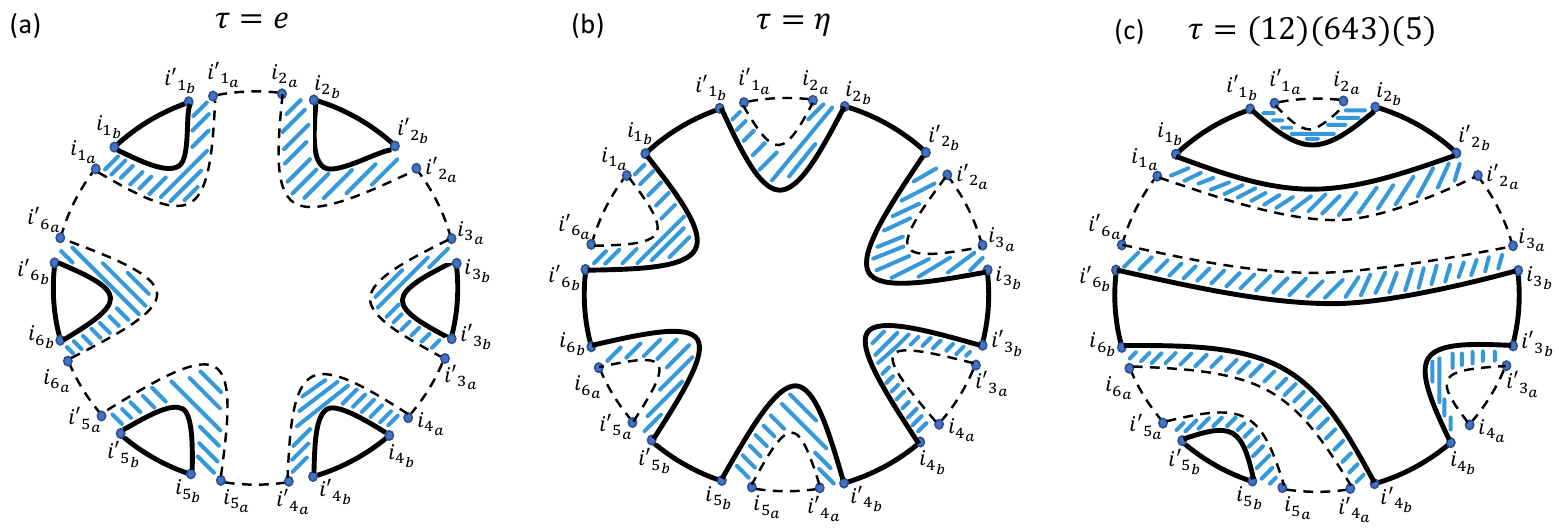}
\caption{Examples of planar diagrams corresponding to different choices of planar permutations $\tau$ that saturate \eqref{yeg}.}
\label{fig:eq}
\end{figure}

\begin{figure} [!h] 
\centering
\includegraphics[width=14cm]{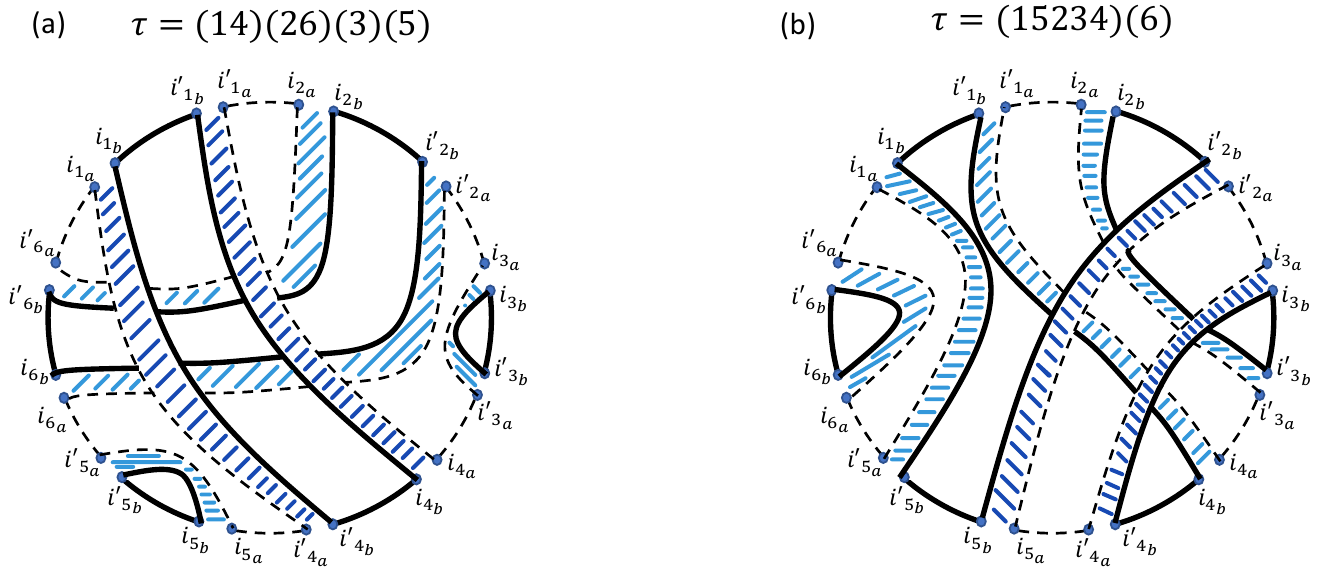}
\caption{Examples of non-planar diagrams corresponding to two choices of $\tau$ that do not saturate \eqref{yeg}.}
\label{fig:rev}
\end{figure}

We now examine more closely the mathematical structure of~\eqref{fen11}. Using~\eqref{aa1}--\eqref{aa2}, the inner product in  $\sZ_{n}^{(A)}  (\tau)$ can be written more explicitly as  
\bega 
   \vev{\eta_A \otimes e_{\bar A} | \sI_\al , \tau} = (\delta_{i_{1_a} i'_{\eta(1)_a}} \cdots \delta_{i_{n_a} i'_{\eta(n)_a}} \delta_{i_{1_b} i'_{1_b}} \cdots \delta_{i_{n_b} i'_{n_b}}) \cr
\times \braket{i_{1_a} i_{1_b}|\sI_{\al} |i'_{\tau(1)_a} i'_{\tau(1)_b}} ~...~ \braket{i_{n_a} i_{n_b}|\sI_{\al} |i'_{\tau(n)_a} i'_{\tau(n)_b}}
\label{zn_t}
\end{gather}
where $\ket{i_{k_a}}, \ket{i'_{k_a}}$  ($\ket{i_{k_b}}, \ket{i'_{k_b}}$) denote basis vectors for subsystem $A$ ($\bar A$) in the $k$-th replica. In the above expression, it should be understood that all indices $i_{k_a}, i_{k_b}, i'_{k_a}, i'_{k_b}$ with $k=1,..., n$, each of which appears twice, are summed  over. Equation~\eqref{zn_t} can be given a diagrammatic representation as in Fig.~\ref{fig:ind_circ}--\ref{fig:rev}. The expression in the parentheses of the first line corresponds to the ``future conditions" indicated in Fig.~\ref{fig:ind_circ} (a), where the Kronecker deltas between indices in $A$ are indicated with dashed lines, while those in $\bar{A}$ are indicated with solid lines. These future conditions are independent of $\tau$, while the $\tau$-dependent factors in the second line of~\eqref{zn_t} correspond to how the indices should be connected to each other in the interior of the diagram. We connect each $i_{k_a}$ to $i'_{\tau(k)_a}$ with a dashed line, and each $i_{k_b}$ to $i'_{\tau(k)_b}$ with a solid line, and read off a factor of $\braket{i_{k_a} i_{k_b}|\sI_{\al} |i'_{\tau(k)_a} i'_{\tau(k)_b}}$ from each such interior connection to obtain \eqref{zn_t}. An example of an interior connection is shown in Fig.~\ref{fig:ind_circ}(b), and some examples of diagrams associated with different $\tau$ are given in Fig.~\ref{fig:eq}-\ref{fig:rev}.

In particular, each loop of solid lines in a diagram for a given permutation corresponds to a  trace in $\bar A$, and each loop of dashed lines to a trace in $A$.  Qualitatively, we expect that a trace in $A$ ($\bar A$) should yield a factor which is of the order of the effective dimension of subsystem $A$ ($\bar A$). The number of $A$ loops in the diagram associated with $\tau$ according to our prescription above is given by $k (\eta^{-1} \tau)$, while the number of $\bar A$ loops is $k(\tau)$, where $k(\sig)$ denotes the number of cycles for a permutation $\sig$. Thus we expect that 
\be \label{bdm}
 \sZ_{n}^{(A)}  (\tau)  \sim {1 \ov Z_1^n} \, d_A^{k (\eta^{-1} \tau)} d_{\bar A}^{k (\tau)} 
\ee
where $d_A, d_{\bar A}$ denote the effective dimension of subsystems $A$ and $\bar A$ respectively. Note that one should view~\eqref{bdm} as a heuristic equation, as in general (for example for both~\eqref{mien} and~\eqref{jen1}) there is no precise definition of effective dimensions for $A$ and $\bar A$.  

\begin{figure}[!h] 
\centering 
\includegraphics[width=7cm]{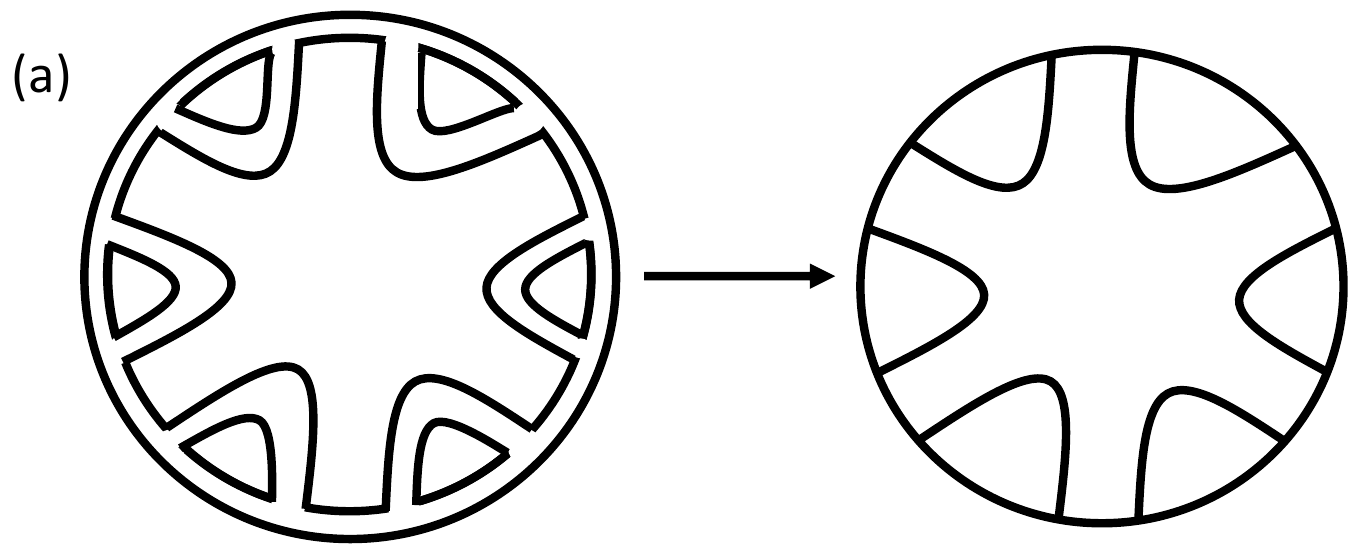}~\includegraphics[width=7cm]{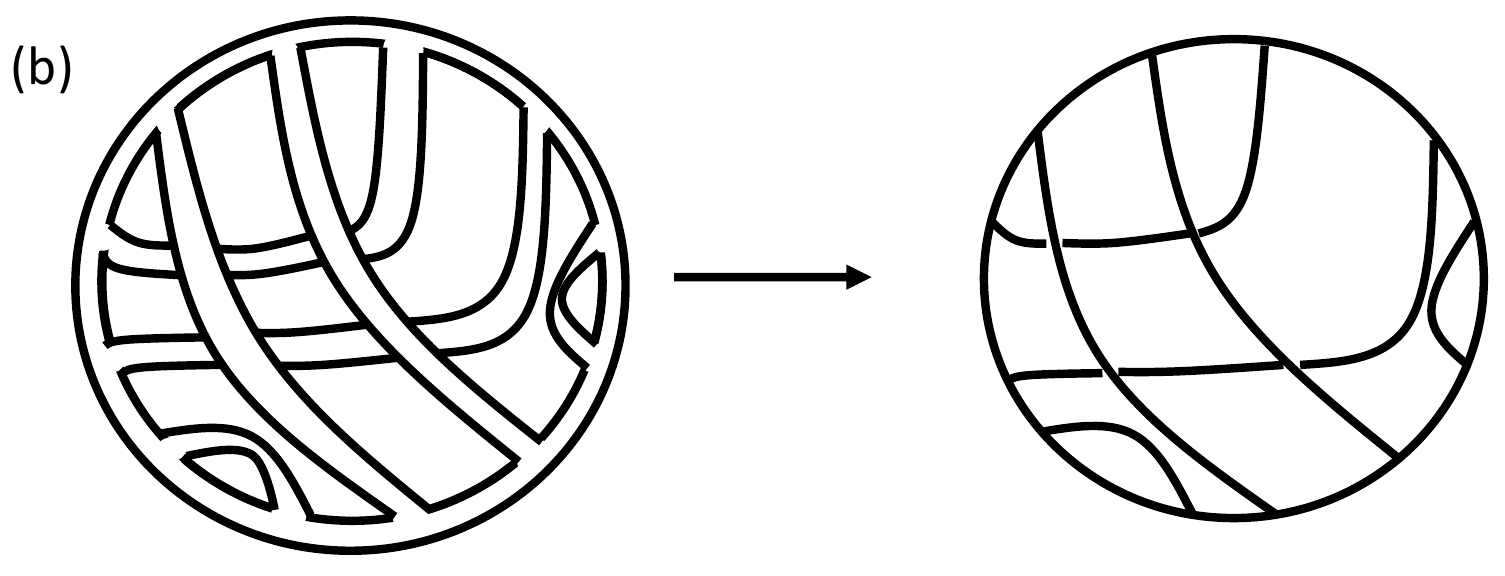}
\caption{Double-line diagrams (and the corresponding polygons) obtained from the diagrams of Fig. \ref{fig:eq}(a) and \ref{fig:rev}(a).}
\label{fig:double_line}
\end{figure}

Note that for any $\tau$,
 \be \label{yeg}
k ( \eta^{-1} \tau ) + k (\tau) \leq n+1  \ .
\ee
We can understand this diagrammatically. 
Let us slightly redraw each of the diagrams in Fig.~\ref{fig:eq}--\ref{fig:rev}, ignoring the distinction between dotted and dashed loops, and adding an extra loop surrounding the diagram. Two examples are shown in Fig.~\ref{fig:double_line}. We then get diagrams similar to 't Hooft's double-line diagrams for large $N$ matrix field theories~\cite{thooft}. With each such diagram, we can associate a polygon by replacing the double lines with single lines. If polygon can be drawn without crossing lines on a surface of minimum genus $h$, then the total number of loops in the double-line diagram is equal to the number of faces $F$ of the polygon on this surface. The total number of loops in the original diagrams in Fig.~\ref{fig:eq}--\ref{fig:rev} is one less than this, so
\be 
k ( \eta^{-1} \tau ) + k (\tau) = F-1 = E-V+ 2-2h-1
\ee
where $E$ is the number of edges of the polygon and $V$ is the number of vertices, and we have used the theorem relating Euler's characteristic to $F$, $E$ and $V$.
But for all diagrams we consider, the number of vertices is $4n$ and the number of edges is $3n$, and hence 
\be 
k ( \eta^{-1} \tau ) + k (\tau) = n +1 -2h.
\ee
The largest value of $k( \eta^{-1} \tau ) + k (\tau)$ is thus $n+1$, corresponding to planar diagrams. 

Two immediate examples of permutations that correspond to planar diagrams are $\tau = e$ and $\tau = \eta$. 
More generally, there is a one-to-one correspondence between such planar permutations  and ``non-crossing partitions'' of $n$ elements.\footnote{Consider a partition of $\{1, 2, \cdots n\}$ and any four elements $a < b < c < d $. The partition is non-crossing if whenever $a, c$ are in the same group and $b, d$ are in the same group,  the two groups coincide.} Given a non-crossing partition $\{ \{a^1_1, a^1_2, ..., a^1_{n_1}\}, \{a^2_1, a^2_2, ..., a^2_{n_2}\},~...~, \{a^k_1, a^k_2, ..., a^k_{n_k}\}\}$ of $\{1,2, \cdots n\}$ into $k$ groups, where the elements of each subset in the partition are arranged in descending order (e.g. $a^1_1> a^1_2 >... >a^1_{n_1}$), 
we obtain a planar permutation with the cycle representation given by  $(a^1_1, a^1_2, ..., a^1_{n_1})~(a^2_1, a^2_2, ..., a^2_{n_2}) ~... ~(a^k_1, a^k_2, ..., a^k_{n_k})$.  

\begin{figure} [!h] 
\centering
\includegraphics[height=8cm]{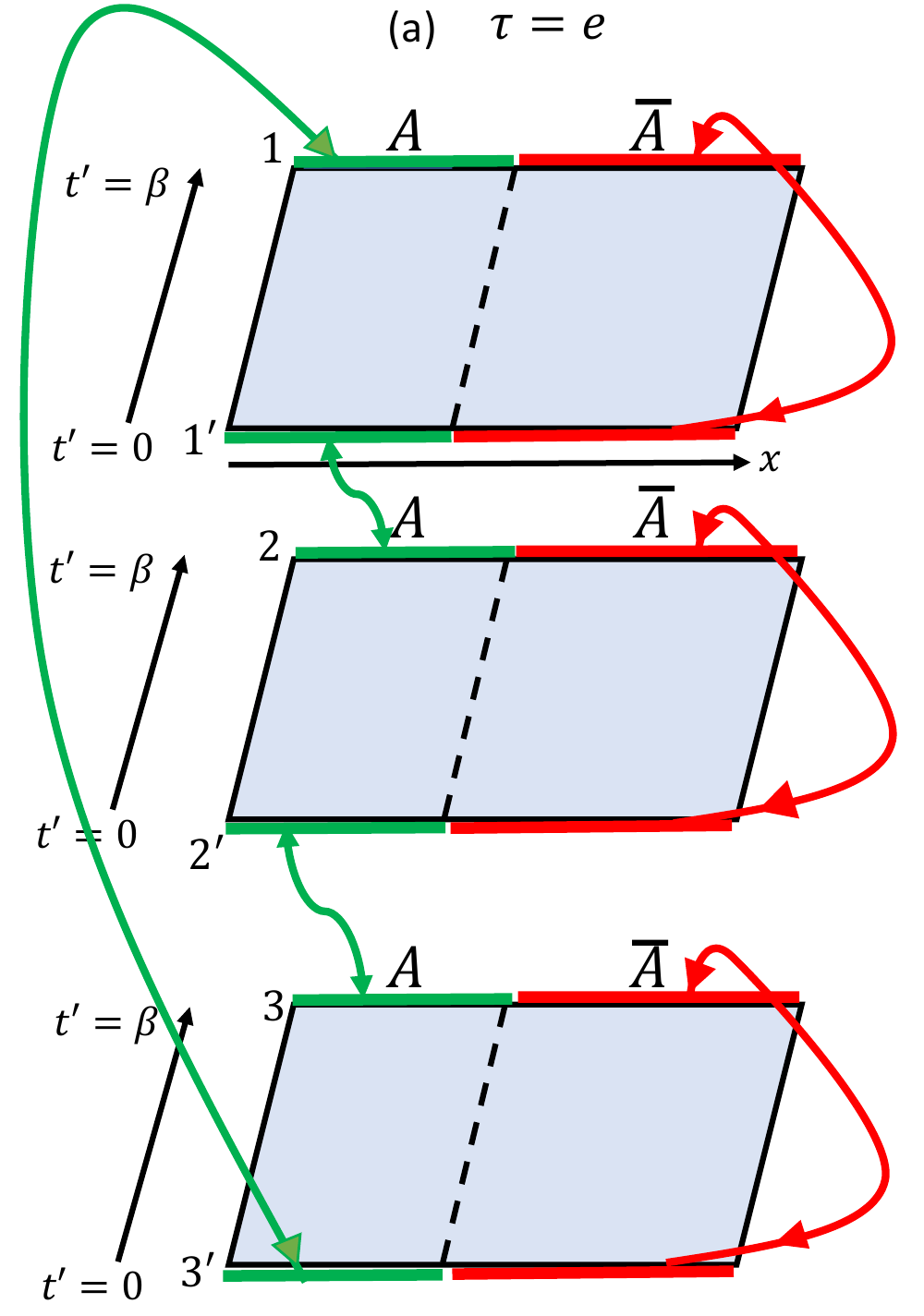}\includegraphics[height=8cm]{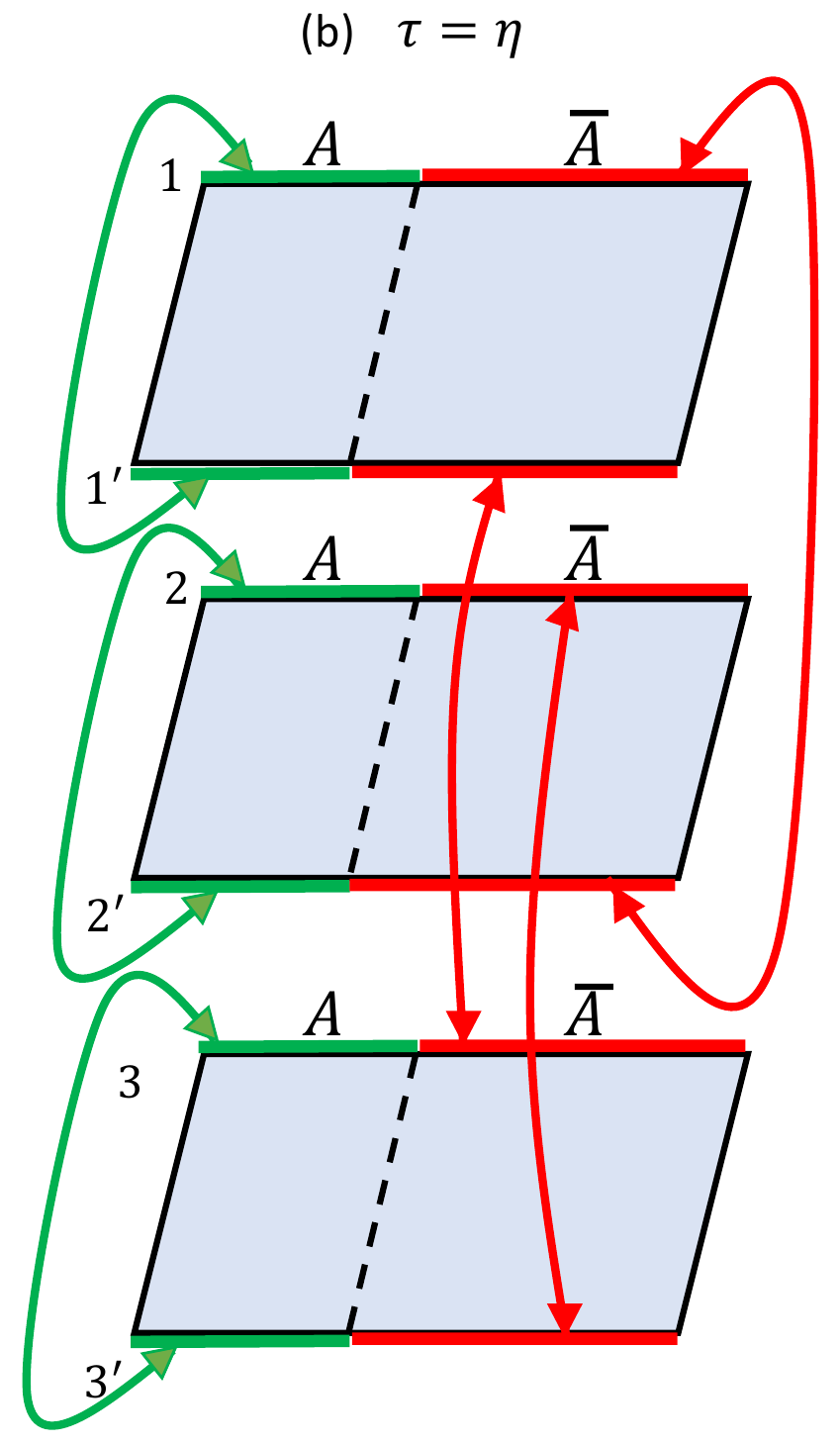}~~\includegraphics[height=8cm]{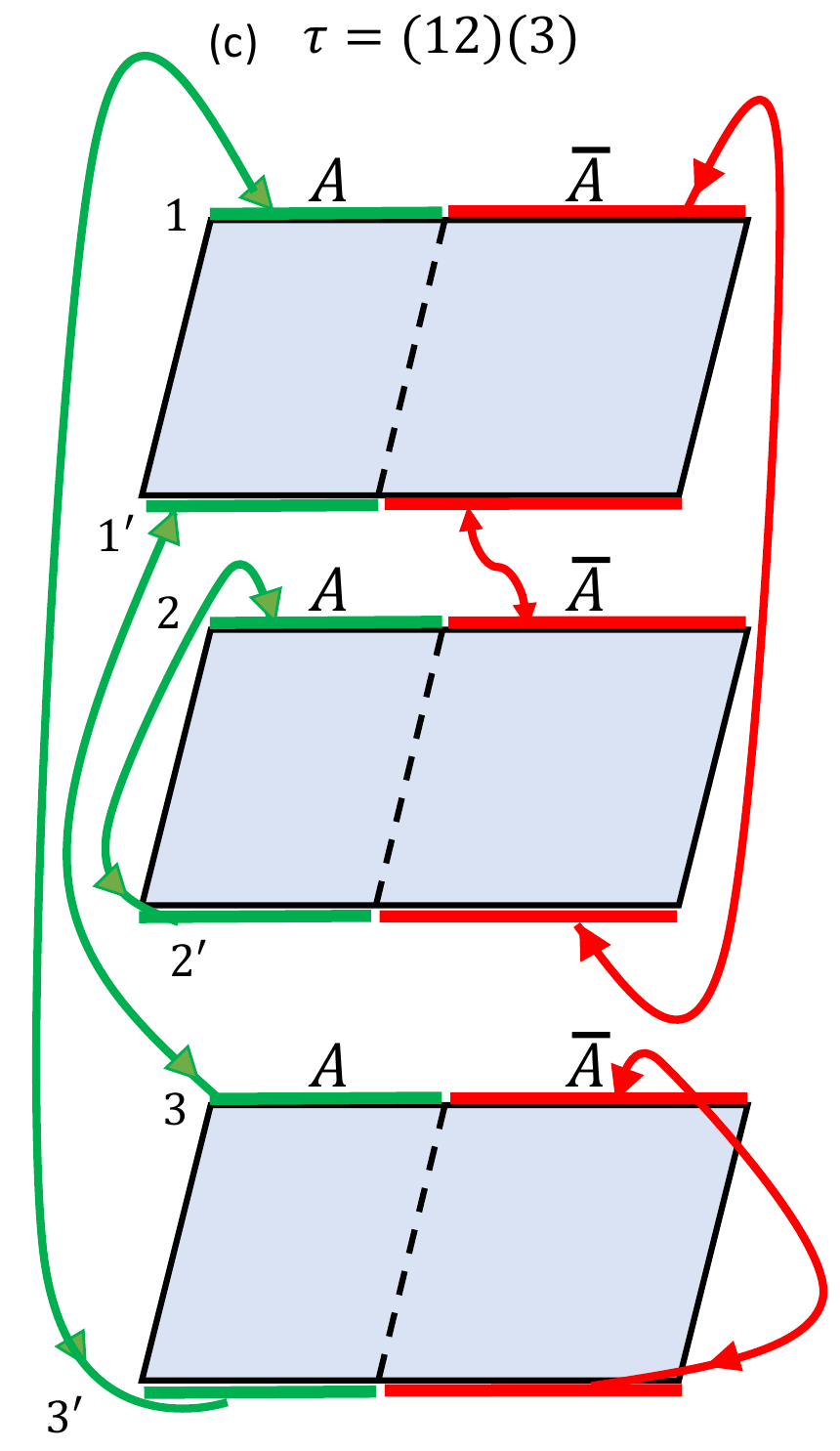}
\caption{Examples of the path integral representation of $\sZ_n^{(A)}(\tau)$ in \eqref{jhq1} for $n=3$, in the case where $\sI_{\al} = e^{-\beta H}$ and the system is (1+1)-dimensional. In the $i$-th replica, the final state is labelled by $\psi_i$ (which we indicated by $1, 2, 3$ in the figure) while the initial state is labelled by $\psi'_{i}$ (which is indicated by $1', 2', 3'$). The shaded regions represent Euclidean path integrals from $t'=0$ to $t'=\beta$ between the initial and final states. For each $\tau$, the final states $\psi_{iA}$ are identified with the initial states $\psi'_{\tau^{-1}\eta(i)A}$, while the final states $\psi_{i\bar{A}}$ are identified with the initial states  $\psi'_{\tau^{-1}(i)\bar{A}}$, as indicated with the arrows in each case.}
\label{fig:pathint}
\end{figure}

Equation~\eqref{zn_t} can be further written as 
\be 
 \begin{gathered} 
  \vev{\eta_A \otimes e_{\bar A} | \sI_\al , \tau}  =   \braket{i_{\eta(1)_a} i_{1_b}| \sI_{\alpha}| i_{\tau(1)_a} i_{\tau(1)_b}} \cdots 
 \braket{i_{\eta(n)_a} i_{n_b}| \sI_{\alpha}| i_{\tau(n)_a} i_{\tau(n)_b}}
 \end{gathered}
 \label{one}
 \ee 
so that~\eqref{fen11} has the form
 \be \label{fen12}
\sZ_{n}^{(A)}  \approx {1 \ov Z_1^n}  \sum_{\tau \in \sS_n}  \braket{i_{\eta(1)_a} i_{1_b}| \sI_{\alpha}| i_{\tau(1)_a} i_{\tau(1)_b}} \cdots 
 \braket{i_{\eta(n)_a} i_{n_b}| \sI_{\alpha}| i_{\tau(n)_a} i_{\tau(n)_b}} \ .
\ee 
In terms of path integrals, the approximation~\eqref{fen12} corresponds to replacing the second line of~\eqref{jh} by a sum of  Euclidean path integrals, each of which involves $n$ copies of that for $\sI_\al$ ``coupled'' together in a certain way specified by permutation $\tau$. More explicitly, we now have 
\bea \label{jhq}
\sZ_n^{(A)}  = \frac{1}{Z_1^n}  \sum_{\tau \in \sS_n} \int \prod_{i=1}^n  D \psi_{i} D \psi'_{i}  \,
 \de (\psi_{iA} - \psi'_{\eta(i)A}) \de (\psi_{i \bar A} - \psi'_{i \bar A}) 
\prod_{i=1}^n  \int_{ \psi'_{\tau (i)}}^{\psi_i}  D \phi_i \, e^{- S_E [\phi_i]} \quad \ \\
 =  \frac{1}{Z_1^n}  \sum_{\tau \in \sS_n} \int \prod_{i=1}^n  
D \psi_{i} D \psi'_{i}  \,
 \de (\psi_{iA} - \psi'_{\tau^{-1}\eta(i)A}) \de (\psi_{i \bar A} - \psi'_{\tau^{-1}(i) \bar A}) 
 \prod_{i=1}^n  \int_{ \psi'_{i}}^{\psi_i}  D \phi_i \, e^{- S_E [\phi_i]} \quad \
\label{jhq1}
\eea
where $S_E [\phi]$ is the Euclidean action for $\sI_\al$, i.e. 
\be 
\Tr \sI_\al = \int D \psi \, \int_{\psi}^\psi D \phi \, e^{- S_E [\phi]}  \ . 
\ee
See Fig.~\ref{fig:pathint} for an illustration. Since our approximation arises from projecting to a subspace of states in the replica theory (with Hilbert space $(\sH \otimes  \sH)^n$), it corresponds to isolating a subset of configurations from the Lorentzian path integrals~\eqref{jh}, which can further be given a Euclidean formulation. We stress, however, that these Euclidean path integrals do not arise from analytically continuing the Lorentzian path integral. Also note that while the Lorentzian replica space has $2n$ copies of the original system, the Euclidean replica space in \eqref{jhq}-\eqref{jhq1} has only $n$ copies.


To conclude this subsection, we make some further observations on the structure of~\eqref{fen12} which will be useful in the later discussion. Since $i_{k_a}, i_{k_b}$ are dummy indices, relabelling $i_{k_b} \rightarrow i_{\nu(k)_b}$, $i_{k_a} \rightarrow i_{\mu(k)_a}$ for any $\mu, \nu \in \sS_n$ leaves~\eqref{one} invariant. We thus have 
\be \label{hty}
\sZ_{n}^{(A)}  (\tau) = {1 \ov Z_1^n}    \vev{\eta_A \otimes e_{\bar A} | \sI_\al , \tau}  =
 {1 \ov Z_1^n}    \vev{(\mu \eta)_A \otimes \nu_{\bar A} | \sI_\al , (\mu \tau)_A (\nu \tau)_{\bar A}} , \quad \mu, \nu \in \sS_n , 
\ee
that is, the inner product is invariant under independent left multiplications for $A$ and $\bar A$. 
The inner product is also invariant under a simultaneous right multiplication for $A$ and $\bar A$ by the same permutation, that is, 
\be \label{hty1}
\sZ_{n}^{(A)}  (\tau) = {1 \ov Z_1^n}    \vev{\eta_A \otimes e_{\bar A} | \sI_\al , \tau}  =
 {1 \ov Z_1^n}    \vev{(\eta \mu)_A \otimes \mu_{\bar A} | \sI_\al , (\tau \mu)_A  (\tau \mu)_{\bar A}} , \quad \mu \in \sS_n , 
\ee
which simply corresponds to a reordering of the $n$ factors the product~\eqref{one}.

\subsection{Unitarity}\label{sec:unit} 

We now examine the physical consequences of \eqref{fen11} further. It is first useful to note that two terms in the sum over $\tau$ in~\eqref{fen11} have a simple physical interpretation 
 \bega \label{ss1}
 \sZ_{n}^{(A)} (e) =  {\text{Tr}}_A \le[\big(\text{Tr}_{\bar{A}}\rho_{\rm eq}\big)^n\ri] = \sZ_n^{(A, {\rm eq})} = e^{-(n-1) S_n^{(A, {\rm eq})}},
 \\
  \sZ_{n}^{(A)} (\eta) =  {\text{Tr}}_{\bar{A}} \le[\big(\text{Tr}_{A}\rho_{\rm eq}\big)^n\ri] =  \sZ_n^{(\bar A, {\rm eq})} = e^{-(n-1) S_n^{(\bar A, {\rm eq})}},
  \label{ss2}
 \end{gather} 
where $S_n^{(A, {\rm eq})}$ and $_n^{(\bar A, {\rm eq})}$ are respectively the $n$-th Renyi entropy
with respect to $A$ and $\bar A$ of the equilibrium density operator $\rho_{\rm eq}$. These two contributions are represented diagrammatically in Fig.~\ref{fig:eq}(a) and Fig.~\ref{fig:eq}(b).  
We can then write~\eqref{fen11} as 
\be \label{fen40}
\sZ_{n}^{(A)}  =  \sZ_n^{(A, {\rm eq})} +  \sZ_n^{(\bar A, {\rm eq})} + \tilde \sZ_{n}^{(A)} , \quad  \tilde \sZ_{n}^{(A)}  = \sum_{\tau \neq e, \eta} \sZ_{n}^{(A)}  (\tau) \ .
\ee

The first two terms in~\eqref{fen40} are together manifestly symmetric under $A \leftrightarrow \bar A$. We now show that $ \tilde \sZ_{n}^{(A)}$ is also invariant under $A \leftrightarrow \bar A$ so that the full expression satisfies the constraint
\be \label{unit1}
 \sZ_{n}^{(A)}  =  \sZ_{n}^{(\bar A)}  \ 
\ee
that must be obeyed in a pure state. For this purpose we write $  \tilde \sZ_{n}^{(A)}$ as 
\be \label{bnm}
  \tilde \sZ_{n}^{(A)}  = \ha \sum_{\tau \neq e, \eta} \le( \sZ_{n}^{(A)}  (\tau) +  \sZ_{n}^{(A)}   ( \sigma^{-1}\eta^{-1} \tau \sigma) \ri),  
\ee
where $\sigma$ is a permutation satisfying $\sigma \eta \sigma^{-1} = \eta^{-1}$, which always exists as $\eta$ and $\eta^{-1}$ are in the same conjugacy class (and is in general non-unique).  Then note that 
\ie \label{bnm1}
\sZ_{n}^{(A)}   (\sigma^{-1}\eta^{-1} \tau \sigma ) & = {1 \ov Z_1^n}  \vev{ (\sigma \eta \sigma^{-1})_{A} \otimes e_{\bar{A}} | \sI_\al , \eta^{-1}\tau} 
= {1 \ov Z_1^n}  \vev{\eta^{-1}_A \otimes e_{\bar A} | \sI_\al , \eta^{-1}\tau}  \cr
& ={1 \ov Z_1^n}  \vev{e_A \otimes \eta_{\bar A} | \sI_\al , \tau} =  \sZ_{n}^{(\bar A)}   (\tau)  
\fe
where in the first and third equalities we have used~\eqref{hty}--\eqref{hty1} repeatedly. Equation~\eqref{bnm} can then be written as 
\be
  \tilde \sZ_{n}^{(A)}  = \ha \sum_{\tau \neq e, \eta} \le( \sZ_{n}^{(A)}  (\tau) +  \sZ_{n}^{(\bar A)}   (\tau) \ri)
\ee
which is manifestly invariant under $A \leftrightarrow \bar A$.

\begin{figure} [!h] 
\centering
\includegraphics[width=15cm]{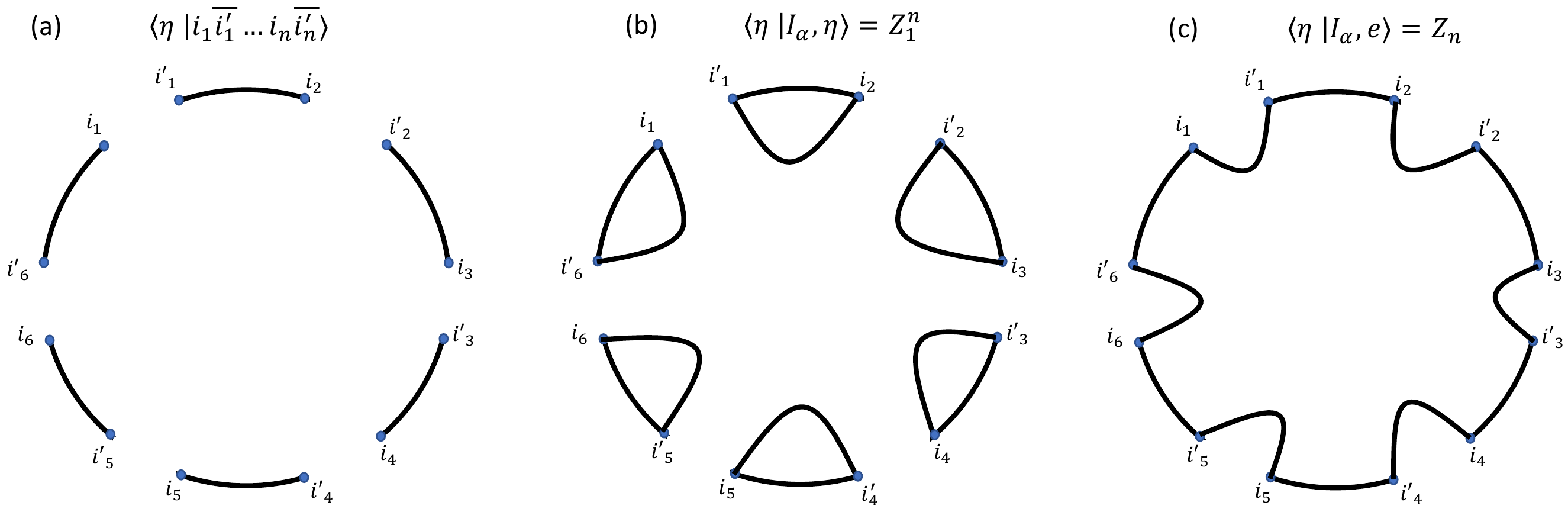}
\caption{(a) shows the ``future conditions" corresponding to $\bra{\eta}$ in the expression \eqref{purity} for $\Tr[\rho^n]$. (b) shows the diagram corresponding to the dominant contribution from $\tau=\eta$, while (c) gives an example of a subleading contribution coming from $\tau=e$.}
\label{fig:purity_bc}
\end{figure}

If the time-evolved state $\rho=\ket{\Psi} \bra{\Psi}$ is pure, we should also have $\Tr \rho^n =1$. 
Let us see how this is realized under our approximation~\eqref{fen}. Following arguments exactly parallel those which led to~\eqref{fen} and further to~\eqref{fen11}, we obtain the following approximation
\be \label{purity}
\Tr \rho^n = {1 \ov Z_1^n}   \sum_{\tau \in \sS_n}  \vev{\eta | \sI_\al , \tau} 
= {1 \ov Z_1^n}  \sum_{\tau \in \sS_n} \Tr \sI_\al^{n_1} \cdots \Tr \sI_\al^{n_{k }}
= {1 \ov Z_1^n}  \sum_{\tau \in \sS_n} Z_{n_1} \cdots Z_{n_{k }} 
\ee 
where $k =k (\tau \eta^{-1})$, and $n_1 , \cdots n_k$, are the lengths of the cycles in $\tau \eta^{-1}$. \eqref{purity} can also be given a diagrammatic representation, as explained in figure \ref{fig:purity_bc}.  
From~\eqref{hebl}, the dominant term is given by $\tau = \eta$ shown in Fig.~\ref{fig:purity_bc}(b),
leading to 
\be \label{rnorm}
\Tr \rho^n ={Z_1^n \ov Z_1^n} + O(Z_1^{-1}) =  1 + O(Z_1^{-1})  \ .
\ee

\subsection{Universal behavior of Renyi entropies for equilibrated pure states} \label{sec:univ}


Before applying~\eqref{fen11} to different classes of systems, here we make some general comments on its implications
in various regimes:
\begin{enumerate}
\item \label{item:b} 
Suppose $A \ll \bar A$ (and hence $d_A \ll d_{\bar A}$). Then, from the argument around \eqref{bdm}, the term with the maximal number of $\bar A$-loops, which corresponds to $\tau = e$, should dominate in \eqref{fen11}. 
In this case, from~\eqref{ss1} we have
\be \label{hei}
\sZ_n^{(A)}\approx  \sZ_n^{(A, {\rm eq})} =  e^{-(n-1) S_n^{(A, {\rm eq})}}\ .
\ee
Similarly, when $\bar A \ll A$ (and hence $d_{\bar A} \ll d_{A}$),
the term with the maximal number of $A$-loops, which corresponds to $\tau = \eta$, should dominate. In this case, from~\eqref{ss2}
\be 
\sZ_n^{(A)} \approx  \sZ_n^{(\bar A, {\rm eq})} =   e^{-(n-1) S_n^{(\bar A, {\rm eq})}}\ . 
\ee
Thus, when one of the effective dimensions of $A$ and $\bar A$ is much larger than the other, we have 
\be \label{hnel0}
S_n^{(A)} =  {\rm min} \le(S^{(A, {\rm eq})}_n , S^{(\bar A, {\rm eq})}_n \ri), \quad n= 2, \cdots 
\ee  
where $S^{(A, {\rm eq})}_n$ and  $S^{(\bar A, {\rm eq})}_n$ are respectively the equilibrium Renyi entropies with respect to $A$ and $\bar A$ of $\rho_{\rm eq}$. Analytically continuing~\eqref{hnel0} to $n=1$,
one then finds that for the von Neumann entropy 
\be \label{hnel}
S_1^{(A)} =  {\rm min} \le(S^{(A, {\rm eq})} , S^{(\bar A, {\rm eq})} \ri),
\ee 
where $S^{(A, {\rm eq})}$ and $S^{(\bar A, {\rm eq})}$ are respectively the ``equilibrium entropy'' for subsystems $A$ and $\bar A$ 
for the system in the equilibrium state $\rho_{\rm eq}$.

\item \label{item:a}
In the regime where $d_A$ and $d_{\bar A}$ are both large and comparable in size, from~\eqref{bdm}, the leading terms in~\eqref{fen11} come from those $\tau$'s which saturate~\eqref{yeg}, that is, from planar diagrams like in Fig.~\ref{fig:eq}.

\item 
In the infinite temperature case, $\sI_\al = \bid = \bid_A \otimes \bid_{\bar A}$. 
In general, $\sI_\al$ does not factorize between $A$ and $\bar A$. 
Nevertheless, as we will see more explicitly below and in Sec.~\ref{sec:holo}, for various situations of physical interest, 
one can have an approximate factorization 
\be \label{gdn}
\sI_\al \approx \sI_{\al_A}^{(A)} \otimes \sI^{(\bar A)}_{\al_{\bar A}} \ . 
\ee
Note that here we have allowed the parameters $\al$ to be different for $A$ and $\bar A$, which can happen if $A$ and $\bar A$ interact only for a finite period of time. Below for notational simplicity, we will simply write $ \sI_{\al_A}^{(A)} , \sI^{(\bar A)}_{\al_{\bar A}}$ as $\sI_A, \sI_{\bar A}$. Define for any integer~$m$
\be  
Z_m^{(A)}  = {\rm Tr}_A  \sI_{A}^{m} , \quad Z_m^{(\bar A) }= {\rm Tr}_{\bar A} \sI_{\bar A}^{m} , 
\quad 
 \hat Z_{m}^{(A)} = {Z_m^{(A)} \ov  \le(Z_1^{(A)}\ri)^m} , \quad 
 \hat Z_{m}^{(\bar A)} 
 = {Z_m^{(\bar A)} \ov  \le(Z_1^{(\bar A)}\ri)^m}  \ .
\ee 
From~\eqref{gdn} we have $Z_1 = \Tr \sI_\al =\le( {\rm Tr}_A  \sI_A \ri) \le( {\rm Tr}_{\bar A}  \sI_{\bar A} \ri)  \equiv Z_1^{(A)}  Z_1^{(\bar A)}$
and~\eqref{fen12} can be expressed as 
\ie \label{fen2}
 \sZ_{n}^{(A)} & \approx  {1 \ov Z_1^n}  \sum_\tau 
 \le(  Z_{m_1}^{(A)}   \cdots   Z_{m_l}^{(A)} \ri) \le(  Z_{n_1}^{(\bar A)}   \cdots   Z_{n_k}^{(\bar A)} \ri) \cr
& =  \sum_\tau 
 \le( \hat Z_{m_1}^{(A)}   \cdots  \hat Z_{m_l}^{(A)} \ri) \le( \hat Z_{n_1}^{(\bar A)}   \cdots  \hat Z_{n_k}^{(\bar A)} \ri),
\fe
where $k$ is the number of cycles of $\tau$ with $n_1, \cdots n_k$ the lengths of the corresponding cycles,
and $l$ is the number of cycles of $\tau \eta^{-1}$ with $m_1, \cdots m_l$ the lengths of the corresponding cycles. 


\een

We now consider more specifically the examples of $\sI_{\al}$ discussed in \eqref{infT}--\eqref{jen1}.

\subsubsection{Infinite temperature}

Let us first consider the case~\eqref{infT}. From~\eqref{fen11} we have 
\ie 
\sZ_{n}^{(A)}   =  {1 \ov d^n} \sum_{\tau} 
 \vev{\eta_A \otimes e_{\bar A} |\tau_A \otimes \tau_{\bar A}}  
 =  {1 \ov d^n} \sum_{\tau} 
d_A^{k  (\tau \eta^{-1})} d_{\bar A}^{k (\tau)} 
\label{infte0}
\fe
where $d_A, d_{\bar A}$ are respectively the dimensions of $A$ and $\bar A$ with $d= d_A d_{\bar A}$.

When one of $d_{A}, d_{\bar A}$ is much greater than the other, we simply find~\eqref{hnel0}--\eqref{hnel} 
with $\sS_n^{(A)} = \log d_A$ and $\sS_n^{(\bar A)} = \log d_{\bar A}$ for all $n$.

Now consider the regime 
\be \label{emm}
d_A, d_{\bar A} \sim d^\ha \to \infty, \quad {d_A \ov d_{\bar A}} = {\rm finite}    \ .
\ee
From~\eqref{yeg} the leading contribution in~\eqref{infte0} comes from permutations $\tau$ corresponding to planar diagrams, which saturate~\eqref{yeg}. All permutations with a given number of cycles $k(\tau) = k$ give the same contribution, which leads to\footnote{This same expression was previously noted for the Haar average in~\cite{canonical}.}
\ie \label{ohn2}
\sZ_{n}^{(A)}  & = {1 \ov d^n} \sum_{k=1}^n N(n,k)  d_A^{n+1 - k} d_{\bar A}^{k } \cr
& = {1 \ov  d_{\bar A}^{n-1}} + \ha{n(n-1) \ov d_A d_{\bar A}^{n-2}} + \cdots +\ha {n(n-1) \ov d_A^{n-2} d_{\bar A}} + {1 \ov d_A^{n-1}} 
\fe
where the coefficients $N(n,k)$ are the number of non-crossing partitions of $n$ objects with $k$ blocks, and are known as the Narayana numbers
\be \label{mem}
N(n,k) = {1 \ov n} {n \choose k} {n \choose k-1} \ .
\ee
In the second line of~\eqref{ohn2}, we have also made explicit that 
$N(n,1) =1$ (this comes from $\tau = \eta$), and $N(n, n) =1$ (from $\tau = e$).  
The von Neumann entropy
can be obtained by analytically continuing~\eqref{ohn2}
to general real values of $n$ and using \eqref{vn_approx}. For this purpose, we note that~\eqref{ohn2} can be written as 
\be \label{kje}
\mathcal{Z}_n^{(A)} =\frac{1}{d_{A}^{n-1}} ~{}_2 F_1\, (1-n \,, \, -n \, ; \, 2; \, d_{A}/d_{\bar{A}}) = \frac{1}{d_{\bar{A}}^{n-1}} ~{}_2 F_1\, (1-n \,, \, -n \, ; \, 2; \, d_{\bar{A}}/d_{A}) 
\ee
which can be continued to general $n$. The derivative with respect to $n$ can be found by expanding the hypergeometric function ${}_2 F_{1} (a,b;c;z)$ as a power series of $z$. Since the power series is convergent for $|z| \leq 1$, we should use the first expression 
in~\eqref{kje} for $d_A< d_{\bar{A}}$, and the second one for $d_A> d_{\bar{A}}$.
We then find that 
\be \label{jhl}
S_1^{(A)} = \bca  \log d_A - \frac{1}{2} \frac{d_A}{d_{\bar{A}}}  & d_A< d_{\bar{A}} \cr
  \log d_{\bar{A}} - \frac{1}{2} \frac{d_{\bar{A}}}{d_{A}}   & d_{\bar{A}}<d_A
  \eca  \ .
\ee
This agrees with the result from Haar averages~\cite{Page_entropy}. As discussed below~\eqref{nel1}, subleading corrections to~\eqref{ohn2} and~\eqref{jhl} beyond the limit of large $d$ will likely not be universal.

\subsubsection{Microcanonical ensemble}

Now we consider the case of the microcanonical ensemble~\eqref{mien}. We expect the result derived below should also apply to a single energy eigenstate of a chaotic system (i.e. to systems satisfying the eigenstate thermalization hypothesis). 


In this case,  $\sI_{E}$ is a projector, $Z_n  = N_I$ for all $n$. For $\ket{\Psi_0}$ lying in the subspace 
defined by the projector $\sI_E$, equation~\eqref{hem} is exactly satisfied.

We can write the Hamiltonian of the system as 
\be \label{ejnl}
H = H_A + H_{\bar A} + H_{A\bar A}
\ee
where $H_{A\bar A}$ denotes the interactions between $A$ and $\bar A$, and $H_{A}, H_{\bar A}$ only involve respectively degrees of freedom of subsystems $A$ and $\bar A$. Let us suppose $H$ is local. Then 
with sufficiently large subsystems $A$, $\bar A$, the contribution of $H_{A \bar A}$ to the energy is much smaller than those of $H_A, H_{\bar A}$ in macroscopic states whose energies are proportional to the volume of the system.

When $A \ll \bar A$ or $\bar A \ll A$, we again have~\eqref{hei}--\eqref{hnel}. Then using the standard argument of statistical mechanics, we can write $\sZ_n^{(A, {\rm eq})}$ and $\sZ_n^{(\bar A, {\rm eq})}$ more explicitly as 
\bega  \label{small_A}
  \sZ_n^{(A)} \approx \sZ_n^{(A, {\rm eq})} \approx
  {\rm Tr}_A \le[ \le(\rho_{\beta}^{(A)}\ri)^n \ri], \qquad \rho_{\beta}^{(A)}  = \frac{e^{-\beta H_A}}{{\rm Tr}_A [e^{-\beta H_A}]} , \quad A \ll \bar A \\ \label{big_A}
 \sZ_n^{(A)} \approx \sZ_n^{(\bar A, {\rm eq})} \approx  {\rm Tr}_{\bar A} \le[\le(\rho_{\beta}^{(\bar A)} \ri)^n\ri], \quad \rho_{\beta}^{(\bar A)}  = \frac{e^{-\beta H_{\bar A}}}{{\rm Tr}_{\bar A} [e^{-\beta H_{\bar A}}]} , \quad \bar A \ll A 
  \end{gather} 
where the inverse temperature $\b$ is determined from the density of states as $\b = {d \log N_I  \ov d E}$. 

Let us now consider the situation in which $A$ and $\bar A$ are comparable in size. 
 More explicitly,  suppose the total system has volume $V$ and $V_A /V = c < 1$, with $V \to \infty$. Using again the fact that the contribution of $H_{A \bar{A}}$ to the energy is small, we can approximate the projector $\sI_E$ as 
\be \label{zwk}
\sI_E \approx \sum_{ E_n^A + E_m^{\bar{A}} \in I} \ket{n}_A \ket{m}_{\bar{A}} \bra{n}_A \bra{m}_{\bar{A}}  
\ee
where $\ket{n}_A, \ket{m}_{\bar A}$ are respectively eigenstates of $H_A $ and $H_{\bar A}$ with energies $E_n^{A}$ and $E_m^{\bar{A}}$. 
 We then have 
\be 
\begin{gathered} 
\sZ_n^{(A)}= \frac{1}{N_I^n} \sum_{\tau}  \braket{\eta_A \otimes e_{\bar{A}}~|~\sI_E, \tau} 
=  \frac{1}{N_I^n} \sum_{\tau}\sum_{\sE} (d_{\sE}^A)^{k(\tau\eta^{-1})} (d^{\bar{A}}_{E-\sE})^{k(\tau)} 
\end{gathered}
\label{micro} 
\ee
where $\sE$ runs over the allowed values of energy in $A$ that can be consistent with total energy $E$, $d^A_{\sE}$ is the dimension of the subspace of $A$ with energy $\sE$, and $d^{\bar{A}}_{E-\sE}$ is the dimension of the subspace of $\bar{A}$ with energy $E-\sE$. 
Let us write 
 \be 
d_\sE^A = e^{V_A D_A (\ep_A)} , \qquad d_{E-\sE}^{\bar A} = e^{V_{\bar A} D_{\bar A} (\ep_{\bar A})} 
\ee
where $\ep_A = {\sE \ov V_A}$, $\ep_{\bar A} = {E- \sE \ov V_{\bar A}}$ are respectively energy densities for $A$ and $\bar A$.  Let $E = \ep V$, so that $\ep_{\bar A} = {\ep - \ep_A c \ov 1-c}$.
Equation~\eqref{micro} can then be written as ($k_1 = k (\tau \eta^{-1})$ and $k_2 = k (\tau)$)
\be 
\sZ_n^{(A)} ={1 \ov N_I^n}  \sum_\tau \sum_{\ep_A} \exp \le[V \le(k_1 c D_A (\ep_A) + (1-c) k_2 D_{\bar A} (\ep_{\bar A}) \ri) \ri] \ .
\ee
 The sum over $\ep_A$ can now be performed by a saddle point approximation with the saddle point $\bar \ep_A$ 
 satisfying the equation 
 \be 
 k_1  D_A' (\bar \ep_A) = k_2 D_{\bar A} ' \le( {\ep - \bar \ep_A c \ov 1-c}\ri) \ .
\ee
 Note that $\bar \ep_A$ depends on $k_1, k_2$. We then find that 
 \be \label{zn_saddle}
 \sZ_n^{(A)} ={1 \ov N_I^n}  \sum_\tau  \exp \le[V \le(k_1 c D_A (\bar \ep_A) + (1-c) k_2 D_{\bar A} \le( {\ep - \bar \ep_A c \ov 1-c}\ri) \ri) \ri] \ .
\ee

To proceed further, let us take the system to be homogenous, so $D_A (\ep) = D_{\bar A} (\ep) = f(\ep)$. We will further take $f(\ep)$ to be described by a power law, i.e. $f(\ep)  = C \ep^\al$ for some exponent $\al$. 
Conventional statistical systems have $\al < 1$ and we will restrict to this case\footnote{$\al =1$ is the so-called Hagedorn spectrum, while 
$\al > 1$ does not correspond to a stable equilibrium as the system has a negative specific heat.}, 
 where we find that the dominant contribution to \eqref{zn_saddle} comes from $\tau = e$ (or equivalently $k_1=1, k_2=n$) when $c<1/2$, and from $\tau = \eta$ (or $k_1=n, k_2=1$) when $c>1/2$. For $c=1/2$, the $\tau = e$ and $\tau = \eta$ contributions are equal and both are dominant.

Note that equation~\eqref{micro} can also be obtained by averaging uniformly over all pure states in the subspace $I$, that is, by taking a fixed $\ket{\psi_0}\in I$ and averaging the value of $\sZ_n^{(A)}$ over states $U \ket{\psi_0}$, where $U$ is a Haar-random unitary matrix acting within the subspace of energy $E$. Equivalently, it can be obtained from an average over the ``ergodic bipartition" states described in \cite{grover}.

 \subsubsection{Canonical ensemble} 
 
 Let us now consider the situation where the effective identity operator $\sI_{\al}$ is given by~\eqref{jen1}. Note that the result \eqref{fen11} with this value of $\sI_{\al}$ can also be obtained by further manipulation of the average over random ``canonical thermal pure quantum states" in \cite{canonical}. 

Let us now consider the result in special regimes. When $A \ll \bar A$ or $\bar A \ll A$, we again have~\eqref{hei}--\eqref{hnel}. Let us now consider the case where $A$ and $\bar A$ are comparable, that is, the total system has volume $V$ and $V_A /V = c < 1$, with $V \to \infty$. With the Hamiltonian~\eqref{ejnl} and assuming local interactions,  we can approximate, within the traces appearing in various quantities, that $\sI_\b$ has a factorized form\footnote{One should not view the equation below as an operator equation, rather as a relation which holds within matrix elements among states who energies are proportional to the volume of the system. 
}, 
 \be
\sI_\b = e^{-\b H} \approx e^{-\b H_A} \otimes e^{-\b H_{\bar A}} =  \sI_\b^{(A)} \otimes \sI^{(\bar A)}_\b  \ . 
\ee
Then $\sZ_n^{(A)}$ is given by \eqref{fen2}, and the quantities appearing  in \eqref{fen2} have the form 
\be
\quad
 \hat Z_{m}^{(A)} =  {{\rm Tr}_A  e^{-m \b H_A}   \ov ({\rm Tr}_A e^{-\b H_A} )^m}, \qquad
 \hat Z_{m}^{(\bar A)} =  {{\rm Tr}_{\bar A}  e^{-m \b H_{\bar A} }  \ov ({\rm Tr}_{\bar A} e^{-\b H_{\bar A}} )^m}  \ .
 \ee
Comparing with~\eqref{zn_saddle}, we see that while for $A\ll \bar A$, the Renyi entropies corresponding to 
 the microcanonical and canonical ensembles have the same form, they differ in the regime where $V_A/V$ is finite. 
 
 
For a homogeneous system, from~\eqref{kjn}, the partition functions for $A$ and $\bar A$ subsystems can be written as 
\be 
{\rm Tr}_A e^{-\b H_A} = e^{N_A \, g(\b)}, \qquad  {\rm Tr}_{\bar A} e^{-\b H_{\bar A}} = e^{N_{\bar A} \, g(\b)}, 
\quad N = N_A + N_{\bar A}  \ .
\ee
We then have 
\be
 \hat Z_{m}^{(A)} =  e^{N_A (g (m \b)- m g(\b))} , \quad
 \hat Z_{m}^{(\bar A)} =  e^{N_{\bar A} (g (m \b)- m  g(\b))} \ .
 \ee
 
 To proceed further let us take $g (\b)$ to be a power law, that is, $g(\b) = \lam \b^{- \al}$ with $\al > 0$. We then find that the $\tau=e$ term is dominant in this expression for $c<1/2$, the $\tau=\eta$ term is dominant for $c>1/2$, and for $c=1/2$, both $\tau=e$ and $\tau=\eta$ give equal contributions, which are dominant. 
 
\subsection{Uncompact systems and subregion equilibration} \label{sec:uncompact}

In our discussion above, we have assumed that the whole system thermalizes after some finite time scale $t_s$. For an uncompact system, such a time scale does not exist. Nevertheless, at a finite time $t$,  subregions of certain finite sizes can thermalize, and we can apply the approximation of Sec.~\ref{sec:sche}--\ref{sec:gen} to such subregions. 

As an illustration, we consider an infinite (1+1)-dimensional system, which can be a spin chain or a quantum field theory. We assume for simplicity that the system is governed by a local Hamiltonian which results in a sharp light-cone, with speed $c=1$.
Suppose we are interested in the entanglement of a finite region $A$ with its complement $\bar A$ at time $t$, as indicated  in Fig.~\ref{fig:causal}. Due to the causality constraint from the sharp light-cone,  the region which is relevant for this purpose 
is $J(A)$, the region at $t=0$ which is causally connected with $A$. Evolution of the system in $\overline{J(A)}$ should not be relevant 
for finding $S_n^{(A)}$ or $S_n^{(\bar A)}$, and we can replace the time-evolution operator in this part of the system with the identity. 
 See Appendix~\ref{app:causal} for a more explicit argument. 

Let us further suppose that the system is sufficiently strongly interacting and chaotic, such that at time $t$, the system is locally equilibrated in region $J(A)$, i.e. the equilibration is maximally efficient as allowed by causality. In this case, we can apply~\eqref{fen11}, treating $J(A)$ as the full system, and the region $J(A)-A$ of length $2t$ as the complement of $A$. From the discussion of item~\ref{item:b} in Sec.~\ref{sec:univ}  we immediately conclude that\footnote{When $t \approx |A|/2$, the expression could be more complicated as other planar permutations besides $\tau=e$ and $\tau=\eta$ may become significant.}
\be 
S_n^{(A)} = s^{\rm eq}_{n} \,  {\rm min} (|A|, 2t) \label{ve1}
\ee
where $s^{\rm eq}_{n}$ is the $n$-th equilibrium Renyi entropy density ($n=1$ being the equilibrium entropy density). Here the entanglement velocity is given by $v_E = c =1$. Since it is expected on general grounds \cite{random_ent, stanford_mezei} that $v_E\leq v_B$, where $v_B\leq 1$ is the velocity associated with the growth of operators, \eqref{ve1} implies that we must also have $v_B=1$, corresponding to the fact that operators grow at the fastest speed allowed by causality. Hence, we can see the maximally fast growth of operators as a necessary condition for the assumption of maximally efficient equilibration that we made above. 



\begin{figure} [!h] 
\centering
\includegraphics[width=12cm]{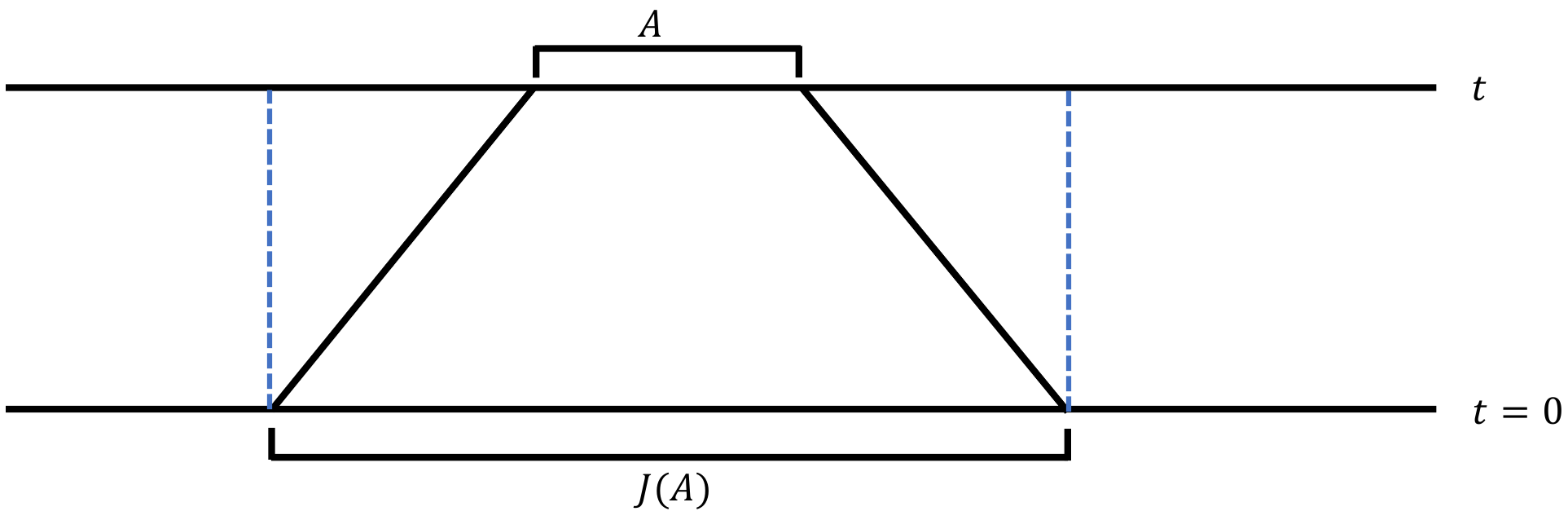}
\caption{Causal constraints and subregion thermalization. To find the entanglement entropies of $A$ in a system with a sharp light-cone structure, it is sufficient to consider the part of the time-evolution operator in $J(A)$, that is, between the dashed lines. }
\label{fig:causal}
\end{figure}





\section{Gravity systems and replica wormholes} \label{sec:holo} 

The  equilibrium approximation discussed in the last section 
can be applied to gravity systems, with the assumption that they follow the usual rules of quantum mechanics. 
In this context, various quantities in~\eqref{fen11} or~\eqref{fen12} should be seen as amplitudes in an exact theory of quantum gravity. In particular, the Euclidean path integrals~\eqref{jhq} emerge universally  
as an approximation to the Lorentzian path integral~\eqref{jh}. Furthermore, different Euclidean 
replica gravity systems have to be ``coupled'' in specific ways. However,  in our current understanding of quantum gravity, gravity path integrals can only be formulated at a semi-classical level, and hence a direct implementation of the prescription~\eqref{jhq1} may be subtle.  For holographic systems, the amplitudes in equations~\eqref{fen11} and~\eqref{jhq1} also have a dual description in terms of the corresponding ones in the boundary system. Here, one has the benefit that the boundary version of path integrals~\eqref{jhq1} can be used to provide  boundary conditions for formulating the corresponding bulk ones by using the standard rules of holography. 

Intuitively, couplings among different replicas could lead to replica wormholes, that is, geometries connecting different replica manifolds. 
In this section, we will make this idea precise by applying the equilibrium approximation to two recently discussed models of  black holes~\cite{replica_1,Almheiri:2019qdq}, and showing that the prescriptions proposed there for including certain replica wormholes in the calculation of entanglement entropies follow from~\eqref{fen11} and~\eqref{jhq1}.  The earlier discussion of Sec.~\ref{sec:unit} then provides an explanation for why including replica wormholes leads to entanglement entropies that are consistent with unitarity.
We will also comment on how the equilibrium approximation provides an alternative to the need for an averaged description discussed in~\cite{replica_1}, and briefly discuss the Renyi entropies for a big black hole in AdS formed from gravitational collapse of a pure state.

\subsection{A model for black hole evaporation} \label{sec:bhe}

Let us first briefly review the model of an evaporating black hole 
discussed in~\cite{replica_1}, where the black hole lives in a (1+1)-D spacetime with JT gravity and has an end-of-the-world (EOW) brane behind the horizon, see Fig.~\ref{fig:bhh}. 
\begin{figure} [!h] 
\centering
\includegraphics[height=3.5cm]{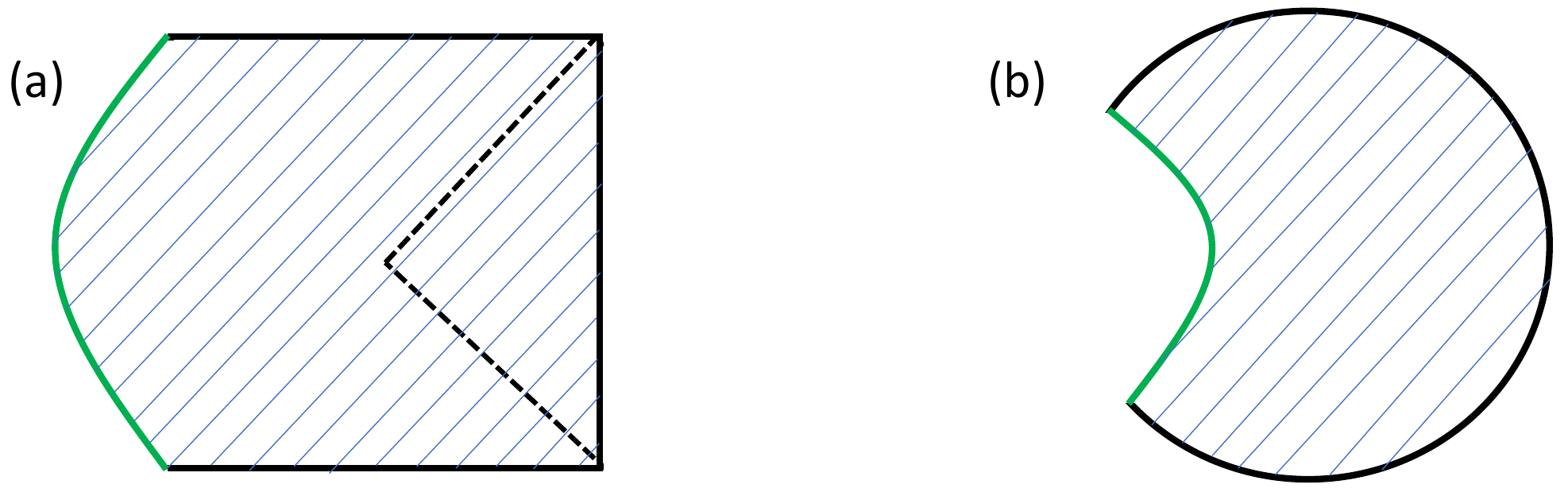}
\caption{Lorentzian and Euclidean geometries for a black hole with an end-of-the-world brane (denoted by the green line). 
}
\label{fig:bhh}
\end{figure}
The state $\ket{\Psi}$ of the full system resulting from the evaporation process is assumed to be such that if $\ket{i}$ is an orthonormal basis of $N$ states for the radiation subsystem $R$, then the matrix element ${(\rho_R)}_{ij}$ of the reduced density matrix for $R$ can be calculated using Euclidean path integrals with the following rules, shown in Fig. \ref{fig:overlap_rules}. The boundary condition is given by a single open asymptotic boundary segment of JT gravity of length $\beta$ for some inverse temperature $\beta$ associated with the state. The endpoints of the segment are  labelled by $i$ and $j$, as shown in Fig. \ref{fig:overlap_rules}(a). In the corresponding bulk path integral in Fig. \ref{fig:overlap_rules}(b),  the two endpoints are connected with an EOW brane. In addition to a gravity path integral indicated by the shaded region, this gives a factor of $\delta_{ij}$, indicated by the dotted line connecting $i$ and $j$. 
\begin{figure} [!h] 
\centering
\includegraphics[height=1.5cm]{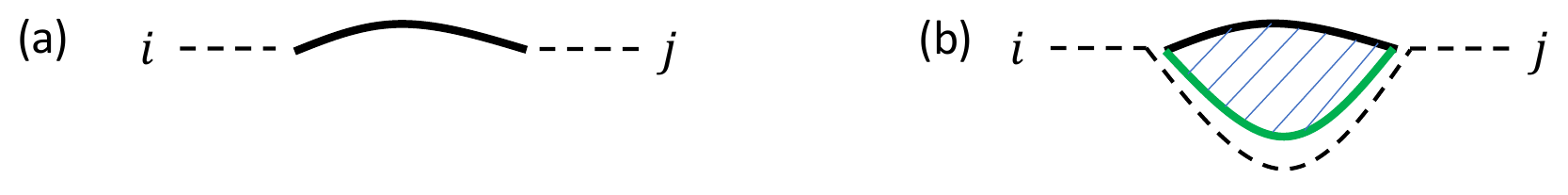}
\caption{(a) shows the boundary conditions and (b) shows the bulk path integral for evaluating ${(\rho_R)}_{ij}$ according to the rules of \cite{replica_1}.}
\label{fig:overlap_rules}
\end{figure}

The boundary conditions for the calculation of $\sZ^{(R)}_n$ involve $n$ open asymptotic boundary segments of length $\beta$ as shown in Fig. \ref{fig:3_rules}(a) for $n=3$, with the dashed lines indicating the contraction of indices in the matrix multiplication. The rule for the corresponding bulk path integral is to sum over all possible ways of connecting the endpoints with EOW branes, like the two examples shown in Fig. \ref{fig:3_rules}(b) and (c). Each resulting loop of dashed lines gives a factor of $N$. Contributions like Fig. \ref{fig:3_rules}(c), where multiple asymptotic boundaries are connected by the bulk geometry, are said to have replica wormholes, and such wormhole contributions are important for giving results for $\sZ^{(R)}_n$ consistent with unitarity.  

\begin{figure} [!h] 
\centering
\includegraphics[height=3cm]{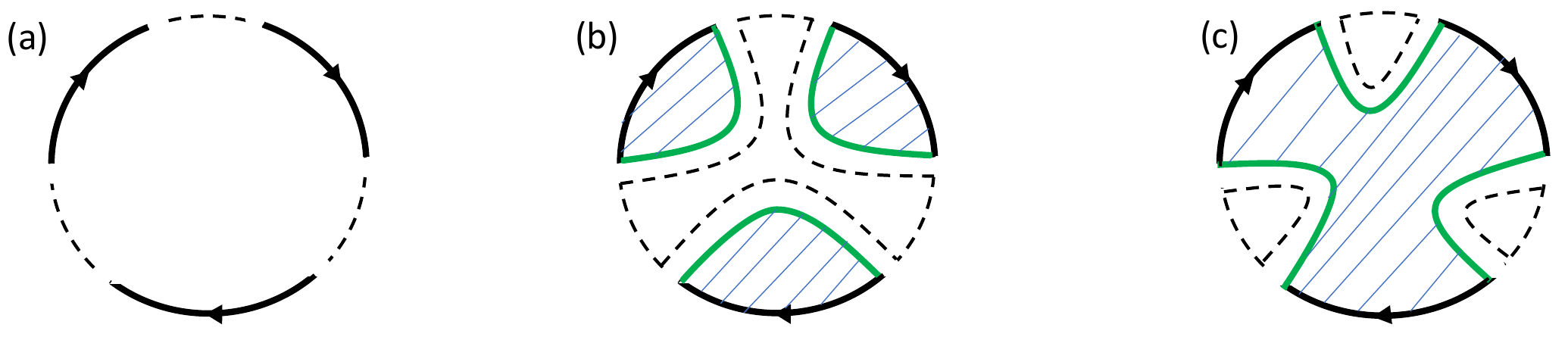}
\caption{(a) shows the boundary conditions, and (b) and (c) show two contributions to the bulk path integral for evaluating $\sZ_n^{(R)}$ for $n=3$ according to the rules of \cite{replica_1}.}
\label{fig:3_rules}
\end{figure}

We will now describe the evaluation of the quantities $\sZ_n^{(A)}$ according to the equilibrium approximation in a more general class of quantum-mechanical systems related to the above model. We will then show that applying the standard rules of holography to the Euclidean path integrals in the resulting expression gives a derivation of the replica wormholes introduced with the {\it ad hoc} rules above. The final result matches precisely with that of \cite{replica_1}.  

Let us consider a situation where the initial state $\ket{\Psi_0}$ in~\eqref{ste} describes a star, which under time-evolution collapses to 
form a black hole and subsequently emits Hawking radiation. The full system at time $t$, described by the state $\ket{\Psi}$, consists of the black hole and the emitted radiation subsystem. 
The radiation subsystem has a Hilbert space of finite dimension $N$ with no energy constraint, while the black hole 
can be associated with an inverse temperature $\b$.  We assume that the radiation separates from and no longer interacts with the black hole after being emitted. We can then write the effective identity operator $\sI_\al$ corresponding to $\ket{\Psi}$ in a factorized form 
\be \label{nel11}
\sI_\al= \bid_R \otimes  \sI^{(B)}_\b, \quad {\rm Tr}_R \bid_R = N 
\ee
where $R$ and $B$ denote respectively the radiation and black hole subsystems. 
 
For an evaporating black hole, the Hamiltonian of the system cannot be strictly time-independent. As a result,~\eqref{nel11} may not strictly satisfy~\eqref{invc}. However,~\eqref{invc} should still be valid to a very good approximation if the evaporation process happens slowly.

For comparison with the discussion of~\cite{replica_1}, we will assume that the black hole subsystem resembles that in Jackiw-Teitelboim (JT) gravity~(or the SYK model at a sufficiently low temperature). That is, it has a large number 
of densely spaced states in the energy range accessible at inverse temperature $\b$, such that 
\be \label{eue}
Z_1^{(B)} = {\rm Tr}_B  \sI_\b^{(B)}  = e^{S_0}  z_1 (\b) , \qquad 
Z_m^{(B)} = {\rm Tr}_B  \le(\sI_\b^{(B)} \ri)^m = e^{S_0}  z_m (\b) 
\ee
where the ``background'' density of states $e^{S_0}$ is large, and  $z_m (\b)$ are $O(1)$ functions. 
We will be interested in the regime 
\be \label{lan}
e^{S_0}, N \to \infty, \qquad N e^{-S_0} = {\rm finite} \ . 
\ee

Applying~\eqref{fen2} to~\eqref{nel11} 
 we find for the radiation subsystem 
\begin{align}
 \sZ_{n}^{(R)} & \approx {1 \ov \le(N Z_1^{(B)}\ri)^n} \sum_\tau N^{k (\tau \eta^{-1})} 
 Z_{n_1}^{(B)}  \cdots  Z_{n_{k(\tau)}}^{(B)}\  \label{fen4} 
 \end{align}
where $n_1$, ..., $n_{k(\tau)}$ are the lengths of the cycles in $\tau$.
In the regime~\eqref{lan}, the leading terms in~\eqref{fen4} are given by those $\tau$'s which saturate~\eqref{yeg}, that is, by the planar diagrams of Sec.~\ref{sec:gen}.  We can write the cycle structure of a permutation $\tau$ as $\tau = (1^{m_1} 2^{m_2} ... n^{m_n})$, which indicates that it has $m_1$ cycles with 1 element, $m_2$ cycles with 2 elements, and so on. By definition we have 
\be 
k (\tau) = m_1 +... +m_n, \qquad n = \sum_{i=1}^n i \, m_i  \ .
\ee
Since the summand in~\eqref{fen4} only depends on the cycle structure of $\tau$, we can write the leading planar contribution as  
\be \label{hej}
 \sZ_{n}^{(R)} \approx  {1 \ov (Z_{1}^{(B)})^{n}}  \sum_{k=1}^n N^{1-k}  \sum_{\sum_i m_i= k}  N (\{m_i\}) \prod_{j=1}^n(Z_{j}^{(B)})^{m_j} 
 \ee
where $ N (\{m_i\}) $ is the number of planar permutations with cycle structure $\{m_i\}$, and from our comments below~\eqref{yeg}, can in turn be understood as the number of non-crossing partitions with $m_i$ blocks of cardinality $i$. The  explicit expression for this number when the total number of blocks $k>1$ is given by~\cite{noncrossing}
 \be \label{hej1}
 N (\{m_i\}) =  \frac{n(n-1)...(n-k+2)}{m_1! ~...~ m_n!} \ . 
 \ee
For $k=1$, we only have one block consisting of all $n$ elements, and $N (\{m_i\})=1$. 

So far, our discussion is general and applies to any system with $\sI_{\al}$ as in \eqref{nel11} and $Z_m^{(B)}$ as in \eqref{eue}. The specific gravity description enters in the explicit evaluation of various partition functions 
$Z_m^{(B)} (\b)$ (or equivalently $z_m (\b)$), which can be expressed in terms of Euclidean gravity path integrals and evaluated using a saddle-point approximation. 
Let us now specify to the gravity system considered in~\cite{replica_1}. This motivates us to write 
\be
 \sI^{(B)}_\b = f(H_B)  e^{-\b H_B} , 
\ee
where $H_B$ is the Hamiltonian for the black hole subsystem and we have included a factor $f(H_B)$ which captures the presence of the EOW brane. The specific form of the function $f$ is not important for our discussion.

The calculation of $Z_m^{(B)} (\b)$ using bulk gravity follows from the standard rules of holography. The partition function $Z_1^{(B)} (\b)$ of $ \sI^{(B)}_\b$ can be obtained from the Euclidean black hole geometry of Fig.~\ref{fig:bhh} (b). The evaluation of $Z_m^{(B)} (\b)$ for $m > 1$ is indicated in Fig.~\ref{fig:part}, and involves replica wormholes. $Z_m^{(B)}$ here should thus be identified with the replica wormhole partition function $Z_m$ with $m$ boundaries given in equation (2.29) or (2.32) of~\cite{replica_1}.\footnote{By definition~\eqref{nel11} we can write $Z_m^{(B)} = \int dE \, \rho (E) e^{-m \b E} f^m (E)$, with $\rho(E)$ the density of states. 
Equation (2.32) of~\cite{replica_1} precisely has this structure.}
The contributions from terms corresponding to different $\tau$ in~\eqref{fen4} saturating \eqref{yeg} can be captured precisely by the planar gravity diagrams of \cite{replica_1} like the ones shown in Fig. \ref{fig:3_rules}(b) and (c), which correspond respectively to $\tau=e$ and $\tau=\eta$.

 \begin{figure} [!h] 
\centering
\includegraphics[width=14cm]{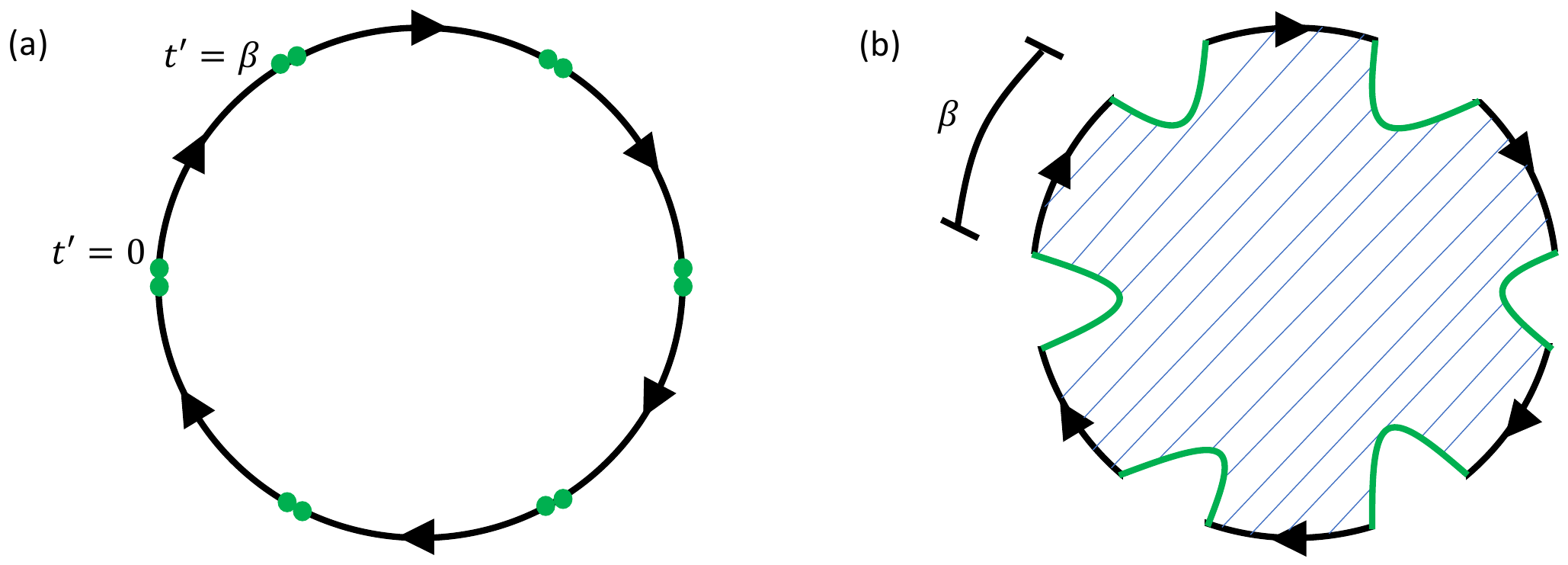}
\caption{The boundary path integrals for $Z_m^{(B)} = {\rm Tr}_B \le(f (H_B) e^{-\b H_B} \ri)^m$ can be represented as in (a) for $n=6$. The Euclidean path integral along each black line from $t'=0$ to $t'=\beta$ represents $e^{-\b H_B}$, and each green dot represents $f^\ha (H_B)$. For convenience of presentation we have arranged the factor $f(H_B)$ symmetrically in each replica. (b) gives the dual gravity description of the boundary path integral, where one should integrate over all bulk configurations with the specified boundary topology, and we have interpreted the green dots representing $ f(H_B)$ as the end-of-the-world branes. 
}
\label{fig:part}
\end{figure}

We can show more explicitly that~\eqref{hej}--\eqref{hej1} agree precisely with the results obtained in the limit \eqref{lan} in~\cite{replica_1}.\footnote{Away from the limit \eqref{lan}, in the equilibrium approximation we get corrections from $\tau$ that do not saturate \eqref{yeg}, and also from higher-order terms in the metric $g_{\sigma\tau}$, which can all be evaluated using Euclidean gravity path integrals.}
In~\cite{replica_1}, the full expression for $\sZ_{n}^{(R)}$ was only given implicitly through 
a generating functional called the resolvent, which can be used to obtain a recursion relation for  $\sZ_{n}^{(R)}$. 
More explicitly,  the trace of equation (2.27) of~\cite{replica_1} can be written in our notation as
\be 
 \sum_{n=1}^{\infty} \frac{1}{\lambda^{n+1}} \sZ_{n}^{(R)} = \frac{1}{\lambda} \sum_{n'=1}^\infty
 \frac{ Z_{n'}^{(B)}}{N^{n'} (Z_1^{(B)})^{n'}} \bigg[\frac{N}{\lambda} + \sum_{m=2}^{\infty}\frac{\sZ_{m-1}^{(R)}}{\lambda^{m}} \bigg]^{n'}  \ .
 \ee
  Equating coefficients of  $1/{\lambda^{n+1}}$ on both sides, we find a recursion relation 
  \be 
\begin{gathered}
\sZ_{n}^{(R)}  = \frac{1}{N} \sZ_{n-1}^{(R)} + \sum_{n'=2}^{n-1} \frac{ Z_{n'}^{(B)}}{N^{n'} (Z_1^{(B)})^{n'}} \sum_{\substack{r_1 + ...+ r_{n} = n'\\ \sum_{i} i r_i = n}} \frac{n'!}{r_1! ~... ~r_{n}!} ~N^{r_1} \prod_{t=2}^{n} (\sZ_{t-1}^{(R)})^{r_t}  +~ \frac{ Z_{n}^{(B)}}{ (Z_1^{(B)})^{n}}  \ .
\end{gathered}
\label{revur}
\ee
One can check that~\eqref{hej}--\eqref{hej1} indeed satisfy~\eqref{revur}.\footnote{We have checked up to $n=12$.} 

\subsection{Comments on averaging and replica wormholes}
 \label{sec:av}
 In the previous subsection, we demonstrated how the Euclidean gravity prescription for computing the Renyi entropies of an evaporating black hole can emerge as an approximation to the Lorentzian path integrals~\eqref{jh}.  
 One important implication of the discussion is that replica wormholes can arise in a system with a fixed 
Hamiltonian, and it not necessary to have an ensemble-averaged theory. We now further clarify an issue raised in~\cite{replica_1}, which was used there to interpret replica wormholes as arising from some averaging procedure.

Let us first recapitulate the issue.  Consider a matrix element of the reduced density matrix for the radiation system (here $\ket{i}$ denotes a basis for the $R$ subsystem),  
\be\label{maw}
(\rho_R)_{ij} = {\rm Tr}_R (\ket{j} \bra{i}  \rho_R) = {\rm Tr}_R (\ket{j} \bra{i}  {\rm Tr}_{B} (U \rho_0 U^\da)) 
= \braket{i_R \bar{j}_R \otimes e_{B}  |U \otimes U^{\da} | \rho_0, e} 
\ee  
Applying the same procedure as in~\eqref{may} to~\eqref{maw}, we find
\be \label{pih}
(\rho_R)_{ij} 
={1 \ov Z_1} \vev{i_R \bar{j}_R  \otimes e_{B} | \sI_\al, e} + \De_{ij}
= \le(\rho^{\rm (eq)}_R \ri)_{ij}  + \De_{ij} 
\ee
where $\rho^{\rm (eq)} = {1 \ov Z_1} \sI_\al$ and $\De_{ij}$ is the contribution from the $Q$ projector. 
Using~\eqref{nel11} and dropping $\De_{ij}$, we obtain the equilibrium approximation for these matrix elements,
\be \label{disg}
(\rho_R)_{ij} \approx   {1 \ov N} \de_{ij}  ,
 \ee
which would imply 
\be \label{emn}
\sZ_n^{(R)} = {\rm Tr}_R \rho_R^n \approx {1 \ov N^{n-1}} \ .
\ee 
 But~\eqref{emn} clearly contradicts~\eqref{hej}. For example, for $n=2$,~\eqref{hej} gives  
 \be 
\sZ_{2}^{(R)}  \approx \frac{1}{N} + \frac{Z_2^{(B)}}{(Z_1^{(B)})^2}  = {1 \ov N} + e^{-S_0} {z_2 (\b) \ov z_1^2 (\b)}  \ .
\label{z2}
\ee
Equations~\eqref{emn} and~\eqref{z2} are compatible only when $N \ll e^{S_0}$, but the derivation of~\eqref{disg} 
uses only $Z_1 \sim N e^{S_0} \gg 1$ and, in particular,  does not 
need to assume any relative magnitude of $e^{S_0}$ and $N$. 
 
In~\cite{replica_1}, the same apparent disagreement was observed, and it was pointed out that~\eqref{disg} and~\eqref{z2} can be compatible if the Euclidean gravity prescription for computing them is interpreted as an average over an ensemble of theories, so that there is a difference between the averages $\overline{(\rho_{R})_{ij}}~\overline{(\rho_{R})_{ij}^{\ast}}$ and $\overline{|(\rho_{R})_{ij}|^2}$.  We now show that 
the conflict between~\eqref{disg} and~\eqref{z2} can be naturally resolved using the equilibrium approximation interpretation of 
 the Euclidean gravity prescription, without the need for any averages.

 For this purpose, let us estimate the term $\De_{ij}$ we dropped in reaching~\eqref{disg} using the analogous procedure 
 to~\eqref{varR}, which  is computed in Appendix~\ref{app:A}. The results are 
 \be \label{sdj}
 (\De_{ij})^2_{\text{eq app}} = {1 \ov N^2}  \de_{ij} \frac{Z_2^{(B)}}{(Z_1^{(B)})^2}  , \qquad 
 |\De_{ij}|^2_{\text{eq app}} = {1 \ov N^2}   \frac{Z_2^{(B)}}{(Z_1^{(B)})^2}\ .
 \ee 
 Comparing~\eqref{sdj} with~\eqref{disg}, we see that $\De_{ij}$ is suppressed by a factor $e^{-\ha S_0}$ compared with the leading order contribution~\eqref{disg}. So the approximation~\eqref{disg} appears to be justified in the limit $e^{S_0} \to \infty$. 
 This still does not say anything about the relative magnitude between $N$ and $e^{S_0}$, so the tension between~\eqref{emn} and~\eqref{z2} remains. But notice that 
 \be \label{yhn}
 \sZ_{2}^{(R)}  = \sum_{i,j} |(\rho_R)_{ij} |^2 = \sum_i   |(\rho_R)_{ii} |^2 + \sum_{i\neq j} |(\rho_R)_{ij} |^2
 = {1 \ov N} + {N^2 - N \ov N^2}  \frac{Z_2^{(B)}}{(Z_1^{(B)})^2} \approx  
  {1 \ov N} +  \frac{Z_2^{(B)}}{(Z_1^{(B)})^2} 
 \ee
which recovers~\eqref{z2}. In~\eqref{yhn}, we have used~\eqref{disg} for the diagonal terms, and the second equation of~\eqref{sdj} for the off-diagonal terms. 

We thus see that although the off--diagonal elements $|\De_{ij}|$ are higher order in $e^{-S_0}$ compared with 
the diagonal elements, there are many more of them ($O(N^2)$) than the number $N$ of the diagonal terms. So $\sZ_{2}^{(R)} $
can receive significant contributions from these off-diagonal terms outside the regime $N \ll e^{S_0}$, and hence the corrections to the equilibrium approximation for the matrix elements are important while estimating $\sZ_2^{(R)}$. Since the first equation of~\eqref{sdj} vanishes for off-diagonal elements, we conclude that 
the off-diagonal elements $\De_{ij}$ must be complex and likely time-dependent.

Our explanation here is consistent with an idea discussed in \cite{replica_1}, that an average over time rather than an average over theories may also explain the disagreement between \eqref{disg} and \eqref{z2}, as the equilibrium approximation should agree with an average over time at late times.

We would like to emphasize that the conceptual picture obtained here is different from certain possibilities proposed in \cite{replica_1}. It was suggested there that bulk geometry may only provide ``an effective, coarse-grained description" of some ``different, more fundamental degrees of freedom" that correspond to the boundary theory, and that such non-geometric degrees of freedom may need to be added to calculate quantities like~\eqref{disg} and~\eqref{z2} to an accuracy that allows us to avoid the apparent disagreement.  Here, we emphasize that the conflict arises from dropping $\De_{ij}$ in~\eqref{pih}, which is equivalent to approximating these quantities with  Euclidean path integrals like in \eqref{jhq}. It therefore arises even before we use  semiclassical gravity to evaluate these Euclidean path integrals, and is thus not linked to using semiclassical gravity. In particular, this means that the approximation can in principle be improved within the framework of semiclassical gravity, if we are able to perform a Lorentzian rather than Euclidean calculation of these quantities. 

For a different perspective on how replica wormholes can emerge without the need for ensemble averaging, see \cite{eth_wormholes}.

\subsection{A model for an eternal black hole coupled to a bath} \label{sec:ebh}

We now consider a quantum-mechanical system that corresponds to the gravitational system discussed in~\cite{Almheiri:2019qdq} (see also~\cite{mathur, Al3, Al4,Rozali:2019day}), which consists of an eternal black hole in AdS$_2$ coupled to a flat (1+1)-dimensional bath system with speed of light $c=1$, see Fig.~\ref{fig:ebh}(a). In the quantum-mechanical system, shown in Fig.~\ref{fig:ebh}(b), the eternal black hole is replaced by a boundary dual. Note that the bath system remains the same in two descriptions.
The whole system is initially put in a pure state which does not have any entanglement 
between the black hole and bath subsystems.\footnote{In~\cite{Almheiri:2019qdq}, the system is initially in a thermal field double state between the left and the right. Our discussion will be insensitive the precise choice of the initial state.} 
We also assume the interactions between the black hole and the bath are local.

\begin{figure} [!h] 
\centering
\includegraphics[height=4.5cm]{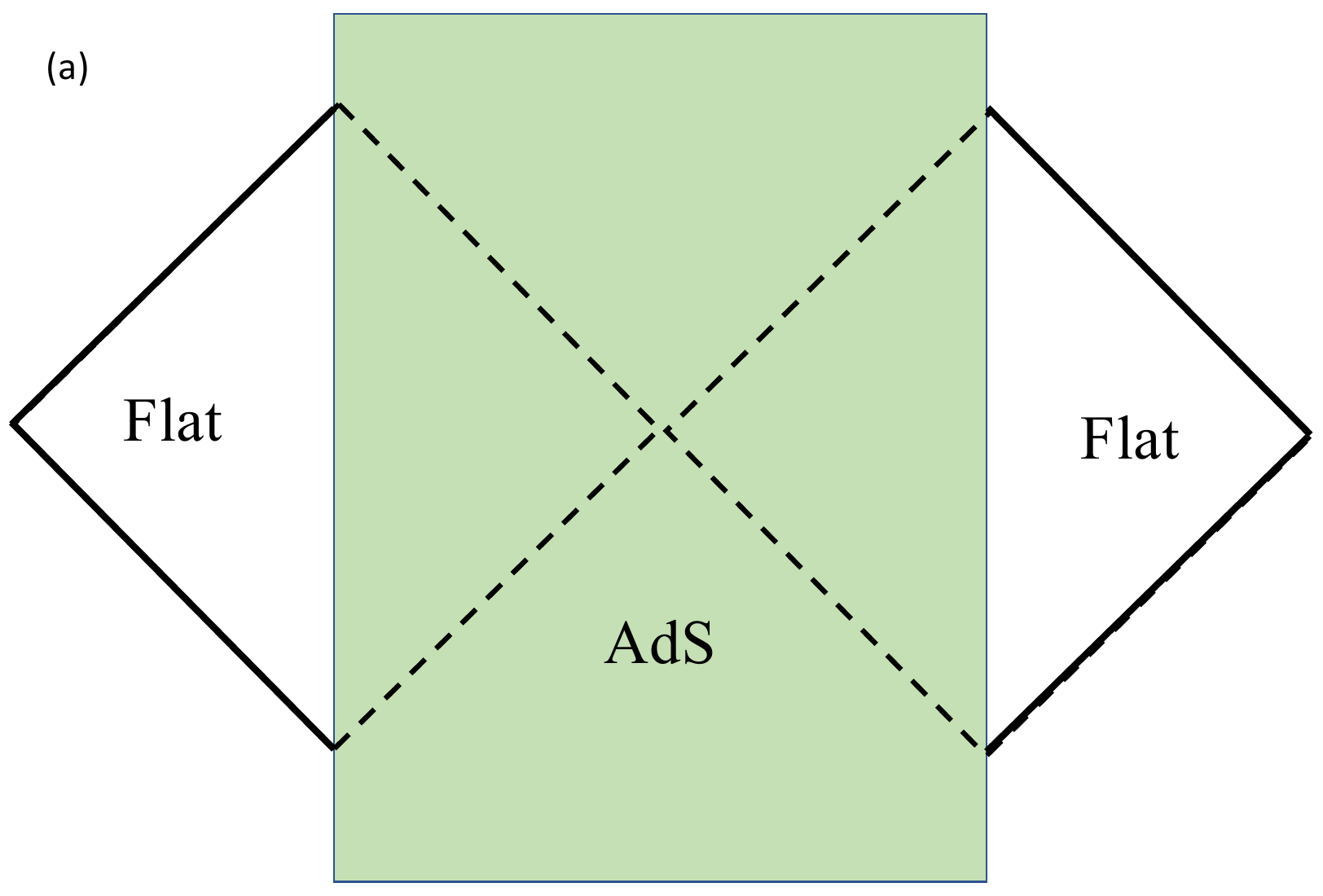}~~ \includegraphics[height=4.5cm]{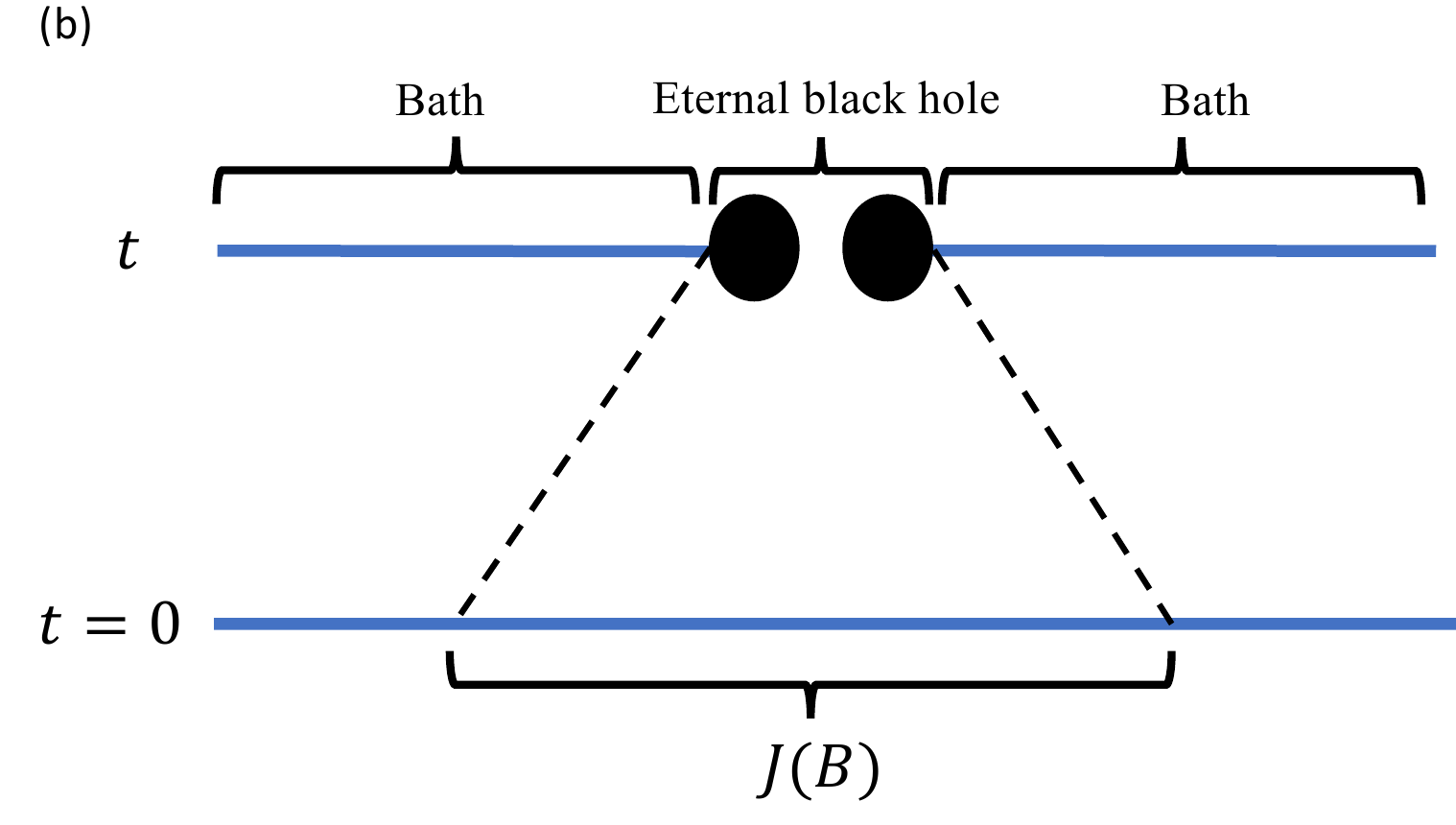} 
\caption{(a) shows the Penrose diagram for an eternal black hole in $AdS_2$ coupled to a bath, the system discussed in 
\cite{Almheiri:2019qdq}. The shaded region coupled to JT gravity corresponds to the black hole, while the unshaded region with flat Minkowski space corresponds to the bath. (b) shows a quantum-mechanical system dual to this gravitational system. In this dual theory, the black hole is a (0+1)-dimensional system while the bath is (1+1)-dimensional. At time $t$, only the part of the time-evolution operator in the region $J(B)$ is relevant for finding the entanglement entropies between the black hole and the bath.}
\label{fig:ebh}
\end{figure} 

The full system is uncompact with local interactions, as in the case discussed in Sec.~\ref{sec:uncompact}. We can therefore apply the discussion of that subsection to the current context, replacing the subsystem $A$ there by
the subsystem describing the eternal black hole.  In particular, for the purpose of studying entanglement between the black hole and the bath at some time $t$, it is enough to consider the finite region $J(B)$ around the black hole determined by causality, as indicated in Fig.~\ref{fig:ebh}(b). When the bath is maximally efficient in thermalization as in our discussion of Sec.~\ref{sec:uncompact},  we can immediately write down the Renyi entropies for either subsystem as a function of time, 
\be \label{ent_bh}
S_n^{(\rm BH)} = S_n^{(\rm bath)} =   {\rm min} (2 \sS^{\rm (BH)}_n, \sS^{(\rm bath)}_n (t) \,), \quad \sS^{(\rm bath)}_n (t) = 2t s^{\rm eq}_{n} 
\ee
where $\sS^{\rm (BH)}_n$ are the thermal entropies for the black hole, and $s^{\rm eq}_{n}$ is the equilibrium  entropy density for the bath. The entropies increase linearly and then saturate at $t_n = \sS^{\rm (BH)}_n / s^{\rm eq}_{n}$. The expression is valid for $t$ not close to $t_n$, and agrees with 
the results for the entanglement entropy in \cite{Almheiri:2019qdq}. 

Using the duality between the black hole and its boundary description, we can understand  the two contributions in~\eqref{ent_bh} more explicitly from the gravity perspective. For this purpose it is convenient to start with the path integral representation~\eqref{jhq1}  for the boundary description of Fig.~\ref{fig:ebh}(b), with  $A = {\rm bath}$ and $\bar A = {\rm BH}$. 
Then the linearly increasing contribution to~\eqref{ent_bh} comes from the configuration in Fig.~\ref{fig:pathint}(a) with $\tau =e$, while the saturation value for $t >t_n$ comes from the configuration in Fig.~\ref{fig:pathint}(b) with $\tau = \eta$. 
For $\tau = e$, one traces over the black hole subsystem within each replica copy. On the gravity side this corresponds to 
the standard evaluation of the black hole partition function using the Euclidean black hole geometry. 
For $\tau = \eta$,  the black hole subsystems for different replica copies are now connected, and on the gravity side 
 this requires the introduction of replica wormholes. 
 The contributions from other values of $\tau$, such as the example in 
Fig.~\ref{fig:pathint}(c), involve other types of replica wormholes, and are only relevant in a relatively short time interval around the transition time $t_n$ between linear growth and saturation.

Our assumption that the bath is maximally efficient in thermalization is not important for obtaining the qualitative features of the results here. The specific physical nature of the bath system may affect the specific form of $\sS^{(\rm bath)}_n (t)$, and the time scale 
for saturation. However, it will not change the fact that entanglement entropies will saturate at $2 \sS^{\rm (BH)}_n$ 
due to the contribution from Fig.~\ref{fig:pathint}(b). In particular, the bath can in principle be a free theory instead of a chaotic system that leads to rapid thermalization, like one of the toy models considered in~\cite{page_void}.

\subsection{Unitarity of Renyi entropies in more general holographic systems} 

In the previous subsections, we considered situations where only one of the subsystems has a gravity description, 
and the path integral involving that subsystem could be turned into a gravity calculation involving 
replica wormholes. However, the unintuitive couplings between replicas should appear in more general contexts. Consider an initial  pure state which subsequently undergoes gravitational collapse to form a big black hole in AdS$_{d+1}$, which is stable and does not evaporate. 
In the boundary language, the system settles into an equilibrated pure state corresponding to the black hole. 
The von Neumann and Renyi entropies of this final state should satisfy the unitarity constraint~\eqref{unit}. 

In holography, the von Neumann entropy $S_1^{(A)}$ for a subregion $A$ is found from the area of the HRT surface~\cite{rt, hrt}. For a black hole formed from collapse, the HRT surface for $\bar A$ is the same as that for $A$, 
and thus the unitarity constraint $S_1^{(A)} = S_1^{(\bar A)}$ is automatically satisfied~\cite{lopez}.  

For the Renyi entropies, it was less well-understood how unitarity is maintained. From our discussion of Sec.~\ref{sec:univ}, the Renyi entropies of a subsystem in the boundary field theory can be obtained by the path integrals in Fig.~\ref{fig:pathint}, and correspondingly the in the bulk calculation, the Renyi entropies can be obtained from 
certain Euclidean black hole geometries, despite the fact that a black hole from collapse does not have a Euclidean analytic continuation. However, the boundary conditions for these gravity path integrals include the unconventional ones specified by permutations $\tau$, as indicated for example in Fig.~\ref{fig:pathint} (b)-(c). From Sec.~\ref{sec:univ}, at leading order in the large $N$ expansion, one should consider 
two types of bulk geometries, one for the boundary conditions with $\tau=e$ and one for those with $\tau =\eta$ in Fig.~\ref{fig:pathint} (a) and (b).   Bulk manifolds with more exotic boundary conditions such as those in Fig.~\ref{fig:pathint} (c)  
provide subdominant corrections which are exponentially suppressed in the large $N$ limit. 


\section{Typicality and the random void distribution} \label{sec:rvd}

In the discussion of the previous sections, we assumed that the time-evolution operator $U$ can take a system from a far-from-equilibrium state to an equilibrated pure state, and furthermore that $U$ is such that the contribution from 
$\sZ_{n,Q}^{(A)}$ in~\eqref{may} can be neglected. For  a finite-dimensional system at infinite temperature, the  approximation yields the same results as those obtained from the Haar-random averages of the quantities $\sZ_{n}^{(A)}$ over the full Hilbert space. Thus, the suppression of $\sZ_{n,Q}^{(A)}$ 
may be viewed as a dynamical criterion for the evolution of a system towards typical states in such systems. 
However, since we expect on general grounds that evolution to typicality should take place in chaotic systems, it would be good to understand more directly which aspects of chaos are responsible for it. In this section, 
we will offer some suggestions from the perspective of operator growth.

We conjectured in \cite{void} that one general feature of operator evolution in chaotic systems at late times is the form of the probability of ``void formation" in them, which we referred to as the random void distribution.  In~\cite{page_void}, based on studies of the second Renyi entropy, we argued that typicality is a direct consequence of the random void distribution.

  In this section, we generalize the notion of the random void distribution to higher moments, and show that the behavior of all higher Renyi entropies of an equilibrated pure state can be seen as special cases of the generalized random void distribution. 
Below, we first review the notion of 
void formation in operator growth, and 
discuss its relevance for typicality. For simplicity, we will only consider a finite-dimensional Hilbert space with no energy constraint (that is, the infinite temperature case). 


\subsection{Random void distribution and typicality} 
Consider the Heisenberg evolution $O(t)$ of an initial operator $O$. Since the identity operator does not evolve, in the discussion below we always assume that $O$ does not have an identity part, i.e. $\Tr O =0$.  With respect to a subsystem $S$, we can decompose $O(t)$ as a sum of two parts, 
\be 
O(t)=O^{(1)}(t) + O^{(2)}(t), \quad  O^{(1)}(t) = \mathbf{1}_S \otimes \tilde{O}_{\bar{S}}, \quad {\rm Tr}_S[O^{(2)}(t)]=0 
\ee
where $\mathbf{1}_S$ is the identity operator for subsystem $S$ and $\tilde{O}_{\bar{S}}$ is some operator in $\bar S$. 

We refer to the presence of $O^{(1)}(t)$ in $O(t)$ as void formation in the subsystem $S$. Using the following inner product between any two operators $A$ and $B$, 
\be 
\braket{A, B} = \frac{1}{d} \Tr[A^{\dagger} B]
\ee
where $d$ is the dimension of the full Hilbert space, we can define the weight (or ``probability'')  that an operator $O$ forms a void in the subsystem $S$ at time $t$ as 
\be \label{heni} 
P_{O,2}^{(S)}(t) = \frac{\braket{O^{(1)}(t), O^{(1)}(t)}}{\braket{O(t), O(t)}}  
\ee 

Moreover, based on studies in random unitary circuits, it was conjectured in~\cite{void} that in a chaotic system, for a generic traceless initial operator $O$, at sufficiently late times the probability $P_{O,2}^{(S)}(t)$ has a simple universal form 
\be 
P_{O,2}^{(S)}(t) =  \frac{1}{d_S^2} , \qquad d_{\bar S} \gg 1, 
\label{rvd}
\ee 
which was referred to as the random void distribution. 
We note that~\eqref{rvd} should apply essentially to all traceless operators, for example, in a spin chain, to local operators, basis operators which cover a finite region, and superpositions of non-trivial basis operators.

We now show that if we assume \eqref{rvd} applies to the non-identity part of a density matrix for a pure state we obtain~\eqref{ohn2} for $n=2$. The discussion below can be seen as a model-independent version of the argument in~\cite{page_void} for the derivation of the Page curve of black hole evaporation using the random void distribution. For this purpose, we decompose the initial density matrix as 
\be 
\rho_0 = \ket{\Psi_0} \bra{\Psi_0} = \frac{1}{d} \mathbf{1} + \hat{\rho}_0,  ~~~ \Tr \hat{\rho}_0 = 0 \  . \label{hatrho}
\ee
Given the initial state is pure,
\be 
\rho_0^2 = \rho_0 \quad \Rightarrow \quad \hat{\rho_0}^2 = \frac{d-1}{d^2} \mathbf{1} + \frac{d-2}{d}\hat{\rho_0} \quad \Rightarrow \quad \Tr{\hat{\rho}_0^2} \approx 1 \label{rho02}
\ee
where the final statement is true in the large $d$ limit. To find the reduced density matrix for a subsystem $A$, we further decompose $\hat{\rho}(t) \equiv U \hat{\rho}_0 U^{\dagger}$ as 
\be 
\hat{\rho}(t) = \hat{\rho}^{(1)} + \hat{\rho}^{(2)}, \quad \hat \rho^{(1)} = {O}_A \otimes \mathbf{1}_{\bar{A}}, \quad {\rm Tr}_{\bar{A}}  \hat{\rho}^{(2)} =0
\ee
for some $O_A$ in subsystem $A$. Tracelessness of $\hat \rho$ implies that $\Tr_A {O}_A = 0$. 
It then follows that the reduced density matrix for $A$ has the form 
\be \label{s19}
\rho_A(t) = \frac{1}{d_A}~ \mathbf{1}_A + \hat \rho_A , \quad {\rm Tr}_A \hat \rho_A = 0, \quad \hat \rho_A = d_{\bar{A}} \, {O}_A
\ee
and the second R\'enyi entropy for $\rho_A (t)$ can be written as 
\be \label{s20}
e^{-S_2^{(A)}} = {\rm Tr}_A \rho_A^2 (t) = \frac{1}{d_A} +  {\rm Tr}_A \hat \rho^2_A = \frac{1}{d_A} + d_{\bar{A}} \Tr[(\hat{\rho}^{(1)})^2]  \ .
\ee
Now assuming the random void distribution~\eqref{rvd} for operator $O = \hat{\rho}_0$ and the subsystem $S=\bar A$, we have 
\be 
 P_{\hat{\rho}_0,2}^{(\bar{A})} = \frac{\Tr[(\hat{\rho}^{(1)})^2]}{\Tr[(\hat{\rho} (t))^2]}= \frac{\Tr[(\hat{\rho}^{(1)})^2]}{\Tr[(\hat{\rho}_0)^2]} = \frac{1}{d_{\bar{A}}^2} \;\; \Rightarrow \;\;
 \Tr[(\hat{\rho}^{(1)})^2] = \frac{1}{d_{\bar{A}}^2}  \;\; \Rightarrow \;\; {\rm Tr}_A \hat \rho^2_A = {1 \ov d_{\bar A}} 
\label{rvd_rho}
\ee
where we have used~\eqref{rho02}.  We thus  find
\be \label{uhn}
e^{-S_2^{(A)}} = \frac{1}{d_A} + \frac{1}{d_{\bar{A}}} 
\ee
Equation~\eqref{sehn} with $n=2$ then follows immediately. 
The first term in~\eqref{uhn} is the contribution from the identity, and the second term comes from processes of void formation in 
$\bar{A}$ under the action of $U$.

\subsection{Higher moments of the random void distribution}

Just like the behavior of the second Renyi entropy may be considered a direct consequence of the random void distribution, our discussion 
of the Renyi entropies in Sec.~\ref{sec:univ} can be considered a special case of the random void distribution for
higher moments of general operators. 

More explicitly, consider the generalization of~\eqref{heni} to higher $n$: 
\be \label{ejk}
P_{O,n}^{(S)} (t) = \frac{\Tr[O_1(t)^n]}{\Tr[O^n]} = \frac{1}{d_S^{n-1}}\frac{1}{\Tr[O^n]}~\text{Tr}_{\bar{S}}[(\text{Tr}_S[O(t)])^n] \ .
\ee
The $n$-th Renyi entropy for $A$ corresponds to taking $O = \rho$ and $S = \bar A$. 
The general approximation scheme we developed in Sec.~\ref{sec:univ} can be used to 
find~\eqref{ejk}, assuming there exists a time scale such that the quantity $P_{O,n}^{(S)} (t) $ saturates and the contribution from projector $Q$ can be neglected.  More explicitly, we find in the large $d$ limit that 
\be 
\text{Tr}_{\bar{S}}[(\text{Tr}_S[O(t)])^n] ={1 \ov d_{\bar S}^{n-1}}(\Tr O)^n + \sum_{\tau \neq e, \eta} d_{\bar S}^{k(\eta^{-1} \tau)-n} d_{S}^{k(\tau)-n}   \braket{\tau| O, e}  + {1 \ov d_{ S}^{n-1}} \Tr[O^n]
  \label{av_tro}
\ee
where we have separated the contribution from $\tau =e $ and $\tau =\eta$ explicitly. Recall that 
\be 
\vev{\tau|O,e} =  \Tr O^{n_1}  \cdots  \Tr O^{n_k}
\ee
where $k = k(\tau)$ and $n_i$ are the lengths of the cycles of $\tau$. If $O$ is traceless, 
\be 
\begin{gathered}
\text{Tr}_{\bar{S}}[(\text{Tr}_S[O(t)])^n] =\sum_{\substack{\sigma \neq e, \eta, \\ \text{$\tau$ has no cycle with}\\ \text{one element}}} d_{\bar S}^{k(\eta^{-1} \tau)-n} d_{S}^{k(\tau)-n}   \braket{\tau| O, e}  +  {1 \ov d_{ S}^{n-1}} \Tr[O^n] \ .
 \end{gathered}
 \label{trle}
\ee
In particular, with $\Tr[O^n] \sim O(1)$ (i.e. independent of $d_S, d_{\bar S}, d$),  then for $d_{S} \ll d_{\bar S}$ we have 
\be \label{nel}
P_{O,n}^{(S)} (t)  = {1 \ov d_S^{2(n-1)}} \ .
\ee
Heuristically, \eqref{rvd} and \eqref{nel} suggest that the operator has become uniformly spread throughout the system, so that the probability that it is localized within any small subsystem is exponentially suppressed in the number of degrees of freedom in that subsystem.

Setting $O$ in~\eqref{av_tro} to be $\rho_0$ and $S= \bar A$, we then recover~\eqref{ohn2}. 
Setting in~\eqref{trle} $O$ to be $\hat \rho_0$ (i.e. the traceless part of $\rho$) and $S= \bar A$, we obtain 
a generalization of~\eqref{rvd_rho} to higher $n$ (in the regime of~\eqref{emm}) 
\be\label{y1}
{\rm Tr}_A [\hat{\rho}_A^n] =  \frac{1}{d_{\bar{A}}^{n-1}} + \sum_{p=2}^{n-1}~ \tilde N(n, p)~ \frac{1}{d_A^{n-p} d_{\bar{A}}^{p-1}} 
\ee
where $\tilde N(n, p)$ is the number of non-crossing partitions of $n$ objects into $p$ blocks such that each block has more than one element.
$\tilde N (n,p)$ are called the Riordan numbers\footnote{See the discussion around equation (2.1) of \cite{sequences}. There does not seem to be a simple closed expression for them, but they can be obtained from a generating function~\cite{sequences}}.
The consistency of~\eqref{y1} with~\eqref{ohn2} leads to a nice relation between the Riordan and Narayan numbers
\be 
N(n,p) = \sum_{k=p}^{n} C_n^k \, \tilde N (k, p)  ={1 \ov n} C_n^p C_n^{p-1} \ .
\ee

It is tempting to conjecture that~\eqref{trle}--\eqref{nel} apply to a general chaotic system. 
In Appendix~\ref{app:D} we show that they hold in the local random unitary circuits of~\cite{nahum1, frank}.

\section{Conclusions and discussion} \label{sec:conc}

In this paper, we developed an approximation to calculate the entanglement entropies of an equilibrated 
pure state.  The resulting expressions can be written solely in terms of the partition functions and thermodynamic entropies of the equilibrium density operator $\rho^{(\rm eq)}$, but at the same time are compatible with unitarity. One immediate implication is that a set of Euclidean path integrals for the equilibrium density operator emerge universally as an approximation to the Lorentzian path integrals for Renyi entropies, with a variety of boundary conditions specified by different permutations of the replica systems. We introduced a criterion for checking that the approximation 
is self-consistent, which at the same time provides an estimate of the contribution  $\sZ_{n,Q}^{(A)}$ we neglected. 
We also extracted the universal behavior of the entanglement entropies for various classes of equilibrated pure states.  

Applied to two recently discussed models of  black holes~\cite{replica_1,Almheiri:2019qdq}, the equilibrium approximation leads to a derivation of the prescriptions proposed in these papers for including replica wormholes in the calculation of entanglement entropies, and provides a general explanation for why such a prescription leads to results compatible with unitarity. Replica wormholes are thus one manifestation of a universal structure which appears in a large variety of thermalizing systems, including quantum-mechanical systems and quantum field theories without holographic duals. Our derivation can be used to see when and how replica wormholes should be included in more general gravity theories, and in particular it shows that they can arise in systems with a fixed Hamiltonian, without any need for an ensemble average. 

We further discussed a mechanism for equilibration in the infinite-temperature case from the perspective of operator growth. The underlying property of operator growth, called the random void distribution, can be seen as a more direct manifestation of quantum chaos than the assumptions that went into the equilibrium approximation.

One important open question for the future is to understand better the role of the contribution $\sZ_{n,Q}^{(A)}$ which we neglected, and to 
 develop a further approximation scheme to capture its effects~(see~\cite{adam} for a discussion in the infinite-temperature case). 
A related question is about the time scales for which our equilibrium approximation should be valid. We expect it to be valid for time scales much longer than the thermalization scale $t_s$, but perhaps not at very long time scales at which large fluctuations could become relevant. 

It is also interesting to consider for which other observables the approximation is expected to work well, and for which observables it is expected to fail. Let us discuss two examples. For correlation functions of the form 
\be 
F = \vev{\psi_0 |\sO_1 (t_1) \sO_2  (t_2)  \cdots \sO_n (t_n) |\psi_0},  
\ee
if the smallest time $t_m$ is much greater than $t_s$, then we can apply the equilibrium approximation, which gives the equilibrium correlation functions associated with $\rho^{\rm (eq)}$ (see Appendix~\ref{app:A} for details):
\be \label{yue}
F \approx \Tr (\rho^{\rm (eq)} \sO_1 (t_1-t_m) \sO_2  (t_2-t_m)  \cdots \sO_n (t_n-t_m)). 
\ee
 Using an analogous criterion to the one discussed in Sec.~\ref{sec:cri}, we show in Appendix~\ref{app:A} that~\eqref{yue} is valid provided that the minimal subsystem $A$ in which $\sO \equiv \sO_1 (t_1-t_m) \sO_2  (t_2-t_m)  \cdots \sO_n (t_n-t_m)$ is contained is much smaller than its complement  
$\bar A$. This is also consistent with what we expect based on the behavior of Renyi entropies in this regime (recall~\eqref{hnel0}--\eqref{hnel}).

As a final example, let us consider the quantity
\be \label{ex1}
\sZ  \equiv  (\Tr U \Tr U^\da)^n  = \hat \Tr (U \otimes U^{\dagger})^{n} 
\ee
where in the second equality we have expressed the quantity in the replica system, with $\hat \Tr$ denoting the trace in $(\sH \otimes  \sH)^{n}$. This is the $n$th power of the spectral form factor studied in \cite{ss1}. This quantity does not correspond  to the equilibration of a far-from-equilibrium initial state or have an equilibrium value. We therefore intuitively expect that the equilibrium approximation should not make sense for \eqref{ex1}. Indeed, for any choice of $\sI_{\al}$, on applying the equilibrium approximation to the above expression, we get $\sZ \approx n!$, which does not satisfy the self-consistency criterion of Sec.~\ref{sec:cri}.\footnote{We can see this immediately by taking $n =2 m$, and noting that $(2m)!-(m!)^2$ is always $\geq (m!)^2$ for $m\geq 1$.} 

Based on studies in random matrix theory and the SYK model, the spectral form factor is expected to have a linear ``ramp" and eventually a constant ``plateau" of order $e^S$ (where $S$ is the thermodynamic entropy) at late times in chaotic systems with conserved energy, both of which are not captured by the equilibrium approximation. In the JT gravity calculations of \cite{ss1, ss2}, the ramp contribution to the spectral form factor is correctly captured by ``trumpet" geometries, which involve bulk connections between disconnected closed boundaries. While these structures geometrically resemble the replica wormholes of \cite{replica_1}, our observations above imply that these two kinds of connected geometries appearing in two different quantities have distinct physical origins. The replica wormholes of \cite{replica_1}, appearing in the calculation of the Renyi entropies, can be fully explained within the framework of the equilibrium approximation. On the other hand, the ramp contributions to the spectral form factor, and hence the trumpet geometries associated with them, appear to be beyond the scope of the equilibrium approximation. The trumpet geometries lead to a disagreement between the direct evaluation of $\Tr[U]\Tr[U^{\dagger}]$ and the product of $\Tr[U]$ and $\Tr[U^{\dagger}]$ in the gravity calculation, which is similar to the conflict between \eqref{disg} and \eqref{z2} discussed in section \ref{sec:av}. However, while the equilibrium approximation can resolve the issue discussed in section \ref{sec:av}, it does not seem to explain the factorization problem of the spectral form factor.

In a different future direction, it would also be interesting to understand how the mechanism for equilibration based on operator growth in the infinite-temperature case in section \ref{sec:rvd} generalizes to other choices of $\sI_{\al}$.  

\vspace{0.2in}   \centerline{\bf{Acknowledgements}} \vspace{0.2in}
We would like to thank Ping Gao and Sam Leutheusser for helpful discussions. 
This work is supported by the Office of High Energy Physics of U.S. Department of Energy under grant Contract Number  DE-SC0012567.

\appendix

\section{Equilibrium approximation for a general initial density matrix}
\label{app:general_state}

Here we discuss the generalization of the equilibrium approximation to a general initial density operator rather than a pure state which has been the focus of the main text. 

Consider the quantities
\be 
z_n = \Tr \rho_0^n = \Tr \rho^n = \Tr (U \rho_0 U^\da)^n = \vev{\eta|(U \otimes U^{\da})^n| \rho_0,e}.  
\ee
Applying the equilibrium approximation to the above expression, we have 
\be 
z_n = {1 \ov Z_2^n} \sum_\tau \vev{\eta|\sI_\al , \tau} \vev{\sI_\al, \tau| \rho_0, e}  \ .
\ee 
In the case where $\rho_0$ is a pure state, recall that $A_n =   \vev{\sI_\al, \tau| \rho_0, e} $ is independent of $\tau$, and 
by imposing~\eqref{r11} we obtain $z_n \approx 1$ as discussed in~\eqref{rnorm}. For a $\rho_0$ which is a mixed state,  requiring the $n=1$ equation to be satisfied again 
fixes 
\be 
\Tr \sI_ \al \rho_0 = {Z_2 \ov Z_1}  \ .
\ee
For $n=2$, 
\be 
z_2 = {1 \ov Z_2^2} \le(\vev{\eta|\sI_\al , e} \vev{\sI_\al, e| \rho_0, e} + \vev{\eta|\sI_\al , \eta} \vev{\sI_\al, \eta| \rho_0, e}\ri)
= {1 \ov Z_2^2} \le(Z_2 {Z_2^2 \ov Z_1^2} + Z_1^2 \Tr (\sI_\al \rho_0)^2  \ri)
\ee
which requires 
\be \label{z2_small}
\Tr (\sI_\al \rho_0)^2  = \le(z_2 - {Z_2\ov Z_1^2} \ri) {Z_2^2 \ov Z_1^2}  \ .
\ee
Since $Z_2/Z_1^2 \sim Z_1^{-1} \ll 1$, for $z_2 \sim O(1)$ we then have
\be 
\Tr (\sI_\al \rho_0)^2  \approx z_2  {Z_2^2 \ov Z_1^2}  \ .
\ee
Continuing this further, the equation for $z_n$ can be used to determine $\Tr [(\sI_\al \rho_0)^n]$ in terms 
$z_1, \cdots z_n$ and $Z_1, \cdots Z_n$. Each of these equations should be imposed as a self-consistency condition 
on $\sI_\al$. In particular for $z_n \sim O(1)$, we have  
\be 
\Tr (\sI_\al \rho_0)^n  \approx z_n  {Z_2^n \ov Z_1^n}  \ .
\ee

Now for a general density operator $\rho_0$ we have in the large $\sZ_1$ limit 
\be \label{fen_large}
\sZ_{n}^{(A)} \approx \sZ_{n,P}^{(A)}  =  {1 \ov Z_2^n}  \sum_{\tau}  \vev{\eta_A \otimes e_{\bar A} | \sI_\al , \tau} \vev{\sI_\al, \tau|\rho_0,e}, \quad n=2,3, \cdots  \ .
\ee
Now using~\eqref{z2_small} and it higher $n$ counterparts we can express~\eqref{fen_large} in terms of $z_1, \cdots z_n$ and 
various partition functions of $\sI_\al$. When $z_n \sim O(1)$ for all $n$, we have
\be
\sZ_{n}^{(A)} \approx \sZ_{n,P}^{(A)}  =  {1 \ov Z_1^n}  \sum_{\tau}  b_\tau \vev{\eta_A \otimes e_{\bar A} | \sI_\al , \tau} , \quad b_\tau = z_{n_1} \cdots z_{n_k} , \quad n=2,3, \cdots  
\ee
where $k$ is the number of cycles of $\tau$ with $n_1, \cdots n_k$ the lengths of the cycles. 
For general $z_n$, the explicit expressions for~\eqref{fen_large} are somewhat complicated. For example, for $n=2$ and $n=3$, we find 
\be 
\begin{gathered} 
\sZ_2^{(A)} 
\approx \sZ_n^{(A, \rm eq)} + \le(z_2- \frac{Z_2}{Z_1^2} \ri) \sZ_n^{(\bar{A}, \rm eq)}
\end{gathered}
\ee
and 
\be 
\begin{gathered}
\sZ_3^{(A)} 
\approx  \sZ_n^{(A, \rm eq)} + 3 \le(z_2- \frac{Z_2}{Z_1^2}\ri) \Tr[(\rho^{\rm (eq) }_A\otimes\rho^{\rm (eq) }_{\bar{A}}) \rho^{\rm (eq) }] + \\\le(z_3-\frac{Z_3}{Z_1^3} - 3 \frac{Z_2}{Z_1^2}\le(z_2-\frac{Z_2}{Z_1^2}\ri)\ri) \le(  \sZ_n^{(\bar{A}, \rm eq)} + \braket{\eta_A\otimes e_{\bar{A}}|\rho^{\rm (eq) }, \eta^{-1}} \ri). 
\end{gathered}
\ee

\section{Estimate of $\sZ_{n,Q}^{(A)}$ in various cases} \label{app:A} 

Here we discuss the calculation of~\eqref{varR} to show the self-consistency of the equilibrium approximation, and also the analogous quantities for the equilibrium approximation of other observables. 

For any quantity $\sT$ that can be written as a transition amplitude in a replica Hilbert space, we can separate $\sT$ as a sum of two parts,
\be 
\sT = \vev{b | (U \otimes U^{\da})^n |a}  = \sT_P + \sT_Q , \qquad \sT_P = \vev{b| P_\al| a} , 
\ee
for some states $\ket{a}, \ket{b} \in (\sH \otimes  \sH)^n$. The definition of $\sT_P, \sT_Q$ are in complete analogue with~\eqref{may} for the Renyi entropies.  Note that in general $\sT$ may not be real. 
To discuss the self-consistency of the equilibrium approximation for the quantity $\sT$, similar to~\eqref{varR}, we will consider
\bega 
\le(|\sT_Q|^2\ri)_{\text{eq app}} = (|\sT|^2)_{\text{eq app}}  - |\sT_P|^2 =\vev{\bar b \otimes b| \tilde P_\al |\bar a \otimes a} - |\sT_P|^2 \\
\le(\sT_Q^2\ri)_{\text{eq app}} = (\sT^2)_{\text{eq app}}  - \sT_P^2 =\vev{b \otimes b| \tilde P_\al |a \otimes a} - \sT_P^2
\end{gather} 
where $\ket{\bar a} , \ket{\bar b} \in (\sH \otimes  \sH)^n$ are defined from $\ket{a}, \ket{b}$ by the following procedure
\be 
\vev{i_1 \bar i_1' i_2 \bar i_2' \cdots i_n \bar i_n' | a} = a_{i_1 i_1' i_2 i_2' \cdots i_n i_n'}  \quad \to \quad \vev{i_1 \bar i_1' i_2 \bar i_2'\cdots i_n \bar i_n' |\bar a} = a^*_{i_1' i_1 i_2' i_2 \cdots i_n' i_n} 
\ee
and 
\be
\tilde P_\al = {1 \ov Z_2^{2n}} \sum_{\tau \in \sS_{2n}} \ket{\sI_\al, \tau} \bra{\sI_\al , \tau}
\ee
 is the projector associated with $\sI_\al$ in $(\sH \otimes  \sH)^n \otimes (\sH \otimes \sH)^n$. 
We have again assumed $Z_1$ is large and~\eqref{idme}. 

We then have 
\bea
\le(|\sT_Q|^2\ri)_{\text{eq app}} &= &{1 \ov Z_2^{2n}} \sum_{\tau \in \sS_{2n}} \vev{\bar b \otimes b| \sI_\al, \tau} \vev{\sI_\al , \tau |\bar a \otimes a} \cr
&& \qquad \qquad - {1 \ov Z_2^{2n}} \sum_{\tau' , \sig' \in \sS_{n}} \vev{\bar b  | \sI_\al, \tau'} \vev{\sI_\al , \tau' | \bar a} 
\vev{b  | \sI_\al, \sig'} \vev{\sI_\al , \sig' | a}  \ .
\label{ueb}
\eea
When $\tau \in \sS_{2n}$ can be written in a factorized form $\tau= \sigma' \otimes \tau'$ for $\sig', \tau' \in \sS_n$ (meaning that it can be seen as a composition of a permutation $\sigma'$ involving only the first $n$ elements with a permutation $\tau'$ involving only the last $n$ elements), 
the contribution from $\tau$ in the first term cancels with the contribution from the associated $\sigma', \tau'$ in the second term. 
One can write down a similar expression for $(\sT_Q^2)_{\text{eq app}}$. Hence, 
 \bega \label{hok}
 \le(|\sT_Q|^2\ri)_{\text{eq app}} = \frac{1}{Z_2^{2n}}\sum_{\tau \neq \sigma' \otimes \tau'}  \vev{\bar b \otimes b| \sI_\al, \tau} \vev{\sI_\al , \tau |\bar a \otimes a}  \\
  \le(\sT_Q^2\ri)_{\text{eq app}} = \frac{1}{Z_2^{2n}}\sum_{\tau \neq \sigma' \otimes \tau'}  \vev{b \otimes b| \sI_\al, \tau} \vev{\sI_\al , \tau |a \otimes a} 
  \label{hok1}
 \end{gather} 
 where $\tau \neq \sigma' \otimes \tau'$ indicates that $\tau \in \sS_{2n}$ cannot be written in a factorized form.

We now consider~\eqref{hok} or~\eqref{hok1} for a few different observables.

\subsection{Renyi entropies} 

For the $n$-th Renyi entropy we take $\ket{a} = \ket{\rho_0, e}$ and $\ket{b} = \ket{\eta_A \otimes e_{\bar A}}$, for which
we have
\be \label{hok2}
\De^2 \equiv \le[\le(\sZ_{n,Q}^{(A)}\ri)^2 \ri]_{\text{eq app}} = \frac{1}{Z_1^{2n}}\sum_{\tau \neq \sigma' \otimes \tau'}  \vev{\eta_A \otimes e_{\bar A} \otimes \eta_A \otimes e_{\bar A}| \sI_\al, \tau} \ .
 \ee
As in~\eqref{bdm},  we can estimate the magnitude of the above expression by counting the number of traces for $A$ and $\bar A$,
with $\Tr_A \propto d_A, \Tr_{\bar A} \propto d_{\bar A}$ 
\be
\vev{\eta_A \otimes e_{\bar A} \otimes \eta_A \otimes e_{\bar A}| \sI_\al, \tau} \sim d_A^{k_1} d_{\bar A}^{k_2} , \quad
k_1 = k (\nu^{-1} \tau), \quad k_2 = k(\tau), \quad \nu = \eta \otimes \eta. \label{dev}
\ee
$k_1$ and $k_2$ can be interpreted respectively as the number of dashed and solid loops in the diagrams shown in Fig.~\ref{fig:m_2}. These diagrams are obtained by following exactly the same rules as those for Fig.~\ref{fig:ind_circ}--\ref{fig:rev}. 
The contractions corresponding to the final condition $\bra{\eta_A \otimes e_{\bar A} \otimes \eta_A \otimes e_{\bar A}}$ 
are given in Fig.~\ref{fig:m_2}(a). Notice that the contractions for the first group of $n$ elements and those for the second group of $n$ elements respectively form separate sub-diagrams in such a way that for any $\tau \neq \sigma' \otimes \tau'$, we always get non-planar diagrams, as such a $\tau$ necessarily connects some elements of the first group to those of the second group. 
An example is Fig.~\ref{fig:m_2}(c), while $\tau$ for Fig.~\ref{fig:m_2}(b) is factorizable and does not contribute to~\eqref{hok2}. 

 Like in the discussion around Fig.~\ref{fig:double_line} in section \ref{sec:gen}, we can obtain a double-line diagram from each of these diagrams by adding an extra surrounding loop. The total number of loops in the double-line diagram is again equal to the number of faces $F$ of the polygon associated with it, when it is placed on a manifold where it does not have crossing lines. In this case, the total number of edges of the polygon is $E=6n-1$ and the total number of vertices is $V=4n-2$, so if the diagram associated with $\tau$ can be drawn without crossings on a surface of minimum genus $h$, then
\be 
k_1+k_2 = F-1 = E-V+2-2h-1= 2n + 2-2h. 
\ee
Since we always get non-planar diagrams with $h\geq 1$ for $\tau \neq \sigma' \otimes \tau'$,
\be 
k_1 + k_2 \leq 2n \ .
 \label{k12}
\ee
Also, since any $\tau \neq \sigma' \otimes \tau'$ is not equal to either $e$ or $\nu$, and for any $\tau \neq e$, $k(\tau)\leq 2n-1$, 
\be 
k_1, k_2 \leq 2n-1. 
\ee

\begin{figure}[!h] 
\centering 
\includegraphics[width=15cm]{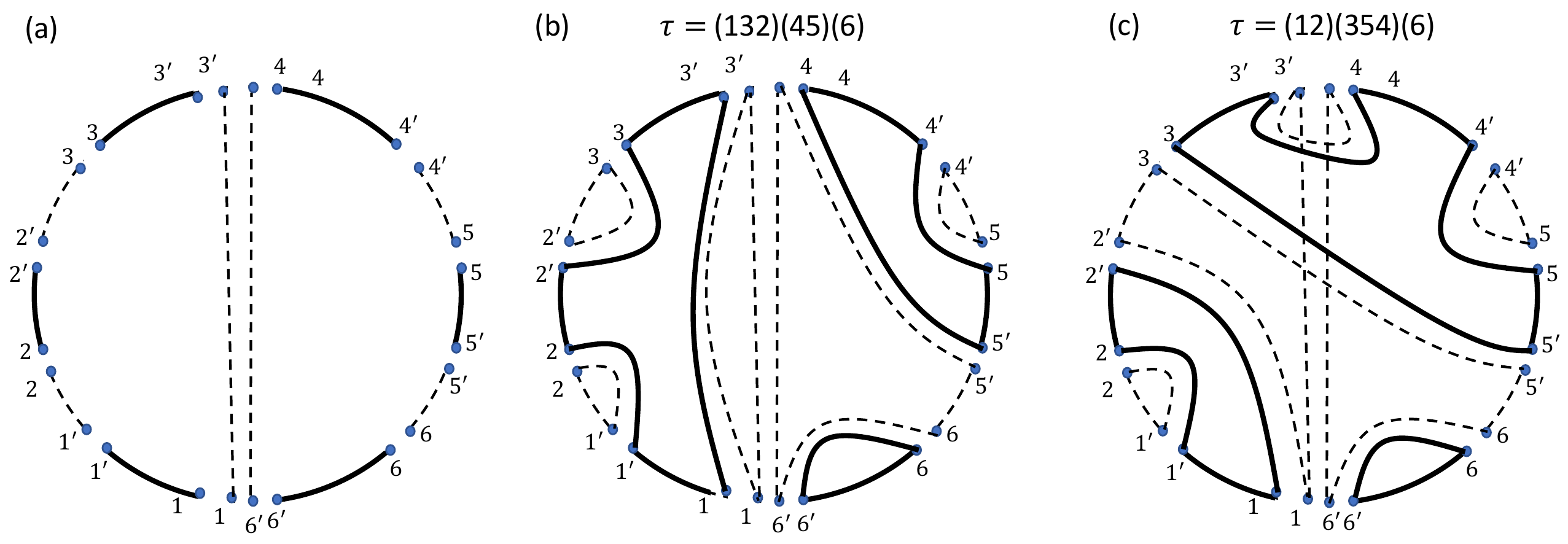}
\caption{Examples of the diagrammatic representation for~\eqref{hok2} for $n=3$.
(a) shows the contractions of indices from ``future conditions" $\bra{\eta_A \otimes e_{\bar A} \otimes \eta_A \otimes e_{\bar A}}$. (b) and (c) show examples of interior connections for two different choices of $\tau$.}
\label{fig:m_2}
\end{figure}

Recall that  for $d_A \sim d_{\bar A} \sim Z_1^\ha$, the leading term of $\sZ_{n,P}^{(A)}$ scales as $Z_1^{\ha (1-n)}$, 
while from the above $\De \sim Z_1^{-\ha n}$. When $d_A \ll d_{\bar A} \sim Z_1$, 
the leading contribution of $\sZ_{n,P}^{(A)}$ scales as $d_A^{1-n}$ while from the above $\De \sim d_A^{1/2-n} d_{\bar A}^{-\ha} $. 
In both cases we have 
\be 
{\De \ov \sZ_{n,P}^{(A)} }\sim Z_1^{-\ha} \ll 1 \ .
\ee
Also note that in both cases $d_A \sim d_{\bar A} \sim Z_1^\ha$ and $d_A \ll d_{\bar A} \sim Z_1$, the next-to-leading order correction to $\sZ_{n,P}$ is suppressed by $Z_1^{-1}$ relative to the leading contribution, so that the contribution from $\sZ_Q$ is larger than this contribution.

For $n=1$, we have (with $\eta =(12)$ below) 
\be 
\De^2 = {1 \ov Z_1^2} \vev{e|\sI_\al, \eta} = {Z_2 \ov Z_1^2} \ll 1
\ee
which is consistent with the other self-consistency condition \eqref{r11} that we introduced earlier. 

We see from the above discussion that the suppression of the correction $\sZ^{(A)}_{n,Q}$ has to do with the trace structure in the definition of the Renyi entropies. 

To see the factorization condition for higher $m$ in \eqref{higher_m}, let us now consider the quantity
\be
\De_m = \le[\le(\sZ_{n}^{(A)}\ri)^m \ri]_{\text{eq app}} - (\sZ_{n,P}^{(A)})^m = \frac{1}{Z_1^{mn}}\sum_{\tau \neq \sigma'_1 \otimes \sigma'_2...\otimes \sigma'_m}  \vev{(\eta_A \otimes e_{\bar A})^m | \sI_\al, \tau} \ . \label{dm_perm}
 \ee
 Where $\tau\neq \sigma'_1 \otimes \sigma'_2...\otimes \sigma'_m$ indicates a permutation in $\sS_{mn}$ that is not factorized among each of the $m$ consecutive sets of $n$ elements. The discussion completely parallels the $m=1,2$ cases discussed earlier, with 
 \be
\vev{(\eta_A \otimes e_{\bar A})^m | \sI_\al, \tau} \sim d_A^{k_1} d_{\bar A}^{k_2} , \quad
k_1 = k (\nu_m^{-1} \tau), \quad k_2 = k(\tau), \quad \nu_m = \underbrace{\eta \otimes \eta \otimes ... \otimes \eta}_\text{$m$ times},
\ee
$k_1$ and $k_2$ can again be seen as the number of dashed and solid loops in the diagrams of Fig.~\ref{fig:m_3} (shown for the $m=3$ case). 
Now for a diagram we have 
$E=3mn-m+1, ~ V=2mn-2m+2$. Hence, 
\be 
k_1+k_2 = F-1 = E-V+2-2h-1= mn + m-2h.
\ee 
Examples of $\tau$ corresponding to $h=0$ and $h=1$ are shown respectively in Fig.~\ref{fig:m_3} (b) and (c). For any $\tau\neq \sigma'_1 \otimes \sigma'_2...\otimes \sigma'_m$, we get a non-planar diagram, so for all terms appearing in \eqref{dm_perm},  
\be 
k_1 + k_2 \leq  mn + m-2.  \label{k122}
\ee
Also, again since any $\tau\neq \sigma'_1 \otimes \sigma'_2...\otimes \sigma'_m$ is not equal to $e$ or $\nu_m$, 
\be 
k_1, k_2 \leq mn-1.  
\ee
\begin{figure}[!h] 
\centering 
\includegraphics[width=15cm]{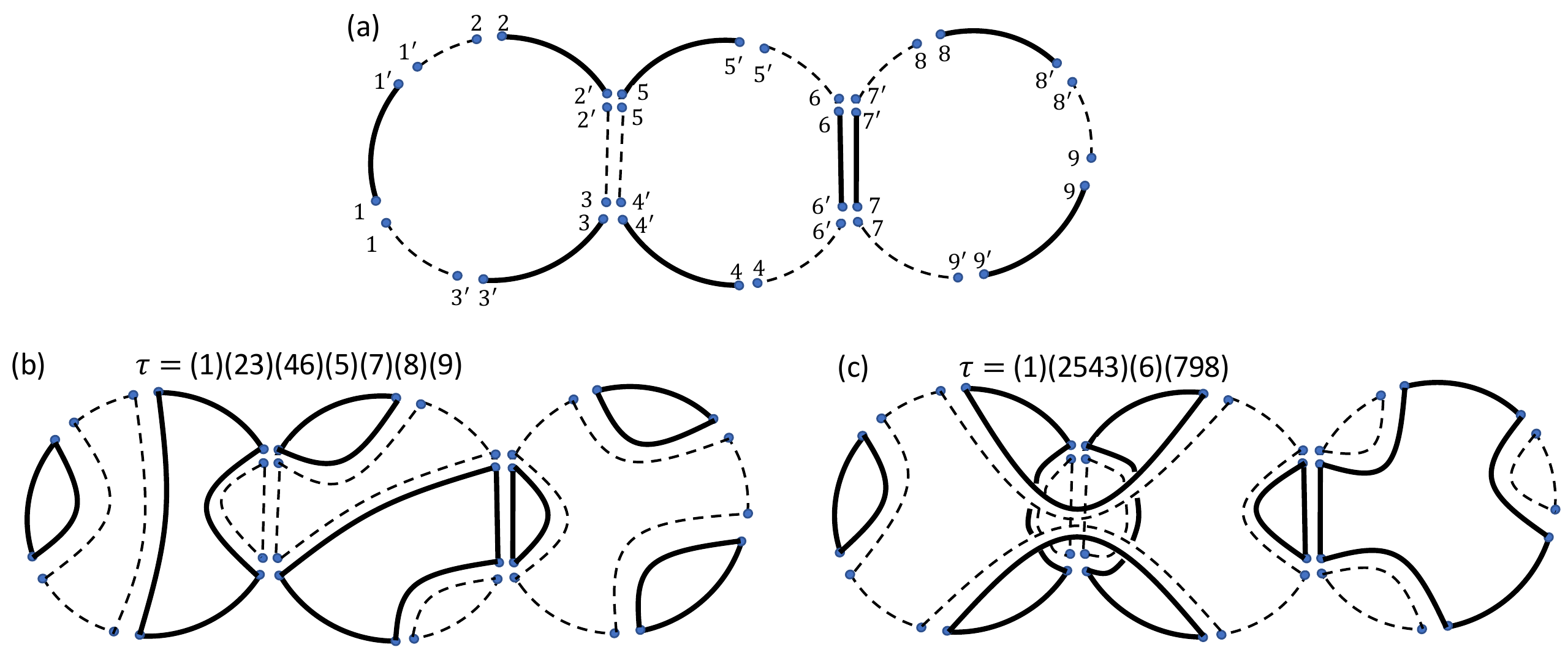}
\caption{Examples of the diagrammatic representation for~\eqref{dm_perm} for $n=3$ and $m=3$. (a) shows the ``future conditions" common to all diagrams for $\le[\le(\sZ_{n}^{(A)}\ri)^m \ri]_{\text{eq app}}$ with $m=3$. (b) and (c) show examples of interior connections for two different choices of $\tau$.}
\label{fig:m_3}
\end{figure}
We then find that in both limits, $d_A\sim d_{\bar{A}}\sim Z_1^{1/2}$ and $d_A\ll d_{\bar A} \sim Z_1$, $\Delta_m$ is suppressed by at least a factor of $Z_1^{-1}$ relative to the leading term in $(\sZ_{n, P})^m$.

\subsection{Matrix elements and correlation functions}

Let us now consider 
\be 
\sT =  \vev{\psi_0 |\sO_1 (t_1) \sO_2  (t_2)  \cdots \sO_n (t_n) |\psi_0} = \Tr (\sO U(t_m) \rho_0 U(t_m)^\da)
\ee
where $\sO \equiv O_1(t_1-t_m) O_2(t_2-t_m) ... O_n(t_n-t_m)$, and $t_m$ is the smallest time among $t_1, ..., t_n$. If $t_m \gg t_s$, then we can apply the equilibrium approximation for $U(t_m)$, with
 $\ket{a} = \ket{\rho_0,e}$,  $\ket{b} = \ket{\sO^\da, e}$ and $n=1$. We find 
 \be \label{gbe}
\vev{\psi_0 |\sO_1 (t_1) \sO_2  (t_2)  \cdots \sO_n (t_n) |\psi_0} \approx \Tr (\sO \rho^{(\rm eq)}) = {1 \ov Z_1} \Tr (\sI_\al \sO) \
\ee
which implies that the expectation value of $\sO$ should be equal to that in the equilibrium density operator $\rho^{(\rm eq)}$. 
For~\eqref{gbe}, equations~\eqref{hok} and~\eqref{hok1} become (below $\eta = (12)$)
 \bega 
 \label{eno0}
 \le(\sT_Q^2\ri)_{\text{eq app}} = {1 \ov Z_1^2} \vev{\sO^\da, e| \sI_\al, \eta} = {1 \ov Z_1^2} \Tr (\sO \sI_\al \sO \sI_{\al}) , \\
  \le(|\sT_Q|^2\ri)_{\text{eq app}} = {1 \ov Z_2^2} \vev{\sO^\da ,e \otimes \sO ,e | \sI_\al, \eta} \vev{\sI_\al, \eta| \rho_0, e} = 
  {1 \ov Z_1^2} \Tr (\sO^\da \sI_\al \sO \sI_\al)   \ . 
 \label{eno}
 \end{gather} 

\subsubsection{Matrix elements of reduced density matrix} 

A special example is $\sO = \ket{j} \bra{i} \otimes \bid_{\bar A}$ where $\ket{i}$ is a basis for a subsystem $A$, in which case 
we have 
\be \label{heh}
(\rho_A)_{ij} = (\rho^{(\rm eq)}_A)_{ij}  + \De_{ij} 
\ee
where $\De_{ij}$ is dropped under the equilibrium approximation. Furthermore, 
\be 
\le(\De_{ij}^2 \ri)_{\text{eq app}} = {1 \ov Z_1^2} \sum_{a,b} (\sI_{\al})_{ia,j b} (\sI_\al)_{ib,ja}, 
\quad 
\le(|\De_{ij}|^2\ri)_{\text{eq app}} = {1 \ov Z_1^2} \sum_{a,b} (\sI_{\al})_{ja,j b} (\sI_\al)_{ib,ia}
\ee
where $a,b$ denote indices for a basis of $\bar A$. 
Now suppose $\sI_\al$ can be factorized  $\sI_\al = \sI_A \otimes \sI_{\bar A}$,  
we then find that 
\be 
\le(\De_{ij}^2 \ri)_{\text{eq app}} =((\rho^{(\rm eq)}_A)_{ij})^2 {\rm Tr}_{\bar A} [(\rho^{(\rm eq)}_{\bar A})^2] , 
\quad 
\le(|\De_{ij}|^2\ri)_{\text{eq app}} = (\rho^{(\rm eq)}_A)_{ii} (\rho^{(\rm eq)}_A)_{jj} {\rm Tr}_{\bar A} [(\rho^{(\rm eq)}_{\bar A})]^2 \ .
\ee

Now as an illustration let us consider the example of Sec.~\ref{sec:bhe} with~\eqref{nel11} and $A = R$. We have 
have 
\be 
 (\rho^{(\rm eq)}_R)_{ij} \approx {1 \ov N} \de_{ij} , \qquad \De_{ij}^2 = {1 \ov N^2}  \de_{ij} {Z_2^{(B)} \ov (Z_1^{(B)})^2}   \qquad
 |\De_{ij}|^2 = {1 \ov N^2}{Z_2^{(B)} \ov (Z_1^{(B)})^2} \ .
 \ee

\subsubsection{Equal-time correlation functions} 

Let us now consider~\eqref{gbe}--\eqref{eno} for some generic operator $\sO$ (which can be a product of observables),  
with $A$ the smallest subsystem containing $\sO$.

 In the case $A \ll \bar A$, from our earlier discussion in Sec.~\ref{sec:univ}, in particular~\eqref{hnel0}-\eqref{hnel},  
the reduced density operator $\rho_A$ of $\rho = U \rho_0 U^\da$ 
is well approximated by $\rho^{(\rm eq)}_A$, and indeed we expect~\eqref{gbe} to hold
as $\vev{\sO (t)}  = \Tr_A (\sO  \rho_A) \approx \Tr_A (\sO (\rho_e)_A) = \Tr (\sO \rho_e)$. 
But going away from the regime $A \ll \bar A$, we expect the approximation~\eqref{gbe} to break down. 
Let us see how this comes about from examining~\eqref{eno0}--\eqref{eno}. 
As illustration we consider a finite dimensional Hilbert space at infinite temperature with $\sI = \bid$. 

For this purpose,  consider 
\be 
\sO = \le(\sum_{i,j} c_{ij} \ket{i} \bra{j}\ri)\otimes\mathbf{1}_{\bar{A}}
\ee
where $\ket{i}$ again denotes a basis for $A$. Suppose $c_{ij}$ are randomly picked numbers with comparable magnitude (which we can take to be $1$) 
but random phases. 
From~\eqref{gbe} 
\be 
\vev{\sO(t)} \approx {1 \ov d_A} \sum_i c_{ii} \sim {1 \ov d_A^\ha} 
\ee
while 
\be 
 \De^2 \equiv \le(|\sT_Q|^2\ri)_{\text{eq app}}  \approx {1 \ov d_{\bar A}} {1 \ov d_A^2} \sum_{ij} |c_{ij}|^2 \sim {1 \ov d_{\bar A}} {1 \ov d_A^2} d_A^2 \sim {1 \ov d_{\bar A}} \ .
\ee
We thus find the approximation~\eqref{gbe} is good if 
\be 
{\De \ov \vev{\sO(t)}} = {d_A^\ha \ov d_{\bar A}^\ha} \ll 1 \quad \iff \quad d_A \ll d_{\bar A}  \ .
\ee

\section{Causality argument} \label{app:causal} 

Here we illustrate the causality argument of Sec.~\ref{sec:uncompact} by modeling the time evolution using discrete steps, that is, by 
a unitary circuit.  More explicitly, we consider an infinite spin chain in 1+1-D with local Hilbert space dimension $q$, and its local time-evolution  $U$ is modeled by a circuit of two-site unitary operators, as shown in Fig.~\ref{fig:circ_a}. We would like to 
show that the entanglement entropies of the region $A$ shown in Fig.~\ref{fig:causal} are independent of the part of the time-evolution operator outside the region $J(A)$ which is in causal contact with $A$.  

 \begin{figure}[!h] 
 \centering
\includegraphics[width=14cm]{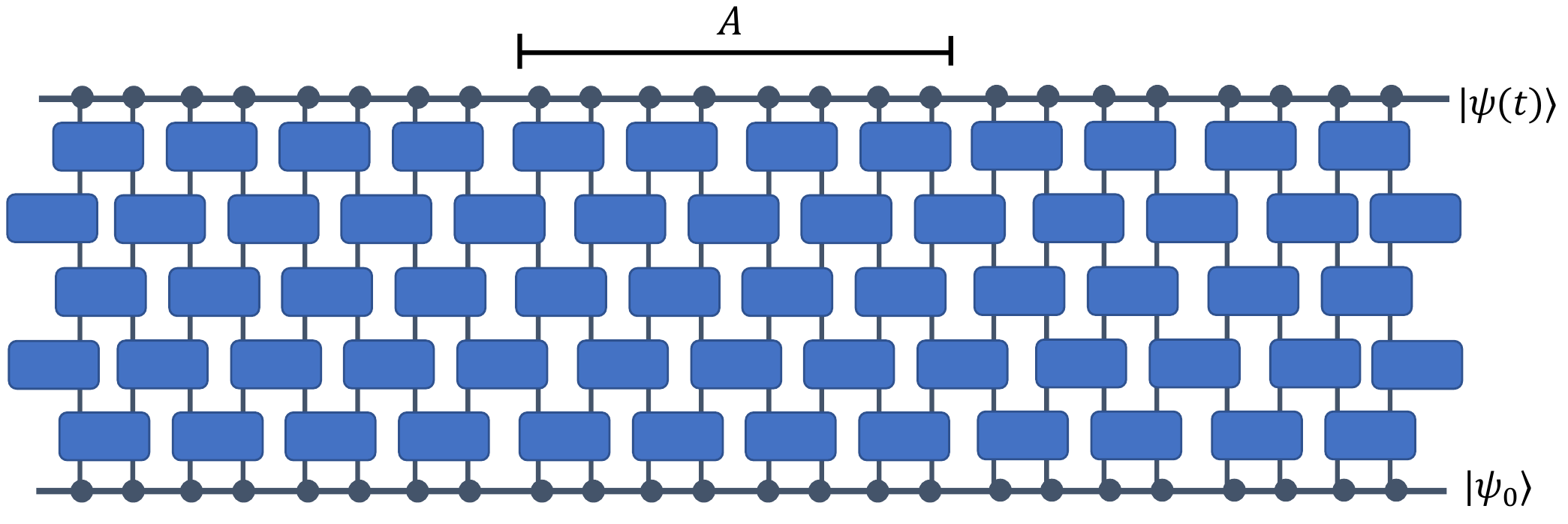}
 \caption{Time-evolution of the initial state $\ket{\psi_0}$ to the state $\ket{\psi(t)} = U \ket{\psi_0}$ at time $t$ by a circuit of local two-site unitary operators in an uncompact system.}
 \label{fig:circ_a}
 \end{figure} 

Recall that for any subsystem $A$, for any unitary operators $U_A$ and $U_{\bar{A}}$ acting only on $A$ and $\bar{A}$ respectively and any state $\rho$,
\begin{align}  
S_n^{(A)}\le((U_A \otimes U_{\bar{A}})\rho (U_A^{\dagger} \otimes U_{\bar{A}}^{\dagger})\ri) &= S_n^{(A)}(\rho) \\
 S_n^{(\bar{A})}\le((U_A \otimes U_{\bar{A}})\rho (U_A^{\dagger} \otimes U_{\bar{A}}^{\dagger})\ri) &= S_n^{(\bar{A})}(\rho)
\end{align} 
for all $n$.  Let us act on the state $\ket{\psi(t)}$ created by the circuit in Fig.~\ref{fig:circ_a} with ${(U^{-})}_{\bar{A}} \otimes \mathbf{1}_A$, where the operator $U^{-}$ acting on $\bar{A}$ is constructed from local unitary operators as shown in Fig.~\ref{fig:circ_b}. $U^{-}$ has $t$ layers of local unitary operators like the original time-evolution operator $U$, and each local unitary in $U^{-}$ is equal to the inverse of the operator in $U$ located at its mirror image with respect to the line at time $t$. 
By construction, the state prepared by the circuit in Fig.~\ref{fig:circ_b} is equal to the state prepared by that of Fig. \ref{fig:circ_c}, where the overall time-evolution operator is of the form $\tilde{U}_{J(A)}\otimes \mathbf{1}_{\overline{J(A)}}$ with $\tilde{U}_{J(A)}$ non-trivial.

 \begin{figure}[!h] 
 \centering
\includegraphics[width=15cm]{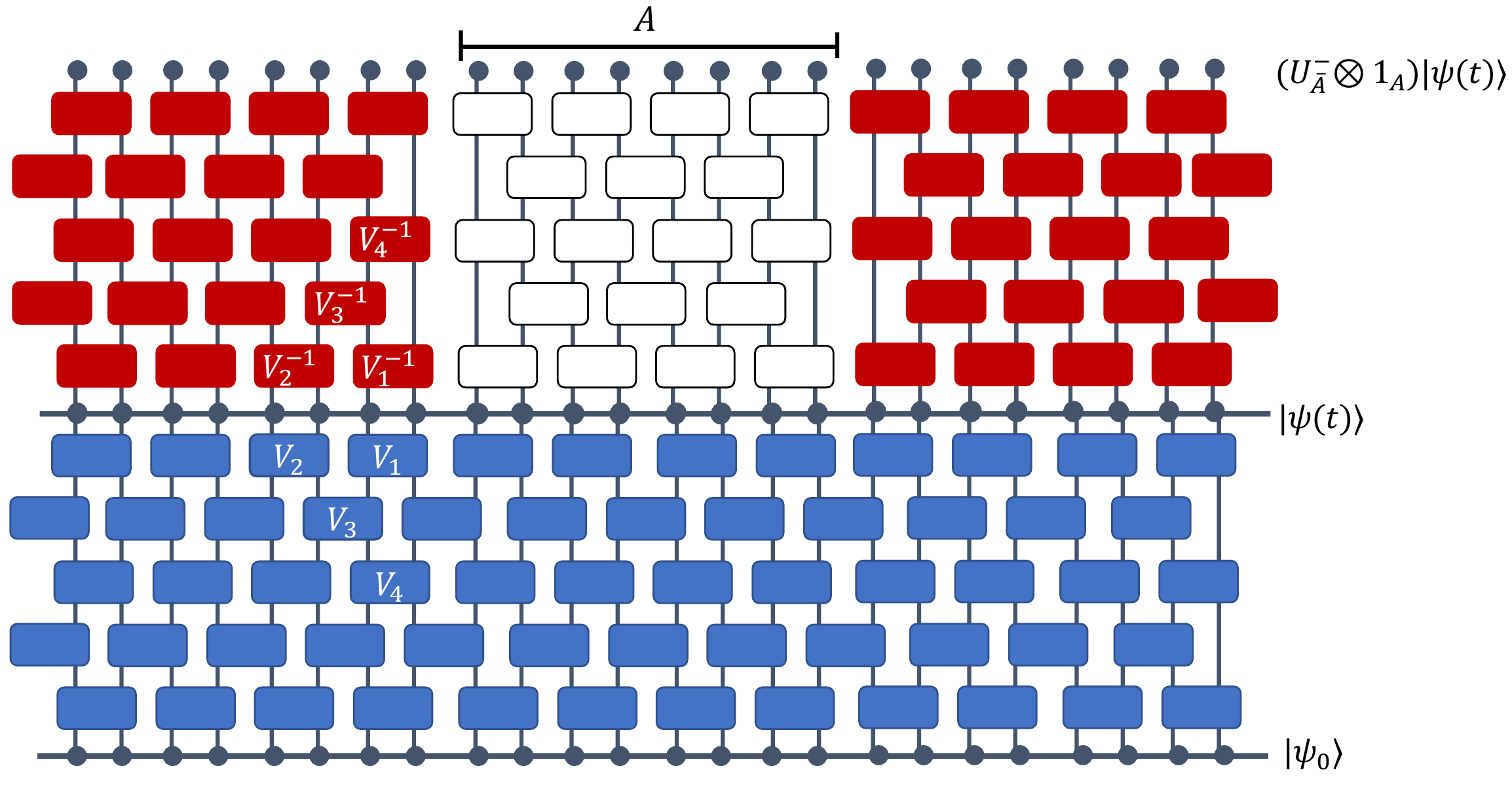}
 \caption{Blue rectangles indicate the local unitaries of the original circuit, red rectangles indicate their inverses used to construct $U^{-}$, and white rectangles indicate the identity operator. We show the action of $U^{-}_{\bar{A}}\otimes \mathbf{1}_A$ on $\ket{\psi(t)}$, and some of the local random unitaries in the definition of $U^{-}$.}
 \label{fig:circ_b}
 \end{figure} 

 \begin{figure}[!h] 
 \centering
 \includegraphics[width=15cm]{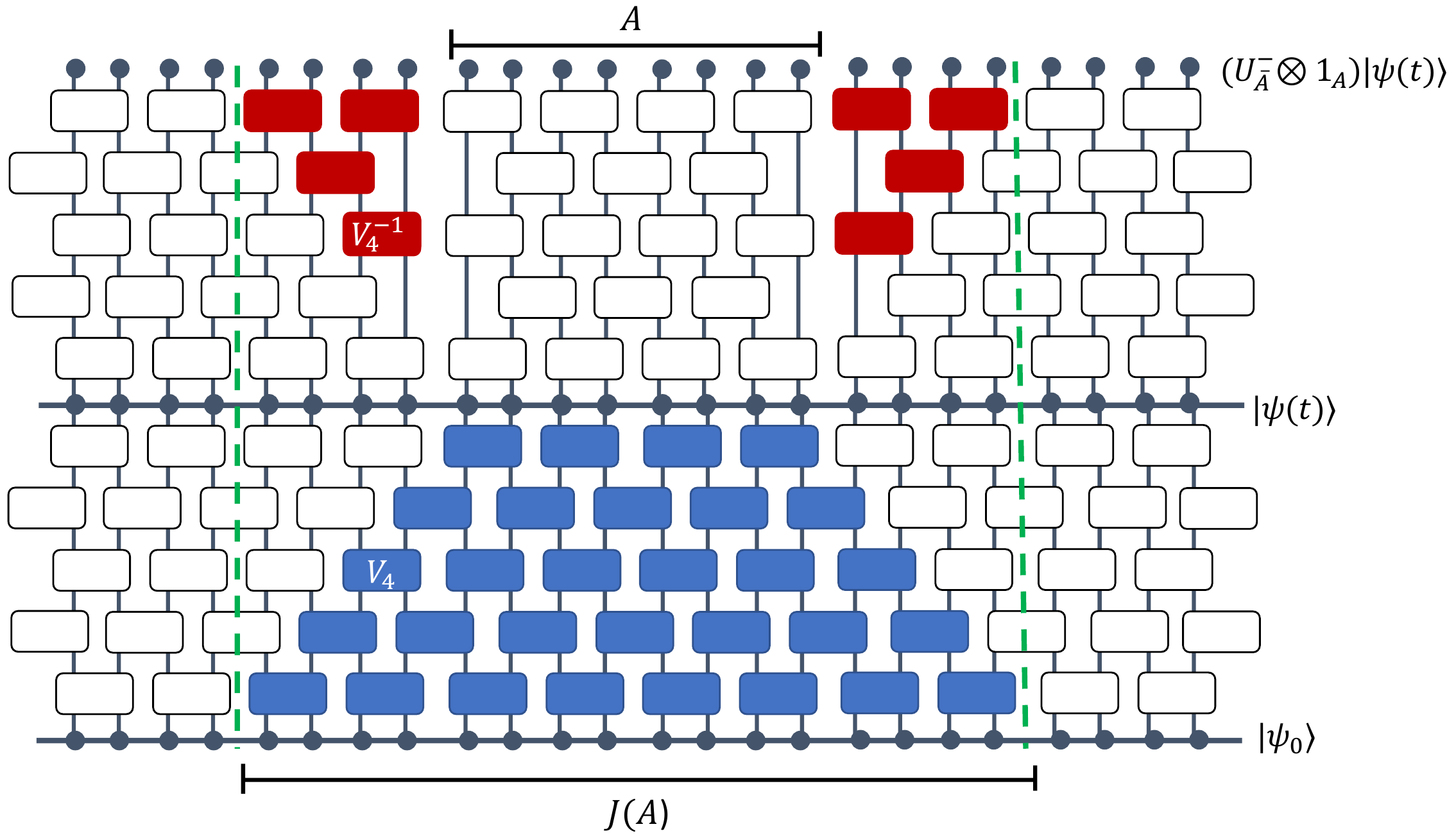}
 \caption{Due to the cancellation of inverses between local operators in $U$ and $U^{-}_{\bar{A}}$, the final state in Fig. \ref{fig:circ_b} can also be produced by the circuit shown above, where the time-evolution operator is of the form $\tilde{U}_{J(A)} \otimes \mathbf{1}_{\overline{J(A)}}$, with $\tilde{U}$ given by the part of the circuit between the green dashed lines.}
 \label{fig:circ_c}
 \end{figure} 

\section{$P_{O, n}^{(S)}$ in random unitary circuits} \label{app:D} 

In this section, we show that the result \eqref{av_tro} obtained from the equilibrium approximation holds 
for sufficiently late times  in local random unitary circuits.
 We use the setup and methods of \cite{nahum1, frank, nahum2}, where the system is a spin chain in (1+1) dimensions with local Hilbert space dimension $q$. The time-evolution operator is constructed from local unitary operators as shown in Fig. \ref{fig:circuit}(a), where each $V$ appearing in the circuit is an independent random unitary matrix acting on two sites, drawn from the Haar measure of $U(q^2)$. We denote the number of sites in the full system and in the subsystem $A$ respectively as $|L|$ and $|A|$. 
Then $d_A = q^{|A|}$ and $d_{\bar{A}}= q^{|L|-|A|}$, with the subsystems $A$ and $\bar{A}=B_1\cup B_2$ as shown in Fig. \ref{fig:circuit}(a).  
We will take both $|A|$ and $|L|-|A|$ to be large, and consider times $t\gg |A|, |\bar{A}|$.

\begin{figure}[!h] 
\centering
\includegraphics[width=15cm]{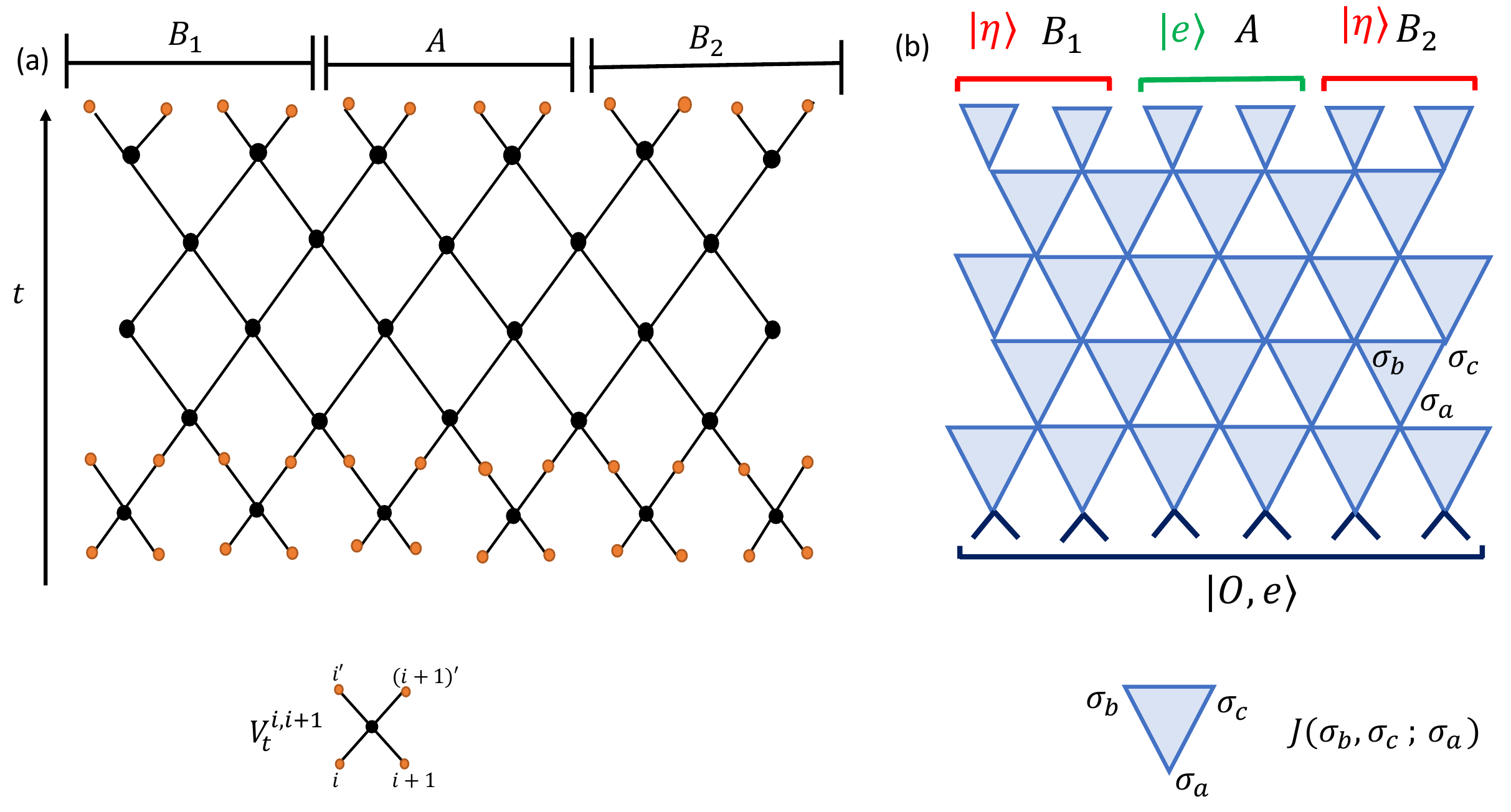}
\caption{(a) shows the tensor network for the time-evolution operator in random unitary circuits and the regions $A$ and $\bar{A}=B_1 \cup B_2$. (b) shows the corresponding triangular lattice for the evaluation of \eqref{2n_rep_1}, which can have a spin labelled by elements of $\mathcal{S}_n$ at each vertex, like $\sigma_{a,b,c}$ shown explicitly in the figure.}
\label{fig:circuit}
\end{figure} 

Now consider 
\be 
\overline{\text{Tr}_{\bar{A}}[(\text{Tr}_A[O(t)])^n]} = \otimes_{m \in \bar{A}} \bra{\eta}_m \otimes_{m \in A} \bra{e}_m ~\overline{(U\otimes U^{\da})^{n}}~\ket{O, e}, \label{2n_rep_1}
\ee
 where we can express $\overline{(U\otimes U^{\dagger})^{n}}$ as a product of $\overline{(V \otimes V^{\dagger})^n}$ for the local unitaries $V$.  As explained in \cite{nahum1, frank, nahum2}, by using the Haar average of $\overline{(V \otimes V^{\da})^n}$, each $V$ in Fig.~\ref{fig:circuit}(a) can be associated with a ``spin" $\sigma$ taking values in the permutation group $\mathcal{S}_n$, and~\eqref{2n_rep_1} becomes the partition function for a classical spin system which has the same lattice structure 
 as the circuit in Fig. \ref{fig:circuit}(a). See Fig. \ref{fig:circuit}(b). The final state and initial state in~\eqref{2n_rep_1} respectively determine the spin configurations at the top and bottom layers of the systems. In the top layer, the $A$ subsystem has $e$ spins while $\bar{A}$ has $\eta$ spins. The bottom layer has a superposition of spin states determined from the matrix elements of operator $O$.  The spin system has interactions among the three spins on each shaded triangle in Fig. \ref{fig:circuit}(b), characterized by 
a factor $J(\sigma_b, \sigma_c; \sigma_a)$. 
The contribution to~\eqref{2n_rep_1} from a given spin configuration 
is given by the product of $J(\sigma_b, \sigma_c; \sigma_a)$ from all 
shaded triangles in Fig. \ref{fig:circuit}(b), and a factor involving $O$ that comes from the bottom layer.

As in the usual Ising model, the partition function for this spin system can be obtained by summing over different 
domain wall configurations. A domain wall between spins $\sig$ and $\tau$ is labeled as $\sig^{-1} \tau$ (See Fig. \ref{fig:general_domain}), and thus there are altogether $n!$ different types of domain walls, one for each element of $\sS_n$. 
We refer to domain walls associated with transpositions of two elements as elementary domain walls. Just like an element of  $\sS_n$ can be decomposed into a product of transpositions, a domain wall can decomposed into compositions of elementary domain walls. 
For example, given that $\eta= (n, ~n-1, ~... ~1)$ can be decomposed as $\eta = (1, 2)~(1, 3)~...  ~(1, n-1)~ (1, n)$, 
an $\eta$-domain wall may be considered as a composite of $n-1$ elementary domain walls associated with these transpositions.

\begin{figure}[!h]
\centering
\includegraphics[width=6cm]{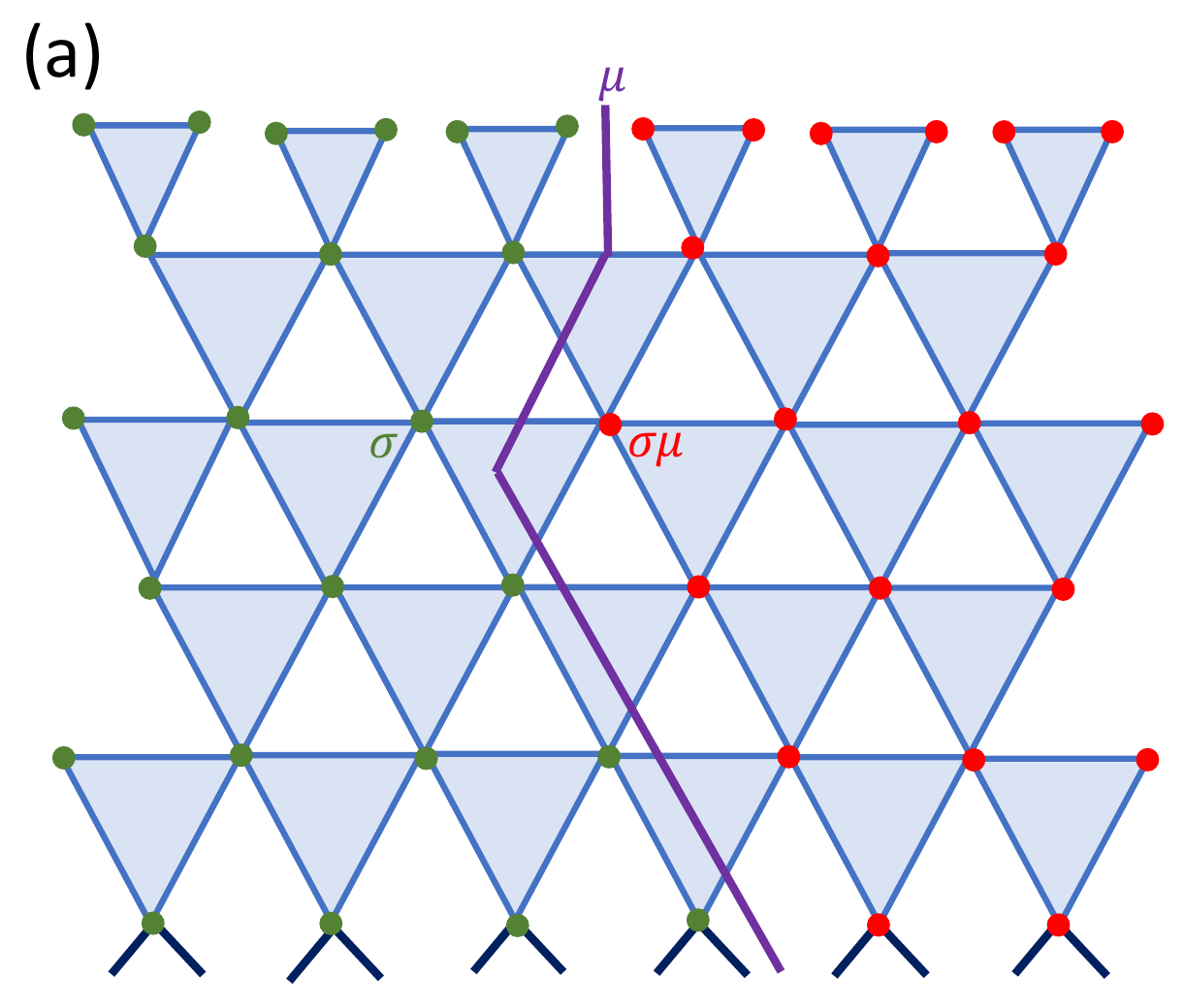}~~\includegraphics[width=8cm]{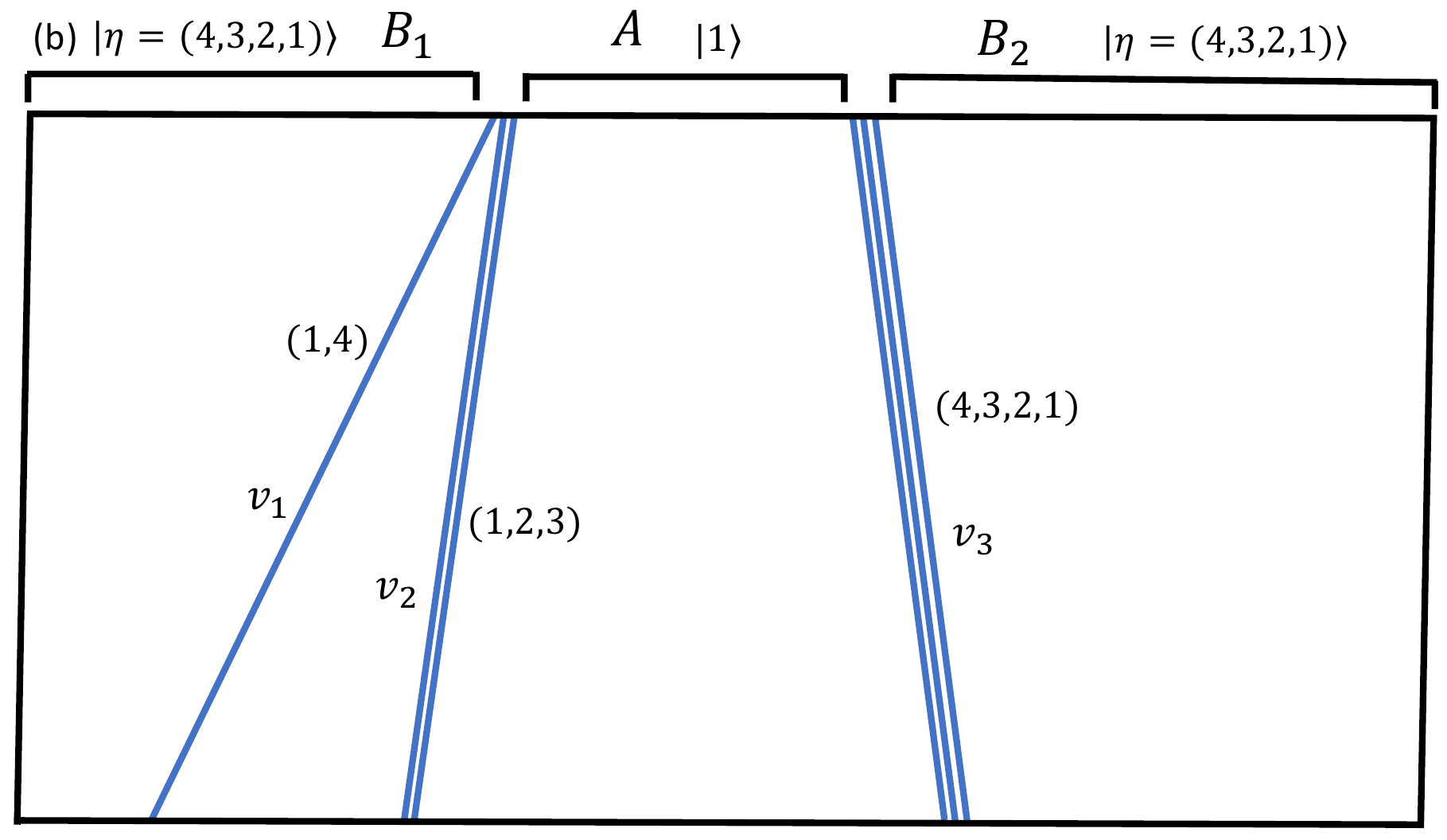}
\caption{A domain wall that separates $\sigma$ spins on the left and $\sigma \mu$ spins on the right is labelled by $\mu$, as shown in (a). For the configuration shown in (b) (where we do not show the details of the lattice), the combined contribution from all domain walls is $q^{-\mathcal{E}^{(2)}(v_1)\, t} ~ q^{-2 \, \mathcal{E}^{(3)}(v_2)\, t} ~q^{-3\, \mathcal{E}^{(4)}(v_3)\, t}$.}  
\label{fig:general_domain}
\end{figure} 

When the difference in position $\Delta x$ and the difference in time $\Delta t$ between the initial and final points are both large for all domain walls relevant for a quantity of interest, there exists a coarse-grained description where we can characterize domain walls by their velocities $\Delta x/ \Delta t$, and collectively take into account the contributions from all detailed configurations that correspond to these velocities \cite{nahum2}. Then a domain wall with velocity $v$ contributes a factor $q^{- \mathcal{E}^{(2)}(v) \De t}$ to the partition function, where $\mathcal{E}^{(2)}(v)$ is independent of the type of the elementary domain wall. Furthermore, if there are  $l-1$ elementary domain walls traveling together at velocity $v$ for time $\Delta t$, we get a factor  $q^{-(l-1)~\mathcal{E}^{(l)}(v) \Delta t}$. With  finite $q$, for general $v$, $\mathcal{E}^{(l)}(v) $ is different from $\mathcal{E}^{(2}(v)$ due to interactions among domain walls (they become equal in the limit $q \to \infty$). 
 It was argued in \cite{membrane} based on the dynamics of entanglement growth in chaotic systems that the conditions 
\be  
 \mathcal{E}^{(l)}(v_B) = v_B,  ~~~~ \mathcal{E}^{(l)}(v)\geq v, ~~~~ {\mathcal{E}^{(l)}}'(v_B)=1 \label{l_conditions}
\ee
should be satisfied for all $l$ in any chaotic system. Here $v_B$ is the butterfly velocity of the system. 
The explicit form of $\mathcal{E}^{(l)}(v)$ in random unitary circuits with finite $q$ is not known for $l>2$, but the condition $\mathcal{E}^{(3)}(v_B) = v_B$ was checked up to next-to-leading order in $1/q$ in \cite{nahum2}.

From the spin configuration indicated at the  top boundary~of Fig. \ref{fig:circuit}(b), we have an $\eta^{-1}$ domain wall at the left edge of $A$, and an $\eta$ domain wall at the right edge. As discussed above, each of these domain walls can be seen as composites of $n-1$ elementary walls  which can in principle travel independently through the lattice. The lower end-points of each of these elementary domain walls in the lattice can be either on the left or right edges, or at the bottom.\footnote{In principle, there can also be cases where $n-1$ elementary domain wall starting at the top boundary ``split" into more than $n-1$ elementary domain walls as they pass through the lattice, but as we explain later, in the limit we are considering, we will not need to take into account such possibilities.}  
 One example of a possible configuration is Fig. \ref{fig:general_domain} (b).

Now suppose $t\gg |A|, |\bar{A}|$. Since the $\mathcal{E}^{(l)}(v)$ are all $O(1)$, in all configurations where any domain walls reach the lower boundary, we get factors exponentially suppressed in $t$ relative to configurations where all domain walls meet in the middle or end at the edges of the system. So at such times, it is sufficient to consider configurations which do not reach the lower boundary. A general example of such a configuration is indicated in Fig.~\ref{fig:gen_config} (a). From~\eqref{l_conditions}, we see that for a pair of domain wall configurations which meet in the middle, the maximal contribution comes from the case where both have velocity $v_B$. Similarly with those ending on the edges. Thus any domain wall which travels to the left or right must have velocity $v_B$ at leading order in our limit. 
For a decomposition of $\eta$  
\be 
\eta= \sigma \times \nu \label{mu_nu}
\ee
where $\sig$ corresponds to the composite domain walls which meet in the middle and $\nu$ to those ending on the edge,  the corresponding domain wall configuration contributes 
\be  
 q^{-|\sigma| \, |A|} q^{-|\nu|\, |\bar{A}|}   = q^{-(n-k(\sigma))\,  |A|} q^{-(n-k(\eta^{-1}\sigma))\, |\bar{A}|}  \ .
 \ee 
In the regime we are interested in, the above expression is maximized when $k (\sig) + k(\eta^{-1} \sig)$ is maximized, that is, they should saturate~\eqref{yeg}. 

 On the lower boundary, $\ket{O,e}$ is attached to a $\sigma^{-1}$ spin at each site, so that we get a factor of $\braket{\sigma^{-1} | O, e}$.   Putting together the contributions from such leading domain wall contributions for different choices of $\sigma$ that saturate \eqref{yeg}, we then obtain the result \eqref{av_tro}. 

\begin{figure}[!h] 
\centering 
\includegraphics[width=15cm]{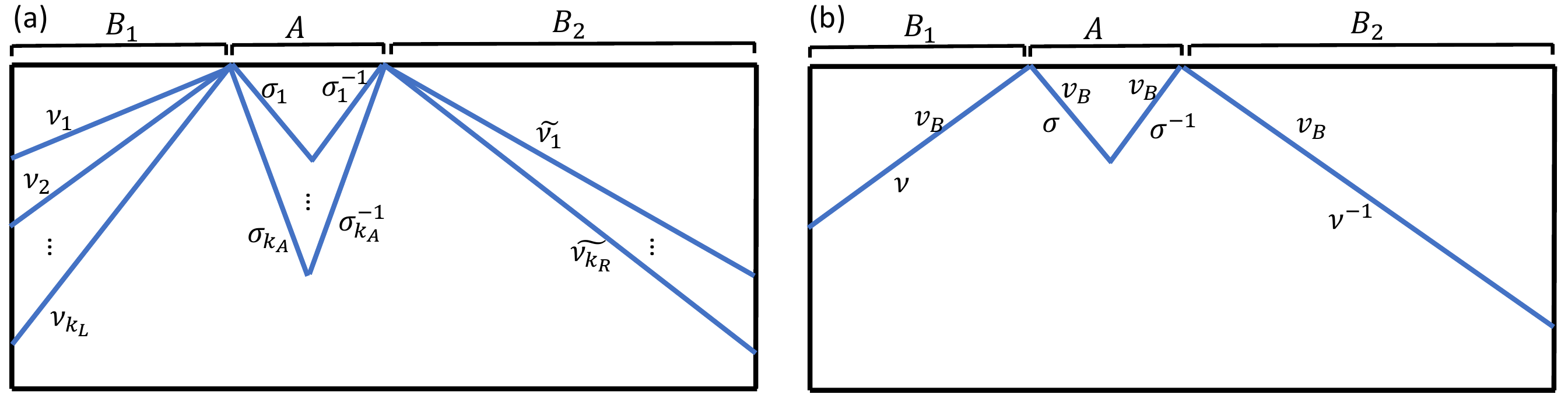}
\caption{(a) shows a general domain wall configuration that can contribute in the scaling limit and at late times. (b) shows the leading contribution in the scaling limit for a particular choice of $\sigma$, where each of the domain walls has velocity $v_B$.}
\label{fig:gen_config}
\end{figure}

In the above derivation, we only made use of the fact that the average over local random unitaries could be expressed in terms of ``spins" associated with permutations, and that there is a membrane tension associated with the domain walls between such spins in the scaling limit, which satisfies the conditions \eqref{l_conditions}.  The discussion of \cite{adam} implies that the above features are also present in the scaling limit in a variety of chaotic systems involving no random averaging, such as the floquet spin chains studied there. We therefore expect that it may be possible to show using the methods of \cite{adam} that the result \eqref{av_tro} holds in the systems considered there.

\end{document}